\documentclass[1p]{elsarticle}
\usepackage[colorlinks]{hyperref} 
\usepackage[latin1]{inputenc}
\usepackage{amsmath}
\usepackage{amsfonts}
\usepackage{amssymb}
\usepackage{graphicx}
\usepackage{mathrsfs}
\usepackage{xcolor}
\usepackage{subfig}

\newcommand\bra[1]{\langle\,#1\,\vert}
\newcommand\ket[1]{\vert\,#1\,\rangle}
\newcommand\braket[2]{\big\langle\,#1\,\big|\,#2\,\big\rangle}
\newcommand\conm[2]{\left[#1,#2\right]}

\let\abs=\envert

\newcommand\bs[1]{{\bf #1}}  
\newcommand\op[1]{{\bf #1}}  
\newcommand\opc[1]{{\cal #1}}

\newcommand\ic{{\rm i}}
\newcommand\un[1]{{\rm\,#1}}

\newcommand\tr[1]{\mathrm{Tr}\left\{#1\right\}}

\newcommand\trs[2]{\mathrm{Tr}_{#1}\!\left\{#2\right\}}
\newcommand\trbs[2]{\mathrm{Tr}_{#1}\!\big\{#2\big\}}

\newcommand\numeq[1]{(\ref{#1})}
\newcommand\eq[1]{Eq.~(\ref{#1})}
\newcommand\eqs[1]{Eqs.~(\ref{#1})}
\newcommand\fig[1]{Fig.~\ref{#1}}

\newcommand\sect[1]{Section~\ref{#1}}
\newcommand\sects[1]{Sections~\ref{#1}}
\newcommand\nsect[1]{\ref{#1}}
\newcommand\apen[1]{\ref{#1}}

\newcommand\hs{{(\!s)}}
\newcommand\hA{{(\!s_{\!A}\!)}}
\newcommand\he{{(\!e)}}
\newcommand\hl{{(\!s_{1}\!)}}
\newcommand\hN{{(\!s_{\!N}\!)}}

\newcommand\hAp{{(\!s_{\!A'}\!)}}
\newcommand\oS{\overline{S}}
\newcommand\cur[1]{\mathfrak{#1}}
\newcommand\csl[1]{\textsl{#1}}

\newcommand\cbb[1]{\mathbb{#1}}

\newcommand\sgn{{\rm sgn}}

\newcommand\mA{m_{\!A}}
\newcommand\mAp{m_{\!A'}}
\newcommand\mAn{m_{\!\bar{A}}}
\newcommand\nA{n_{\!A}}

\newcommand\nAn{n_{\!\bar{A}}}
\newcommand\ml{m_{1}}
\newcommand\mi{m_{i}}
\newcommand\mN{m_{\!N}}
\newcommand\nl{n_{1}}
\newcommand\nii{n_{i}}
\newcommand\nN{n_{\!N}}

\newcommand\Cmn{\big|_{\{\mAn=\nAn\}}}
\newcommand\Cmnb{\bigg|_{\{\mAn=\nAn\}}}
\newcommand\rhoini{\rho_{\mA,\nA}^{\{\!\mAn\!\}}\!(0)}
\newcommand\Drhoini{\Delta\rho_{\mA,\nA}^{\{\!\mAn\!\}}\!(0)}
\newcommand\nuD{\tilde{\nu}}
\newcommand\vs{c_s}

\newcommand\shS{\mathscr{H}_{\cbb{S}}}
\newcommand\shSl{\mathscr{H}_{\cbb{S}_{1}}}
\newcommand\shSN{\mathscr{H}_{\cbb{S}_{N}}}
\newcommand\shSi{\mathscr{H}_{\cbb{S}_{i}}}
\newcommand\shSA{\mathscr{H}_{\cbb{S}_{\!A}}}
\newcommand\shSAn{\mathscr{H}_{\,\overline{\cbb{S}}_{\!A}}}
\newcommand\shE{\mathscr{H}_{\cbb{E}}}

\newcommand\HamS{\op{H}_\op{S}}
\newcommand\HamE{\op{H}_\op{E}}
\newcommand\HamI{\op{H}_\op{I}}

\newcommand\Iec{\mathcal{I}}

\newcommand\FC{\mathscr{C}}
\newcommand\FS{\mathscr{S}}
\newcommand\sB{\triangleleft}

\newcommand\fB{\delta}
\newcommand\eB{\cdot}

\newcommand\insertfig[6]
{\begin{figure}[#1]
    \begin{center}
        \includegraphics[bb = 0 0 #4cm #5cm]{#3.jpg}
    \end{center}
    \caption{#6}
    \label{#2}
\end{figure}}

\begin{document}

\begin{frontmatter}

\title{Adiabatic quantum decoherence in many non-interacting subsystems induced by the coupling with a common boson bath.}
\author{H. H. Segnorile\corref{cor1}}
\ead{segnorile@famaf.unc.edu.ar}
\author{C. E. González}
\ead{ceciliae.gonzalez@unc.edu.ar}
\author{R. C. Zamar}
\ead{ricardo.zamar@unc.edu.ar}
\cortext[cor1]{Corresponding author}
\address{Facultad de Matemática, Astronomía, Física y Computación (FAMAF),
Universidad Nacional de Córdoba,  M. Allende y H. de la Torre - Ciudad Universitaria, X5016LAE - Córdoba, Argentina. \\
Instituto de Física Enrique Gaviola - CONICET - Córdoba, Argentina.}

\begin{abstract}
    This work addresses quantum adiabatic decoherence of many-body spin systems coupled with a boson field in the framework of open quantum systems theory.
    We generalize the traditional spin-boson model by considering a system-environment interaction Hamiltonian that represents a partition of non-interacting subsystems and highlights the collective correlation that appears exclusively due to the coupling with a common environment.
    Remarkably, this simple, exactly solvable model encompasses relevant aspects of a many-body open quantum system and features the subtle quantum effects that arise when the size scales up to a macroscopic level.
    We derive an analytical expression for the time dependence of the density matrix elements (in the preferred basis) without assuming coarse-graining.
    The resulting decoherence function is eigen-selective and is a complex exponential whose exponent has a real part that introduces a decay similar to that in the spin-boson model. On the contrary, the imaginary part depends on the quantum numbers and geometry of the whole partition and does not reflect the system temperature.
    Motivated by decoherence in solid-state NMR, and in search of realistic numerical estimations, we apply the theoretical results to a partition of dipole-coupled spin pairs in contact with a common phonon bath, using typical parameters of hydrated salts. The proposal allows estimating the decoherence time scale in terms of the system physical constants: sound velocity and eigenvalue distribution width. As a significant novelty, the decoherence function phase depends on the eigenvalue distribution throughout the sample. It plays the leading role, overshadowing the mechanism associated with the bath thermal state.
    Finally, we apply the formalism to describe decoherence in the ``magic echo'' NMR reversal experiment. We find that the system-environment correlation explains the origin of irreversibility, and both the decoherence rate value and its dependence on the dipolar frequency, are remarkably similar to the experiment.
\end{abstract}

\end{frontmatter}

\section{Introduction} \label{sec:intro}

Realistic quantum systems can rarely be strictly isolated since they intertwine with some environment, and may be theoretically treated as open quantum systems \cite{BreuerPetruccione, vernon63, Yukalov12}. Accordingly, the system dynamics is represented by the density operator reduced over the environment degrees of freedom.
Physical processes that drive a system to final equilibrium with the environment are, in some cases, very slow in the time scale of the system dynamics. Nevertheless,  even a weak system-environment coupling can rapidly correlate their degrees of freedom, giving rise to irreversible adiabatic processes capable of eroding the  quantum coherence over an early time scale before relaxation has any effect \cite{schlosshauer_19,OxmanVillar_18}.
This phenomenon of adiabatic quantum decoherence is of great importance in quantum information processing technology. It involves a rapid and irreversible loss of available quantum information in many-body systems, induced by coupling with the environment, even when the energy can in practice be conserved \cite{Palma96,Merkli_10}.

From the basic point of view, adiabatic decoherence is a stimulating challenge, as it is a purely quantum process linked to the very foundations of the theory. It arises in fields as varied as quantum measurement \cite{Olli_Zurek_01,Zurek_RevModPhys}, the black hole information problem \cite{Unruh_2017,Unruh_2012},  environment-induced decoherence in condensed matter \cite{MorelloStamp_06}, etc. Therefore, establishing the characteristics that the system-environment coupling must have to stand as the universal vehicle for quantum decoherence appears necessary  \cite{Kuml_06}. In this sense, both the fundamentals and applications of many-body physics require a quantitative calculation of the decoherence rates that allows comparison with the experiment. This work, inscribed within this purpose, aims to derive the decoherence rate of a many-body system coupled with a common boson environment, in terms of the system and environment physical properties.

\subsection{Background} \label{sec:backg}

A widely used model for studying adiabatic decoherence in open quantum systems is a qubit or spin coupled with a boson bath (spin-boson model). It provides an analytic solution that can be studied in different regimes \cite{BreuerPetruccione,Luczka_90}. When the many-body character of the observed system can affect the system-environment correlation, such a single-spin model has to be generalized beyond of simply considering a collection of two level systems. In this work, we introduce the many-body nature in the system-environment  Hamiltonian.
We model the environment by a boson bath with long-range-correlation modes, having the effect of interrelating the different parts of the system through a common field.
	
The NMR experimental setup allows tracking the spin dynamics along different timescales imposed by the system-environment Hamiltonian. The long term spin-lattice relaxation processes that bring the spin system to thermal equilibrium with the environment ($T_1$ processes) are well understood in terms of thermal fluctuations of the environment (or lattice) \cite{abragam61}. The same kind of fluctuations give rise to coherence loss over a similar timescale in liquids and gases, that is, the ``transverse'' $T_2$ process  \cite{abragam61,footnote-dephasing_A}.
In contrast, the description of irreversible spin dynamics in the early time scale  in solids (and liquid crystals) is, at present, not mature.
The fact that the energy transfer involved in spin-lattice relaxation is a slow mechanism compared to the characteristic time scale of the NMR signals led many to associate the early spin dynamics to a genuinely closed system that undergoes a unitary evolution \cite{PRB_Cory-etal,multispBoutis12,SkrebZaripov_99}.
However, there is no physical reason for disregarding the rapid and strong correlation (without energy exchange) between the system and environment degrees of freedom \cite{schlosshauer_19} within this time scale.

NMR experiments can also scrutinize within this time scale.
Notably, there is a set of pulse sequences with the effect of reversing the evolution under the dipole-dipole interaction Hamiltonian (sometimes called ``time-reversal'' sequences) \cite{RhimPinesVaugh_71}.
The refocused signal amplitude attenuates irreversibly for increasing refocusing times, and such decay cannot be ascribed to experimental misadjustments, nor non-reversed terms of the evolution Hamiltonian \cite{SkrebZaripov_99,GzzSegZam11,DomZamSegGzz17,LevsHirsc_PhA2000,SanchezCapellaro_PRL20,Buljubasichetal_JCP15,AlvarezSuter_RevModP16}.
This phenomenon, called ``decoherence'' in the NMR literature for it concerns the decay of the density matrix off-diagonal elements, takes place over a time range much shorter than $T_1$ (but longer than the typical NMR signal decay time).

The proposal addressed in ref.\cite{DomGzzSegZam16} describes the NMR decoherence in a many-body observed system coupled with a quantum bath. In such a proposal, a system of weakly interacting spin pairs is adiabatically coupled with a phonon bath, where the ``flip-flop'' terms in the dipolar Hamiltonian allow the inter-pair correlation growth \cite{Keller_88}.
The system-bath interaction, attributed to small shifts of dipole intra-pair distances due to a common phonon environment, correlates different pairs through their collective dynamics.
This treatment yielded an irreversible decay of the off-diagonal matrix elements (in the interaction Hamiltonian basis). Physically, this dynamics reflects the interplay of two opposite tendencies: the dipole Hamiltonian  promotes the build-up of multiple quantum coherences, while adiabatic decoherence degrades them with an efficiency that increases with  coherence order.
This result allowed estimating an upper bound for the refocused NMR signal attenuation rate \cite{DomZamSegGzz17} that agrees qualitatively with experimental results, suggesting that the spin-phonon coupling provides a relevant decoherence mechanism. The magnitude of this upper bound strongly depends on the number of correlated spins at a given time, which in turn is decidedly driven by the 
flip-flop terms of the dipolar Hamiltonian \cite{footnote-scrambling}. Its calculation would imply solving the exact dynamics of an unmanageable number of interacting particles; however, an estimate was obtained by resorting to reasonable statistical hypotheses based on multiple quantum coherence experiments in crystals with uniformly distributed spins \cite{LevyGleason_92}.

Despite this success, the irreversible attenuation of coherences is a prevalent phenomenon. It is also present in materials where the symmetry and interactions may not ensure the long-range ``quantum connectivity'' needed for the former mechanism to be dominant.
For example, partly deuterated crystals \cite{BaumMunowPines_85}, and hydrated crystals like gypsum (dihydrated calcium sulfate) or POMH (hydrated phosphate oxalate), may present much of these characteristics.
For this reason, we aim to widen the study to another possible decoherence mechanism.

\subsection{Aim and description} \label{sec:descrip}

In this work we focus on decoherence in what we call {\it partitionable} systems; that is, open quantum systems where a partition in equivalent elements can be clearly identified, and where the interactions within each element are much stronger than between different elements. In this way, we explore conditions that stand as an opposite limit of that analyzed in ref.\cite{DomGzzSegZam16}.
That is, the decoherence mechanism that appears exclusively due to coupling the partition elements with a common environment.

The key feature of our proposal is to introduce the many-body character through a system-environment Hamiltonian that accounts for the correlation between the partition elements in spite that they do not interact directly. Keeping the site dependence of the system-environment couplings introduces correlation between elements mediated by the boson modes.	

We first derive the decoherence function of a generic partitionable observed system, and then particularize to a concrete example. We apply  this result
to estimate the decoherence time scale of a (realistic) model which shares many features with real samples of hydrated crystals. In order to contrast the theoretical approach, we analyze the dynamics of the open spin  system in a reversal experiment (magic echo) and derive the corresponding decoherence function. This analysis allows to realize that irreversible decoherence is inherent to the system-environment quantum correlation. Finally, we contrast our analytical and numerical outcomes with previous experimental data.

The work is organized as follows:
In \sect{sec:partition}, we define the Hilbert space of a partitionable observed system and the {\it local} observables.
We write the density operators involved in the local-observable dynamics by reducing both over the environment variables as well as to a single partition element. Auxiliary equations are summarized in \apen{app:reduc}.
\sect{sec:Hamiltonians} defines the Hamiltonians that govern the dynamics of a general partitionable observed system coupled with a common boson bath, in the framework of open quantum systems theory, in the adiabatic case.
In Section \ref{sec:din_dens_red_obs} we follow the strategy of ref.\cite{priv98} to derive the time dependence of the density matrix reduced to the Hilbert space of the observed system.
An eigen-selective decoherence function emerges that attenuates the different matrix elements according to the difference of eigenvalues.
Section \ref{sec:dyn_sigmaA} shows the calculation of the partial trace over the states of all the partition elements but a representative one.
Since the decoherence function is central in our work, we dedicate \apen{app:calc_Smnkl} to show the emergence and properties of the phase factor in the decoherence function.
In \apen{app:coherent_states}, we summarize some aspects of the coherent state formalism used in \apen{app:calc_Int_EC} to calculate the trace over the boson variables.

Motivated by solid-state NMR phenomenology, in \sect{sec:app_pair_phon} we analyze an example that closely approximates the case of weakly interacting spin-pairs in hydrated crystals.
Section \ref{sec:app_sysm_sigma} introduces the pair-phonon model: a system of non-interacting spin-pairs in a solid array, coupled with a phonon bath (spin-flips between pairs are ruled out). The quantization of spin positions that allows identifying the system-environment interaction Hamiltonian is described in \apen{app:quant_intra-pair_dist}.
In \sect{sec:app_sigma} we write the reduced density matrix and the decoherence function for this example.
Then, under several non-restrictive assumptions and using approximations detailed in \apen{app:calc_Deco_func_1Dk} and  \apen{app:dist_eig}, \sect{sec:app_calc_tdeco} explores the plausibility of this proposal by estimating the decoherence time scale yielded by the pair-phonon model.
The result reveals that the dominant decoherence mechanism generates in the interference of phase factors, reflecting the eigenvalue distribution throughout the sample.
In \sect{sec:U_ME} we describe an NMR reversion experiment from the viewpoint of open-quantum systems. Using a similar strategy to  \sect{sec:din_dens_red_obs}, in \sect{app:deco_rev_exp} we obtain the decoherence function under reversion and in \sect{sec:sigma_ME} the corresponding reduced density matrix for the pair-phonon model of \sect{sec:app_sysm_sigma}, where the calculations are detailed in \apen{app:rev_p-p_model}. We use the obtained result to calculate the signal amplitude in the reversion experiment and compare with experimental data in \sect{sec:ME_amp_exp}. The results keep qualitative and quantitative similarity with the experiment.
Finally, in \sect{sec:Disc}, we remark the main results and conclusions of this work.

\section{Dynamics of an observed system coupled with a boson bath} \label{sec:din_sist-boson}

\subsection{Partitionable systems and local operators} \label{sec:partition}
We consider an open quantum system $\cbb{S}$ which can in principle be addressed, prepared, and measured, in contact with an environment or bath $\cbb{E}$ that cannot be observed nor controlled. This work focuses on observable systems $\cbb{S}$, which we call \emph{partitionable} systems, that can be viewed as a partition of equivalent elements where the interactions within each element are much stronger than between different elements.

Let us call $\shS$ and $\shE$ to the Hilbert spaces of the system and environment, respectively.  A basis for the compound system can be written as
\begin{equation}\label{eq:eigsta_full}
\left\{ \ket{m,e} \equiv \ket{m}\otimes\ket{e} \right\},
\end{equation}
where $\{\ket{m}\} \in \shS$ and $\{\ket{e}\} \in \shE$.\\
Since only $\cbb{S}$ is accessible to measurement, the relevant observables have the form $\op{O} = \op{O}^\hs \otimes \op{1}^\he$, where the  superscripts indicate the space ($\shS$ or $\shE$) where each operator acts on.
Hereafter, if not otherwise stated, we assume that both $\cbb{S}$ and $\cbb{E}$ are characterized by a discrete (though possibly infinite) set of states.

The expectation value of $\op{O}$ on a state defined by the density operator $\rho(t)$ of the compound system, can be written in terms of $\rho^\hs(t)$, the density operator reduced over the environment variables (see \apen{app:reduc_E})
\begin{equation}\label{eq:valexp_sist}
\langle \op{O} \rangle(t) = \tr{\op{O}\,\rho(t)} = \trs{S}{\op{O}^\hs \rho^\hs (t)},
\end{equation}
with
\begin{equation}\label{eq:rho_sist_red}
\rho^\hs(t) \equiv \sum_{e}\bra{e}\rho(t)\ket{e} = \trs{E}{\rho(t)},
\end{equation}
where $\trs{S}{\cdot}$ and $\trs{E}{\cdot}$ stand for the trace over the $\cbb{S}$ and the $\cbb{E}$ variables, respectively.

We are interested in describing observed systems which admit a partition on $N$ equivalent elements. In consonance with this attribute, we now consider a kind of observables on $\shS$,  which may be called ``local observables'', having the form
\begin{equation}\label{eq:obs_partS}
\op{O}^\hs = \sum_A \op{O}_A^\hs =  \sum_A \op{1}^\hl\otimes\cdots\otimes \op{O}^\hA \otimes\cdots\otimes\op{1}^\hN,
\end{equation}
where $A=1,\cdots,N$ labels the partition elements, and the superscripts $(s_i)$ indicate the subspace $\shSA$ where each operator belongs.

Accordingly, the observed Hilbert space is a product of the $N$ subspaces into which $\shS$ is partitioned
\begin{equation} \label{eq:Hilbert_A_negA}
\shS = \shSl\otimes\cdots\otimes\shSA\otimes\cdots\otimes\shSN.
\end{equation}
For convenience, we define the complementary Hilbert space
\begin{equation} \label{eq:Hilbert_negA}
\shSAn \equiv \shSl\otimes\cdots\otimes\shSi\otimes\cdots\otimes\shSN,\;\forall\, i \neq A,
\end{equation}
which involves all the $N - 1$ partition elements except $A$.

Experimental techniques frequently deal with observables like \numeq{eq:obs_partS} which represent the same magnitude on each partition element, that is $\op{O}^\hl=\cdots=\op{O}^\hA=\cdots=\op{O}^\hN$. An example of a local observable is the transverse magnetization in NMR experiments.

Let
\begin{equation} \label{eq:eigsta_partS}
\begin{split}
\{\ket{m} &\equiv \ket{\ml \cdots \mA \cdots \mN}\\
&\equiv \ket{\ml}\otimes\cdots\otimes\ket{\mA}\otimes\cdots\otimes\ket{\mN} \},
\end{split}
\end{equation}
be a product basis of $\shS$,  where eigenvectors $\{\ket{\mA}\}$ span $\shSA$.
Using this basis, it is simple to see (\apen{app:reduc_complem}) that expectation values of local observables like \numeq{eq:obs_partS} are then a sum over the partition elements $A$ of the individual expectation values
\begin{equation}\label{eq:redu1}
\begin{split}
\langle \op{O}^\hs \rangle(t) = \sum_A \langle \op{O}_A^\hs\rangle(t)
= \sum_A \trs{S}{\op{O}_A^\hs\;\rho^\hs(t)} = \sum_A \trs{S_A}{\op{O}^\hA\;\sigma^\hA(t)},
\end{split}
\end{equation}
where we define the density operator $\sigma^\hA(t)$ reduced to a single partition element as
\begin{equation}\label{eq:rho_sist_red_SA}
\sigma^\hA(t) \equiv \trs{\oS_A}{\rho^\hs(t)},
\end{equation}
with $\trs{\oS_A}{\cdot}$ the trace over the complementary space $\shSAn$.
Besides, we call $\sigma_{\mA,\nA}(t)$ to its matrix elements and write
\begin{equation}\label{eq:rho_sist_red_SA_elem}
\sigma_{\mA,\nA}(t) \equiv \sum_{\{\!\mAn\!\}} \rho_{m,n}(t)\Cmn,
\end{equation}
where the symbol $\Cmn$ indicates that the matrix elements it affects have the same indices within the complementary subspace $\shSAn$, while the numbers $\mA$ and $\nA$ in $\shSA$ have no restriction. We will frequently use this notation in the rest of this work whenever sums involving the delta function $\delta_{\{\mAn,\nAn\}}$ appear (see \eq{eq:redu3} in \apen{app:reduc_complem}).

It is worth to mention that the ``condensed'' density operator $\sigma^\hA$ is the result of tracing over both the bath variables \emph{and} over $\shSAn$. It contains all the relevant information needed for calculating expectation values of local operators of the form \numeq{eq:obs_partS}. As indicated in \eqs{eq:valexp_sist} and \numeq{eq:redu1} it is formally equivalent to use either the complete or the condensed density operators to calculate expectation values dynamics; however, the second choice allows an explicit reference to the symmetry of the particular problem.

\sect{sec:dyn_sigmaA} concerns the evolution of the doubly reduced density operator \numeq{eq:rho_sist_red_SA} imposed by the coupling with an environment. Before that, let us define the general form of the Hamiltonians of our system of interest.

\subsection{Model Hamiltonians} \label{sec:Hamiltonians}

 We aim to describe the decoherence  of a spin system, induced by the coupling  to a common boson bath. The complete Hamiltonian has the general form
\begin{equation}\label{eq:openQS}
\op{H}=\HamS + \HamI + \HamE,
\end{equation}
and the characteristic features of the system, environment, and system-environment coupling  we are interested in are:
\begin{description}
	\item [system Hamiltoninan $\HamS$] Let us assume that the observed system admits a partition, as declared above Eqs.\numeq{eq:eigsta_full} and \numeq{eq:Hilbert_A_negA}.
        Particularly, we consider the case in which the partition elements do not interact directly with each other, but through a common environment represented by the boson bath. Accordingly, we write the system Hamiltonian as the sum of contributions that account for the interaction within each partition element
        \begin{equation}\label{eq:Ham_sist_obs_ad}
        \HamS^\hs \equiv \sum_A \op{H}_{\op{S},A}^\hs .
        \end{equation}
        Since the partition elements are assumed equivalent, we can write
        \begin{equation}\label{eq:Ham_sist_obs_SA_sep}
        \op{H}_{\op{S},A}^\hs = \op{1}^\hl  \otimes \cdots \otimes \HamS^\hA  \otimes \cdots \otimes\op{1}^\hN ,
        \end{equation}
        where
        $\HamS^\hA =\HamS^\hAp,\;\forall\,A,A'$. \\
        Formally, the complete system Hamiltonian appearing in \numeq{eq:openQS} is $ \HamS \equiv \HamS^\hs\otimes \op{1}^\he$.
	
	\item[environment Hamiltonian $\HamE$] We represent the environment as a sum of uncoupled harmonic oscillators with frequencies  $\omega_{\bf{k}}$, where $\bf{k}$ stands for wavenumber vectors $\vec{k}$ and normal mode $l$
        \begin{equation}\label{eq:Ham_bos}
        \HamE \equiv  \op{1}^\hs\otimes \sum_{\bf k}
        \left[ \omega\,\bs{b}^{\dagger}\;\bs{b}\right]_{\bf{k}}^\he,
        \end{equation}
        with the symbol  $\left[\cdot\right]_{\bf{k}}$ indicating that all the operators and constants inside the brackets have the label $\vec{k},l$, and the index $\he$ emphasizes that the boson creation and annihilation operators $\bs{b}^{\dagger}$, $\bs{b}$ refer to the environment degrees of freedom. These operators satisfy the usual commutation relations
        \begin{subequations}\label{eq:rel_conm_b}
            \begin{equation}\label{eq:rel_conm_b-bd}
            \left[ \bs{b}_{\bf{k}}, \bs{b}_{\bs{k'}}^{\dagger}\right] ^\he = \delta_{\bs{k},\bs{k'}},
            \end{equation}
            \begin{equation}\label{eq:rel_conm_b-b}
            \left[ \bs{b}_{\bf{k}}, \bs{b}_{\bs{k'}}\right]^\he = 0, \quad
            \left[ \bs{b}_{\bf{k}}^{\dagger}, \bs{b}_{\bs{k'}}^{\dagger}\right]^\he = 0.
            \end{equation}
            \end{subequations}
        It is convenient to write the environment Hamiltonian \numeq{eq:Ham_bos} as
        \begin{equation}\label{eq:Ham_bos_M}
        \HamE =  \op{1}^\hs\otimes \sum_{\bf{k}} \op{M}_{\bf{k}}^\he ,
        \end{equation}
        where we defined $\op{M} \equiv \omega\,\op{b}^{\dagger}\;\op{b}$.
	
	\item [system-environment interaction $\HamI$] We define an interaction Hamiltonian that accounts for the coupling of the spin system to the boson bath, where each partition element interacts individually with the common environment. The proposed Hamiltonian has the form
        \begin{equation}\label{eq:Ham_int_ad}
        \HamI \equiv \sum_A \op{H}_{\op{I},A}^\hs\otimes\sum_{\bf{k}}
        \left[g^{A*}\,\op{b} + g^A\,\op{b}^{\dagger}\right]^\he_{\bf{k}}
        \end{equation}
        where the Hermitian spin operators $\op{H}_{\op{I},A}^\hs$ act on $\shS$, and the system-environment coupling strength is represented by the complex coefficients $g_{\bf{k}}^{A}$.
        We assume that the spin operators $\op{H}_{\op{I},A}^\hs$ are ``local'' in the sense of \eq{eq:obs_partS}, that is
        \begin{equation}\label{eq:OpLambda_SA_sep}
        \op{H}_{\op{I},A}^\hs \equiv \op{1}^\hl\otimes\cdots\otimes\HamI^\hA\otimes\cdots\otimes\op{1}^\hN,
        \end{equation}
        which also implies that
        \begin{equation}\label{eq:conm_Lambda_A}
        \conm{\op{H}_{\op{I},A}\,}{\op{H}_{\op{I},A'}}^\hs = 0,\;\forall\,A,A'.
        \end{equation}

        We find it convenient to define a global spin operator
        \begin{equation}\label{eq:OpLambda_S}
        \op{\Lambda}_{\bf{k}}^\hs \equiv \sum_A \op{H}_{\op{I},A}^\hs\;g_{\bf{k}}^{A},
        \end{equation}
        that allows writing
        \begin{equation}\label{eq:Ham_int_ad_reesc}
        \HamI \equiv \sum_{\bf{k}}\big[\op{\Lambda}^{\dagger\hs}\otimes\op{b}^\he + \op{\Lambda}^\hs\otimes\op{b}^{\dagger\he}\big]_{\bf{k}}.
        \end{equation}
        This form makes it evident that the interaction Hamiltonian of our model differs from separable expressions of the type
        \begin{equation} \label{eq:separableHI}
        \op{\Lambda}^\hs\otimes\sum_{\bf{k}}\left(g^*\,\op{b} + g\,\op{b}^{\dagger}\right)_{\bf{k}}^\he,
        \end{equation}
        that are frequently studied in the literature \cite{priv98}.
        It should be stressed that $\HamI$ in \eqs{eq:Ham_int_ad} or  \numeq{eq:Ham_int_ad_reesc} represents non-interacting partition elements that couple independently with the environment, even so, they become correlated through the common bath. Such a form emerges, when each element of the partition couples with the bosons differently, for example by depending on the phase of collective vibrations, as in the example of \sect{sec:app_pair_phon}.
\end{description}

The physical model of non-interacting partition elements represented by these Hamiltonians allows bringing to the forefront how decoherence arises when the only correlation among partition elements comes from their coupling with the common environment. However, the present approach can be extended to include weak interaction between the partition elements.

Up to this point, we have not imposed any restrictions on the form of $\HamS$ and $\HamI$. Let us now focus the analysis on Hamiltonians that satisfy the adiabatic condition, characterized by the following relationships
\begin{equation}\label{eq:cond_adiab}
\conm{\HamS}{\HamE} = 0,\;\; \conm{\HamI}{\HamS} = 0,\;\; \conm{\HamI}{\HamE} \neq 0.
\end{equation}
Consequently, the observed system Hamiltonian commutes with the total Hamiltonian, $\conm{\op{H}}{\HamS} = 0$, which means that the mean value of the energy $\langle \HamS \rangle$ is time-independent; thus, the system-environment coupling does not involve energy exchange. Though the adiabatic condition may seem restrictive, it can be seen  that in a variety of real cases \cite{DomGzzSegZam16,SegZam11} the system dynamics can be safely represented by the part of the interaction Hamiltonian that satisfies \numeq{eq:cond_adiab} within the earlier timescale, where the relaxation effects are negligible.

Having defined an interaction Hamiltonian $\HamI$ where the spin part is a local operator, the adiabatic condition  \numeq{eq:cond_adiab} also implies
$ [\op{H}_{\op{I},A}^\hs,\HamS^\hs]=0, \: \forall A$, and thus $[\HamI^\hA,\HamS^\hA]=0$.
Then, the common basis can be written as a product of eigenstates of the single partition elements, like in \eq{eq:eigsta_partS}, by assuming that  $\{\ket{\mA}\}$ are the common eigenbasis of $\HamS^\hA$ and $\HamI^\hA$. Hence,
\begin{equation}\label{eq:Ham_sys_eigen}
\HamS^\hs\ket{m} = E_m\ket{m},
\end{equation}
and particulary
\begin{subequations}
\begin{equation}\label{eq:Ham_sys_A_eigen}
\HamS^\hA\ket{\mA} = E_{\mA}\ket{\mA},
\end{equation}
\begin{equation}\label{eq:Ham_int_A_eigen}
\HamI^\hA\ket{\mA} = \lambda_{\mA}\ket{\mA},
\end{equation}
\end{subequations}
with $E_{\mA}$ and $\lambda_{\mA}$ the corresponding eigenvalues, and the energy $E_m$ of the whole observed system is
\begin{equation}\label{eq:HS_eigen_des}
E_m = \sum_A E_{\mA}.
\end{equation}

The action of the global spin operator, defined in \numeq{eq:OpLambda_S}, on this basis is
\begin{equation}\label{eq:Lambda_S_eigen}
\op{\Lambda}_{\bf{k}}^\hs\ket{m} = \lambda_{m}^{\bf{k}}\ket{m},
\end{equation}
where we defined
\begin{equation}\label{eq:Lambda_S_HS_eigen}
\lambda_{m}^{\bf{k}} \equiv \sum_A \lambda_{\mA}\,g_{\bf{k}}^{A}.
\end{equation}
Notice that the eigenvalues \numeq{eq:Lambda_S_HS_eigen} may not be real (if for example the coefficients $g_{\bf{k}}^{A} \in \cbb{C}$ as happens in the example of Section \ref{sec:app_pair_phon}) and the spin operators $\op{\Lambda}_{\bf{k}}^\hs$ may not be Hermitian.

Since the Hamiltonian $\op{H}$ of the whole system is time independent, the evolution operator is $\op{U}(t) \equiv e^{-\ic\op{H}\,t}$, and under the adiabatic condition it can be factorized as
\begin{equation}\label{eq:op_evol_full_ad}
\op{U}(t) = e^{-\ic\HamS\,t}\,e^{-\ic\left(\HamI + \HamE\right)\,t},
\end{equation}
 where the Hamiltonians are written in units of $\hbar$, thus the eigenvalues \numeq{eq:HS_eigen_des} are expressed in frequency units.\\
The action of the operator \numeq{eq:op_evol_full_ad} on the spin eigenstates $\ket{m}$ yields
\begin{equation}\label{eq:OpEvol_eigen}
\op{U}(t)\ket{m} = e^{-\ic E_m\,t}\,\prod_{\bf{k}}e^{-\ic\left(\op{M} + \op{J}_m\right)_{\bf{k}}^\he t}\ket{m},
\end{equation}
where we defined an operator
\[\op{J}_{m \bf k}^\he= \lambda^{*\bf k}_m \:\op{b}^\he_{\bf{k}} + \lambda^{\bf k}_m \: \op{b}^{\dagger\he}_{\bf{k}},\]
which acts on $\shE$.\\
Notice in \eq{eq:OpEvol_eigen} that operators with different index $\bf{k}$ commute, i.e.
\[\conm{\op{M}_{\bf{k}} + \op{J}_{m \bf{k}}}{\op{M}_{\bf{k}'} + \op{J}_{m \bf{k}'}}^\he = 0, \; \forall\,\bf{k},\bf{k}',\]
and also that $\op{J}_{m \bf{k}}^\he$ involves the eigenvalues \numeq{eq:Ham_int_A_eigen} associated with the states of the observed system.

\subsection{Dynamics of the density operator reduced over the environment} \label{sec:din_dens_red_obs}

In this section we calculate the density operator reduced to the observed Hilbert space, defined as the partial trace
$\rho^\hs(t) \equiv \trs{E}{\rho(t)}$ (see \eq{eq:rho_sist_red}),
and its time evolution. In the time evolution operator of \eq{eq:op_evol_full_ad} we use the Hamiltonians defined in \sect{sec:Hamiltonians}, which satisfy the adiabatic condition stated in \eq{eq:cond_adiab}.
This derivation follows the strategy of ref.\cite{priv98} and generalizes that presented in \cite{DomGzzSegZam16}, the deep physical meaning of some steps merit displaying details of the calculation.

Let us assume a separable initial state
\begin{equation}\label{eq:rho0_sep}
\rho(0) = \rho^\hs(0)\otimes\rho_{eq}^\he,
\end{equation}
where $\rho^\hs(0)$ is an arbitrary initial state of the observed system,
and the environment state corresponds to the boson bath at thermal equilibrium,
\begin{equation}\label{eq:rho0_Beq}
\rho_{eq}^\he \equiv \frac{e^{-\beta\,\HamE^\he}}{\trs{E}{e^{-\beta\,\HamE^\he}}}
\equiv \prod_{\bf{k}}\op{\Theta}_{\bf{k}}^\he,
\end{equation}
with
$\beta \equiv \hbar/\left(K_B\,T\right)$, $K_B$ the Boltzmann constant, $T$ the absolute bath temperature and
\begin{equation*}
\op{\Theta}_{\bf{k}}^\he \equiv \frac{e^{-\beta\,\op{M}_{\bf{k}}^\he}}{Z_{\bf{k}}}, \qquad Z_{\bf{k}} \equiv \trs{E}{e^{-\beta\,\op{M}_{\bf{k}}^\he}},
\end{equation*}
where $Z_{\bf{k}}$ is the partition function.
The time evolved density matrix elements (reduced over the environment), in the eigenbasis \numeq{eq:eigsta_partS}, are
\begin{equation}\label{eq:rho_sist_red_ad}
\rho_{m,n}(t) \equiv \trs{E}{\bra{m} \op{U}(t)\,\rho(0)\,\op{U}^{\dagger}(t) \ket{n}}
= \rho_{m,n}(0)\;e^{-\ic(E_m - E_n)\,t}\,\prod_{\bf{k}} S_{m,n}^{\bf{k}}(t),
\end{equation}
where we used
\begin{equation}\label{eq:rho_cerrado}
\bra{m} e^{-\ic\HamS^\hs t}\,\rho^\hs(0)\,e^{\,\ic\HamS^\hs t} \ket{n} = \rho_{m,n}(0)\,
e^{-\ic(E_m - E_n)\,t},
\end{equation}
and defined
\begin{equation}\label{eq:fundeco}
S_{m,n}^{\bf{k}}(t) \equiv \trs{E}{e^{-\ic\left(\op{M} + \op{J}_m\right)\,t}\,\op{\Theta}\,
	e^{\,\ic\left(\op{M} + \op{J}_n\right)\,t}}_{\bf{k}}^\he.
\end{equation}
In this way, two factors with essentially different quality contribute to the time evolution of each matrix element in \numeq{eq:rho_sist_red_ad}: \eq{eq:rho_cerrado} is related to the evolution under $\HamS$ only; while \eq{eq:fundeco} stands for the dynamics imposed by the coupling with the boson bath.

The trace over the environment variables is calculated in \apen{app:calc_Smnkl} where we show that the functions $S_{m,n}^{\bf{k}}(t)$ can be written as
\begin{equation}\label{eq:Func_Smnkl_des}
S_{m,n}^{\bf{k}}(t) = e^{-\Gamma_{m,n}^{\bf{k}}(t)}\,e^{-\ic\Upsilon_{m,n}^{\bf{k}}(t)},
\end{equation}
with
\begin{subequations}\label{eq:Func_Gmnkl_Umnkl}
\begin{equation}\label{eq:Func_Gmnkl}
    \Gamma_{m,n}^{\bf{k}}(t) \equiv \bigg\{\frac{2\abs{\lambda_{m}-\lambda_{n}}^2}{\omega^2}
    \sin^2\left(\omega\,t/2\right) \coth\left(\beta\,\omega/2\right)\bigg\}_{\bf{k}},
\end{equation}
\begin{equation}\label{eq:Func_Umnkl}
\begin{split}
    \Upsilon_{m,n}^{\bf{k}}(t)
    &\equiv \bigg\{\frac{\left(\lambda_m-\lambda_n\right)
    	\left(\lambda_m+\lambda_n\right)^*}{\omega^2}\left[\sin\left(\omega\,t\right)-\omega\,t\right]\\
    &\quad-\frac{2\,\Im\left\{\lambda_m\lambda_n^*\right\}}{\omega^2}\bigg[\left[1-\cos(\omega\,t)\right]\;
    +\ic\left[\sin\left(\omega\,t\right)-\omega\,t\right]\bigg]\bigg\}_{\bf{k}},
\end{split}
\end{equation}
\end{subequations}
where $\Im\left\{z\right\}$ is the imaginary part of $z$. Notice that in \numeq{eq:Func_Gmnkl_Umnkl}, the symbol $\left\{\cdot\right\}_{\bf{k}}$ affects the frequency and also the eigenvalues $\lambda_m$ ($\lambda_m^{\bf{k}}$).

Finally, by replacing Eqs.\numeq{eq:Func_Gmnkl_Umnkl} in \eq{eq:rho_sist_red_ad}, we can write a condensed expression
\begin{equation}\label{eq:rho_sist_red_ad_Smn}
\rho_{m,n}(t) = \rho_{m,n}(0)\;e^{-\ic\left(E_m - E_n\right)\,t}\,S_{m,n}(t),
\end{equation}
where we define
\begin{equation}\label{eq:Func_Smn}
S_{m,n}(t) \equiv \prod_{\bf{k}} S_{m,n}^{\bf{k}}(t)
= e^{-\Gamma_{m,n}(t)}\,e^{-\ic\Upsilon_{m,n}(t)},
\end{equation}
and
\begin{equation}\label{eq:def_Gmn_Umn}
\Gamma_{m,n}(t) \equiv \sum_{\bf{k}}\Gamma_{m,n}^{\bf{k}}(t),\quad\text{and}\quad
\Upsilon_{m,n}(t) \equiv \sum_{\bf{k}}\Upsilon_{m,n}^{\bf{k}}(t).
\end{equation}
The time dependence of the reduced density matrix in \eq{eq:rho_sist_red_ad_Smn} comes from an oscillating factor with frequency $(E_m-E_n)$ and also by the function $S_{m,n}(t) $, which is called \emph{decoherence function} \cite{priv98}.
The non separable character of the interaction Hamiltonian \numeq{eq:Ham_int_ad_reesc} makes the dynamics implied by $S_{m,n}(t) $ in equations \numeq{eq:Func_Smnkl_des} and  \numeq{eq:Func_Gmnkl_Umnkl} essentially different from the result in ref.\cite{priv98}. In our case the eigenvalues $\lambda_{m}^{\bf{k}}$ can be complex because of their dependence on the spin-environment coupling coefficients $g_{\bf{k}}^{A}$.
Physically, considering that all the partition elements $A$ interact with a common boson bath introduces the many-body quality in the description of decoherence.

Notice in \eq{eq:Func_Gmnkl} that the function  $\Gamma_{m,n}^{\bs{k}}(t)$ is always real and positive, it can also be seen \cite{BreuerPetruccione,Palma96} that $\Gamma_{m,n}(t)$ in \eq{eq:def_Gmn_Umn}
diverges as $t\rightarrow \infty$. Consequently, the modulus of the reduced density matrix elements attenuate irreversibly in time as
\begin{equation}\label{eq:rho_sist_red_ad_Smn_mod}
\abs{\rho_{m,n}(t)} = \abs{\rho_{m,n}(0)}\;e^{-\Gamma_{m,n}(t)},
\end{equation}
which means that $\rho_{m,n}(t)$ undergoes a non unitary dynamics.
On the other hand, the phase factors $e^{-\ic\Upsilon_{m,n}^{\bs{k}}(t)}$ in \eq{eq:Func_Smn} are relevant when calculating expectation values using \eq{eq:rho_sist_red_ad_Smn} even when they do not participate in \numeq{eq:rho_sist_red_ad_Smn_mod}.
The calculation of these phase factors, and their influence over the expectation values, generalizes the results of ref.\cite{DomGzzSegZam16}.

We also see in \eq{eq:Func_Gmnkl_Umnkl}  that both $\Gamma_{m,n}^{\bf{k}}(t)$ and $\Upsilon_{m,n}^{\bf{k}}(t)$ depend on the differences between the eigenvalues $\lambda_m^{\bf{k}}$ and $\lambda_n^{\bf{k}}$ (or their moduli) therefore they do not contribute to the decay of diagonal elements of the reduced density matrix (in this case also $\Im\left\{\lambda_m\lambda_n^*\right\} = \Im\{\abs{\lambda_m}^2\} = 0$). On the contrary, as seen in \numeq{eq:rho_sist_red_ad_Smn_mod} and \numeq{eq:Func_Gmnkl}, the efficiency of the decoherence function goes as $\abs{\lambda_{m}-\lambda_{n}}_{\bf{k}}$. This selective action that depends on the eigenvalues of the interaction Hamiltonian is a fingerprint of adiabatic decoherence in open quantum systems and has been referred to as \emph{eigen-selection} \cite{DomGzzSegZam16,SegZam11,eigenselection,SegZam13}.
Even when the density operator can be expressed on any basis, the effect of selective decoherence becomes evident when using $\{\ket{\mA}\}$, the common eigenbasis of $\HamS^\hA$ and $\HamI^\hA$.
This characteristic allows calling $\{\ket{\mA}\}$ as the \emph{preferred basis} \cite{PazZurek99} since an arbitrary initial state transforms into one having a diagonal-in-blocks density matrix \cite{SegZam11,SegZam13}.

\subsection{Dynamics of the density operator reduced to one element of the partition} \label{sec:dyn_sigmaA}

As seen in \eq{eq:redu1},  the expectation value of a local observable like \numeq{eq:obs_partS} is related to that of a single partition element.
The reduced density operator $\sigma^\hA(t)$ entering in such expectation value  involves the whole observed system through the trace over the complementary Hilbert space, as in \eq{eq:rho_sist_red_SA} ($\sigma^\hA(t) \equiv \trs{\oS_A}{\rho^\hs(t)}$). In order to calculate the matrix elements of $\sigma^\hA(t)$ we use $\rho_{m,n}(t)$ from \eq{eq:rho_sist_red_ad_Smn} into \eq{eq:rho_sist_red_SA_elem}, as follows
\begin{equation}\label{eq:rho_sist_red_SA_ad}
\sigma_{\mA,\nA}(t)  = \sum_{\{\!\mAn\!\}}\left[\rho_{m,n}(0)\;e^{-\ic\left(E_m - E_n\right)\,t} S_{m,n}(t)\right]\!\!\Cmnb.
\end{equation}
Notice that the symbol $\Cmn$ applies to all the elements within the bracket, i.e.
$\rho_{m,n}(0)\Cmn$, $\left(E_m - E_n\right)\Cmn$, and $S_{m,n}(t)\Cmn$; so, it entails evaluating the functions \numeq{eq:Func_Gmnkl} and \numeq{eq:Func_Umnkl} under the same conditions.
Let us first analyze the oscillatory factor. We begin by separating the $A$-th contribution to the system energy in \eq{eq:HS_eigen_des}, as
\[E_m = E_{\mA} + \sum_{A' \neq A} E_{\mAp},\]
hence, the difference between eigenvalues of the system Hamiltonian becomes
\begin{equation}\label{eq:HS_eigen_difEm}
\left(E_m - E_n\right)\Cmn = \left(E_{\mA} - E_{\nA}\right).
\end{equation}
Therefore this factor is out of the partial trace.
In order to evaluate  $\Gamma_{m,n}^{\bf{k}}(t)\Cmn$ and $\Upsilon_{m,n}^{\bf{k}}(t)\Cmn$, we also rewrite the eigenvalues $\lambda_m$ defined in \eq{eq:Lambda_S_HS_eigen} as
\begin{equation} \label{eq:lambda_nuevo}
\lambda _m^{\bf k}= \lambda_{\mA}\; g_{\bf k}^A+ \sum_{A' \neq A} \lambda_{\mAp}\; g_{\bf k}^{A'},
\end{equation}
and get
\begin{equation}\label{eq:Func_eigenval_GU}
\begin{split}
\left(\lambda_{m}^{\bf k}-\lambda_{n}^{\bf k}\right)\Cmn &=
\left(\lambda_{\mA}-\lambda_{\nA}\right)\,g_{\bf{k}}^{A}, \\
\left(\lambda_{m}^{\bf{k}}\!+\!\lambda_{n}^{\bf{k}}\right)^*\Cmn &= \left(\lambda_{\mA} + \lambda_{\nA}\right)g_{\bf{k}}^{A*} +2 \sum_{A' \neq A} \lambda_{\mAp}\;g_{\bf{k}}^{A'*},\\
\Im\left\{\lambda_{m}^{\bf{k}}\lambda_{n}^{\bf{k}*}\right\}\Cmn
&= \left(\lambda_{\mA}-\lambda_{\nA}\right)
\sum_{A' \neq A}\lambda_{\mAp}\Im\left\{g_{\bf k}^{A}\,g_{\bf k}^{A'*}\right\}.
\end{split}
\end{equation}

After using \numeq{eq:Func_eigenval_GU} in \eqs{eq:Func_Gmnkl} and \numeq{eq:Func_Umnkl}, the functions included in the decoherence function can be written as
\begin{subequations}\label{eq:Gamma_Upsilon_mn-bar}
\begin{equation}\label{eq:Gamma_mn-bar}
\Gamma_{m,n}(t)\Cmn = (\lambda_{\mA}-\lambda_{\nA})^2 \; \gamma_A(\beta,t),
\end{equation}
\begin{equation}\label{eq:Upsilon_mn-bar}
\Upsilon_{m,n}(t)\Cmn = (\lambda_{\mA}^2 - \lambda_{\nA}^2)\,\varepsilon_A(t)
+ (\lambda_{\mA} - \lambda_{\nA})\,\chi_A^{\{\!\mAn\!\}}(t),
\end{equation}
\end{subequations}
where we defined
\begin{subequations}\label{eq:gama-eps-chi}
\begin{align}
\gamma_A(\beta,t) &\equiv 2\sum_{\bf k}\frac{\abs{g_{\bf k}^A}^2}{\omega_{\bf k}^2}\,\sin^2\!\left(\omega_{\bf k}\,t/2\right)
\coth\left(\beta\,\omega_{\bf k}/2\right),\label{eq:Fga_boson-sys}\\
\varepsilon_A(t) &\equiv \sum_{\bf k}  \frac{\abs{g_{\bf k}^A}^2}{\omega_{\bf k}^2} \left[\sin(\omega_{\bf k}\,t) -\omega_{\bf k}\,t \right],\label{eq:Fep_boson-sys}\\
\chi_A^{\{\!\mAn\!\}}(t) &\equiv  \sum_{A' \neq A} \lambda_{\mAp}\,\zeta_{A,A'}(t),\label{eq:Fchi_boson-sys}
\end{align}
\end{subequations}
with the superscript $\{\mAn\}$ emphasizing the dependence on all the quantum numbers $\mAp$ with $A' \neq A$ (hereafter we use this notation), and
\begin{equation}\label{eq:zeta}
\zeta_{A,A'}(t) \equiv 2 \sum_{\bf k} \frac{1}{\omega_{\bf k}^2} \left\{\Re\{g_{\bf k}^{A}\,g_{\bf k}^{A'*}\}
\left[\sin(\omega_{\bf k}\,t) - \omega_{\bf k}\,t\right]\right.
- \left.\Im\{g_{\bf k}^{A}\,g_{\bf k}^{A'*}\} \left[1-\cos(\omega_{\bf k}\,t)\right]\right\},
\end{equation}
with $\Re\left\{z\right\}$ the real part of $z$.
It is worth to remark that the exponential factors in $S_{m,n}(t)\Cmn$ involving $\gamma_A(\beta,t)$ and $\varepsilon_A(t)$ do not depend on the quantum numbers $\{\mAn\}$ and therefore do not contribute to the partial trace. On the contrary, the factor with $\chi_A^{\{\!\mAn\!\}}(t)$ depends explicitly on the set of eigenvalues $\lambda_{\mAp}$, with $A' \neq A$. The occurrence of a contribution like this one is a consequence of the form \numeq{eq:Ham_int_ad} of the interaction Hamiltonian. The sum over $A'$ in \eq{eq:Fchi_boson-sys} can be interpreted as a correlation between partition elements mediated by the phonon environment, and is precisely an indicator of the many-body nature of the quantum decoherence phenomenon.
Finally, the matrix elements of the density operator reduced over $\shSAn$ are
\begin{equation} \label{eq:sigma_elem}
\begin{split}
\sigma_{\mA,\nA}(t) &= e^{-\ic\,\left(E_{\mA}\! - E_{\nA}\!\right)\,t}\,e^{-(\lambda_{\mA}\!-\lambda_{\nA}\!)^2 \, \gamma_A(\beta,t)}
\,e^{-\ic\,(\lambda_{\mA}^2\!-\lambda_{\nA}^2\!)\,\varepsilon_A(t)}\\
&\quad\times \sum_{\{\!\mAn\!\}} \rhoini\;e^{-\ic\,(\lambda_{\mA}\!-\lambda_{\nA}\!)\,\chi_A^{\{\!\mAn\!\}}\!(t)},
\end{split}
\end{equation}
where we wrote in a more compact way
\[\rhoini \equiv \rho_{m,n}(0)\Cmn.\]
Here we see that the matrix elements certainly depend on the initial state, and also there is a manifest dependence on the difference of eigenvalues $(\lambda_{\mA}\!-\lambda_{\nA})$ (eigen-selection). The sums over $\bf k$ contribute to the time dependence through the functions $\varepsilon_A(t)$, $\zeta_{A,A'}(t) $ and  $\gamma_A(t)$, the last one also depends on the environment temperature. The new contribution that arises from the partial trace includes the function $\chi_A^{\{\!\mAn\!\}}(t)$ which, as seen in \eq{eq:gama-eps-chi}, endows the reduced density matrix with a reference to the entire observed system.

The significance of the doubly reduced or condensed density matrix $\sigma^\hA(t)$ derived so far lies in that it allows calculating an analytic expression of local-observable expectation values of a many-body system.  The strategy is valid both in non-interacting (as this work) or weakly interacting elements, provided the Hamiltonians $\HamS$ and $\HamI$ share a common partitionable eigenbasis. It is worth to remark that even in the non-interacting case treated here, the dynamics of a single partition element $\sigma^\hA(t)$ reflects in a simple way the existence of the rest of the elements.
Notice that the absolute value of the matrix elements $\abs{\sigma_{\mA,\nA}(t)}$ depend on the function $\gamma_A(\beta,t)$ coming from $\Gamma$, {\it and also} on $\chi_A^{\{\!\mAn\!\}}(t)$ which derives from the phase factor $\Upsilon$. \sect{sec:app_pair_phon} shows how decisive is the phase factor in the decoherence rate in a concrete example.

\section{Application: ensemble of non-interacting spin pairs coupled with a common phonon bath} \label{sec:app_pair_phon}

In this section we apply the theoretical approach derived so far to explore the plausibility of our approach by estimating the decoherence time scale in a system model that shares many features with a real solid.
With this aim, we consider an observed system composed of nuclear spin-pairs coupled with a phonon bath. The partition elements are dipole-coupled pairs of spins 1/2 with no inter-pair interaction, arranged at the sites of a lattice, as depicted in \fig{fig:Fig_1}. In this way, we focus on the decoherent dynamics of an ensemble of pairs due exclusively to their contact with a common phonon bath.

We are concerned with the spin dynamics in NMR experiments; thus, we assume that the spins are immersed in a strong external magnetic field $\vec{B} = B_0 \hat{z}$, which defines the $z$-axis of the laboratory.
In \sect{sec:app_sysm_sigma} we present the details of the spin-phonon interaction for the model system, and in \sect{sec:app_sigma} we derive the reduced density matrix. In \sect{sec:app_calc_tdeco} we estimate the characteristic times involved in the decoherence dynamics.

\insertfig{ht}{fig:Fig_1}{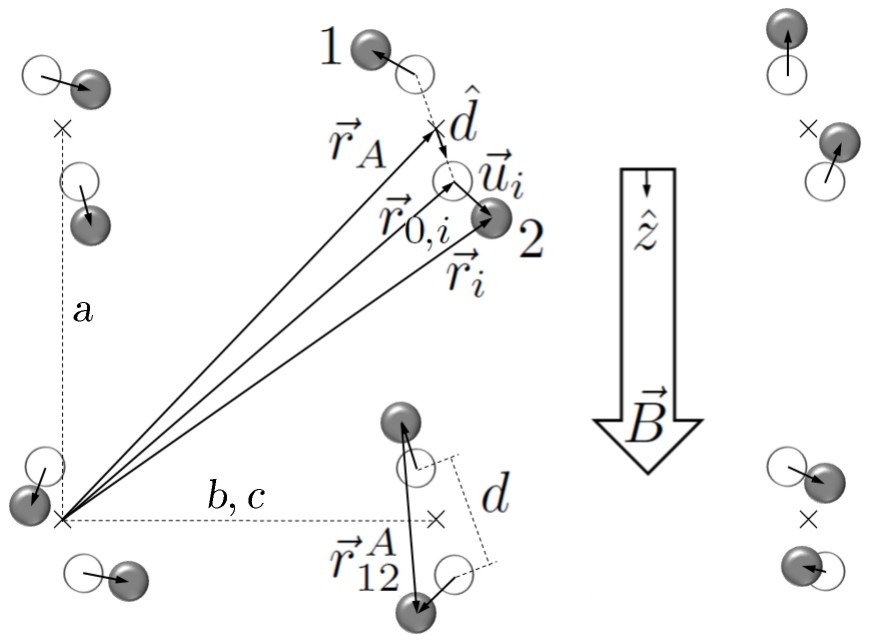}{6.75}{5}{Dipole interacting spin pairs at the sites of a lattice (x). Open circles: equilibrium positions; solid circles: spins displaced by their interaction with the phonon field.}

\subsection{Pair-phonon model} \label{sec:app_sysm_sigma}

We work in the high field approximation \cite{slichbook}, where the relevant dynamics is driven only by the secular part of the dipolar Hamiltonian. Besides, we describe the spin dynamics in a rotating frame (on resonance at the Larmor frequency $\omega_0 \equiv \gamma_p\,B_0$, where $\gamma_p$ is the proton gyromagnetic ratio). Then the appropriate spin Hamiltonian of this example can be written as the sum of the internal secular dipolar energy of the pairs (labeled as $A$) \cite{slichbook},
$\op{H}_\op{D}^\hs \equiv \sum_A \op{H}_{\op{D},A}^\hs$, with
\begin{equation}\label{eq:def_Ham_Da0_A}
\op{H}_{\op{D},A}^\hs \equiv \sqrt{6}\;\Omega_0(r^A_{12}) \left[\op{1}^\hl\otimes\cdots\otimes\op{T}_{2,0}^\hA\otimes\cdots\otimes\op{1}^\hN\right],
\end{equation}
where the dipolar coupling (in frequency units) is
\begin{equation}\label{eq:def_Omega0}
\Omega_0(r^A_{12}) \equiv \frac{\mu_0\gamma_p^{2}\hbar}{8\pi}\,\frac{\left[1-3\cos^2\left(\theta_{\hat{r},\hat{z}}\right)\right]}{(r^A_{12})^{3}},
\end{equation}
with $\theta_{\hat{r},\hat{z}}$ the angle between the laboratory $z$-axis and the direction of vector $\vec{r}_{12}^{\,A}$ (see \fig{fig:Fig_1}), and
\begin{equation}\label{eq:def_T20_A}
\op{T}_{2,0}^\hA \equiv \frac{1}{\sqrt{6}}\big[3\,\op{I}_\op{z}^{(s_{\!A},1)} \otimes \op{I}_\op{z}^{(s_{\!A},2)} - \vec{\op{I}}_\op{1}^{\,\hA} \cdot \vec{\op{I}}_\op{2}^{\,\hA} \big]\\
\end{equation}
is the secular component of the second order spherical tensor which involves the spin operators of spins at pair $A$.

We assume that the interaction between the spin pairs and the phonons originates in the small shift of the spin positions from those corresponding to a rigid lattice, caused by the phonons throughout the lattice. The strategy we follow to set up the system and interaction Hamiltonians is to assume that the position variables involved in $\op{H}_\op{D}^\hs$ are quantum variables. Here we outline the main steps of the derivation and leave the details to \apen{app:quant_intra-pair_dist}.
As a consequence of quantizing $\vec{r}_{12}^{\,A}$, the real constant $\Omega_0(r^A_{12})$ in \eq{eq:def_Omega0} becomes a quantum operator in $\shE$, namely
\[\Omega_0(r^A_{12}) \quad \rightarrow \quad \op{\Omega}^{(e)}.\]
The factor $(r^A_{12})^{-3}$ in \eq{eq:def_Omega0} can be written as an expansion in powers of the displacement from the equilibrium positions of the nuclei bearing the observed spins. Assuming small deviations, we keep only the linear term of the expansion, then, the newly defined operator $\op{\Omega}^{(e)}$ can be written as
\[\op{\Omega}^{(e)} = \frac{D}{d^3} \left[  \op{1}^\he - \frac{3}{d} \sum_{\bf k} \left(g^{A*}\,\op{b} + g^{A}\,\op{b}^{\dagger}\right)_{\bf k}^\he\right],\]
where $D \equiv \mu_0\gamma_p^{2}\hbar\,\left[1-3\cos^2\left(\theta_{\hat{r},\hat{z}}\right)\right]/8\pi$,
and the system-environment coupling coefficients
\begin{equation}\label{eq:def_gk_A}
g_{\bf k}^{A} \equiv e^{-\ic\,\vec{k}\cdot\vec{r}_{A}}\,g_{\bf k},
\end{equation}
are the product of a phase factor that depends on the position of the $A$-th pair and a factor $g_{\bf k}$, defined in \eq{eq:def_gk_desc}, which is common for all the pairs.

In this way, the full quantum dipolar Hamiltonian $\op H_\op{D}$ (see \eq{eq:HDip_qq}) consists of two terms: one of them represents the dipole interaction energy of a system that is isolated from the lattice, while the other amounts for the variation of dipolar interaction due to the lattice motion. Therefore, in this model, we identify the first term with the system Hamiltonian $\op{H_S}$, as introduced in \sect{sec:Hamiltonians}, and the second with the interaction Hamiltonian  $\op{H_I}$. The expressions for these Hamiltonians are
\begin{equation}\label{eq:def_Ham_sys_onres}
\HamS \equiv \HamS^\hs \otimes \op{1}^\he = \sum_A \op{H}_{\op{S},A}^\hs \otimes\op{1}^\he,
\end{equation}
with
\begin{equation}\label{eq:def_Ham_sys_onres_A}
\op{H}_{\op{S},A}^\hs \equiv \sqrt{6}\;\Omega_0(d)\;\left[ \op{1}^\hl \otimes \cdots \otimes \op{T}_{2,0}^\hA \otimes \cdots \otimes  \op{1}^\hN\right] ,
\end{equation}
where the dipolar coupling $\Omega_0(d) = D/d^3 $ is evaluated at the equilibrium intra-pair distance $d$, and the second contribution is
\begin{equation}\label{eq:def_Ham_Ia0_intra}
\begin{split}
\HamI \equiv -\frac{3}{d}\,\sum_A \op{H}_{\op{S},A}^\hs
\otimes \sum_{\bf k} \left(g^{A*}\,\op{b} + g^{A}\,\op{b}^{\dagger}\right)_{\bf k}^\he.
\end{split}
\end{equation}

We may now identify the basis $\{\ket{\mA}\}$ for the spin pair with the triplet-singlet basis, which is an eigenbasis of the secular dipolar tensor
\begin{equation}\label{eq:eigen_T20}
\op{T}_{2,0}^\hA\ket{\mA} = \frac{1}{2\sqrt{6}}\;\kappa_{\mA}\,\ket{\mA},
\end{equation}
where the corresponding eigenvalues are
\begin{equation}\label{eq:def_mA_kappaA}
\ket{\mA} = \left\{\begin{array}{l}
\ket{1,1} \equiv \ket{++}\\
\ket{1,0} \equiv \frac{\ket{+-}+\ket{-+}}{\sqrt{2}}\\
\ket{1,-1} \equiv \ket{--}\\
\ket{0,0} \equiv \frac{\ket{+-}-\ket{-+}}{\sqrt{2}}
\end{array}\right.\;
; \qquad \kappa_{\mA} = \left\{\begin{array}{c}
1 \\
-2 \\
1\\
0
\end{array}\right..
\end{equation}
Notice also from \numeq{eq:Ham_sys_eigen} and \numeq{eq:Ham_sys_A_eigen} that
\begin{equation*}
\HamS^\hs\ket{m} = E_m\ket{m} = \frac{\Omega_0(d\,)}{2}\sum_A \kappa_{\mA} \ket{m};
\qquad E_{\mA} = \frac{\Omega_0(d\,)}{2}\,\kappa_{\mA},
\end{equation*}
and consequently the system energy difference involved in \eq{eq:HS_eigen_difEm} is
\begin{equation*}\label{eq:HS_eigen_difEm_paresNI}
\left(E_m - E_n\right)\Cmn = \frac{\Omega_0(d\,)}{2}\,\left(\kappa_{\mA}-\kappa_{\nA}\right).
\end{equation*}
In accordance with the definitions \numeq{eq:Ham_int_A_eigen}, \numeq{eq:def_Ham_Ia0_intra} and \numeq{eq:eigen_T20}, we have
\begin{equation}\label{eq:def_lambda-kappa}
\lambda_{\mA}\equiv -\frac{3}{2}\,\frac{\Omega_0(d\,)}{d}\,\kappa_{\mA}.
\end{equation}
Substituting \eqs{eq:def_gk_A} and \numeq{eq:def_lambda-kappa} in \eq{eq:lambda_nuevo}, under the same procedures used to obtain \eqs{eq:Gamma_mn-bar} and \numeq{eq:Upsilon_mn-bar}, we get
\begin{subequations}\label{eq:Gmn_Umn_boson-spin}
\begin{equation}\label{eq:Gmn_boson-spin}
\Gamma_{m,n}(t)\Cmn =
\frac{9}{4}\left(\frac{\Omega_0(d\,)}{d}\right)^2\left(\kappa_{\mA} - \kappa_{\nA}\right)^2\gamma(\beta,t),
\end{equation}
\begin{equation}\label{eq:Umn_boson-spin}
\Upsilon_{m,n}(t)\Cmn = \frac{9}{4}\left(\frac{\Omega_0(d\,)}{d}\right)^2\;
\bigg[\left(\kappa_{\mA}^2 - \kappa_{\nA}^2\right)\,\varepsilon(t)
+\left(\kappa_{\mA} - \kappa_{\nA}\right)\,\widehat{\chi}_{A}^{\{\!\mAn\!\}}(t)\bigg],
\end{equation}
\end{subequations}
where the functions $\gamma(\beta,t)$, $\varepsilon(t)$ and $\widehat{\chi}_{A}^{\{\!\mAn\!\}}(t)$ correspond to \eq{eq:gama-eps-chi} with the coupling coefficients given by \eq{eq:def_gk_A}
\begin{subequations}\label{eq:Fga_Fep_Fze_boson-spin}
\begin{align}
\gamma(\beta,t) &\equiv 2\sum_{\bf k} \frac{\abs{g_{\bf k}}^2}{\omega_{\bf k}^2} \,\sin^2\!\left(\omega_{\bf k}\,t/2\right)\,
\coth\left(\beta\,\omega_{\bf k}/2\right),\label{eq:Fga_boson-spin}\\
\varepsilon(t) &\equiv \sum_{\bf k} \frac{\abs{g_{\bf k}}^2}{\omega_{\bf k}^2}\left[\sin(\omega_{\bf k}\,t)-\omega_{\bf k}\,t\right],\label{eq:Fep_boson-spin}\\
\widehat{\chi}_{A}^{\{\!\mAn\!\}}(t) &\equiv \sum_{A' \neq A}\kappa_{\mAp}\,\zeta_{A,A'}(t).\label{eq:Fchi_boson-spin}
\end{align}
\end{subequations}	
In this example  $\zeta_{A,A'}(t)$ adopts the form
\begin{equation}\label{eq:Fze_boson-spin}
\begin{split}
\zeta_{A,A'}(t) &\equiv 2 \sum_{\bf k} \frac{\abs{g_{\bf k}}^2}{\omega_{\bf k}^2} \left\{\cos\left(\vec{k}\cdot\vec{r}_{A,A'}\right)
\left[\sin(\omega_{\bf k}\,t)-\omega_{\bf k}\,t\right]\right.\\
&\qquad\qquad\qquad\qquad  + \left.\sin\left(\vec{k}\cdot\vec{r}_{A,A'}\right)\left[1-\cos(\omega_{\bf k}\,t)\right]\right\},
\end{split}
\end{equation}
where $\vec{r}_{A,A'} \equiv \vec{r}_{A} - \vec r_{A'}$ (see \fig{fig:Fig_1}) and $\vec{k}$ the wavenumber vector, then  $\vec{k}\cdot\vec{r}_{A,A'} = \vec{\abs{k\,}}\,r_{A,A'}\cos\left(\theta_{A,A'}\right)$ with $r_{A,A'} \equiv \abs{\vec{r}_{A,A'}}$ and $\theta_{A,A'}$ the angle between these vectors.
Notice that $\gamma(\beta,t)$, $\varepsilon(t)$ in  \numeq{eq:Fga_boson-spin} and \numeq{eq:Fep_boson-spin}  became independent of the position index $A$ because of the form of $g_{\bf k}^{A} $ of the model. Besides, \eq{eq:Fchi_boson-spin} differs from \eq{eq:Fchi_boson-sys} in just a constant factor.

By comparing equations (\ref{eq:Gmn_boson-spin}) and (\ref{eq:Umn_boson-spin}) one can notice that the real and imaginary parts of the exponent of the decoherence function are radically different in nature. Indeed, the real part depends only on the state of the target pair $A$
and practically coincides with the result of the usual spin-boson theory.
Contrastingly, the imaginary part contributes a noteworthy characteristic since it depends on the state of all the neighbors of pair $A$ through the sum over $A'$ in \eq{eq:Fchi_boson-spin}. Also, it is strictly \emph{temperature independent}, namely, it does not reflect any thermal property of the environment, such as the amplitude of the proton-pair motions. This contribution instead has a purely Hamiltonian origin.

\subsection{Reduced density matrix, initial condition} \label{sec:app_sigma}

At this point we can use the results derived in \sect{sec:app_sysm_sigma} to write the density matrix reduced to a single pair in the pair-phonon case. It still remains discussing about the initial condition $\rhoini$ in \eq{eq:sigma_elem}.

Let us first mention that the typical NMR experiment starts from a spin system in thermal equilibrium with the external magnetic field, at temperature $T$, that is
\[\rho_{eq}^\hs \propto \exp[-\hbar\,\omega_0\,\op{I}_\op{z}^\hs /K_B T],\]
where $\op{I}_\op{z}^\hs$ is the $\op{z}$ component of the spin operator, and $\omega_0$ the Larmor frequency.
The usual high temperature approximation allows keeping up to first order in the expansion of $\rho_{eq}^\hs$, and the pulse sequences applied to the sample prepare the spin system in a state having the general form
\begin{equation}\label{eq:rho_inicial_NMR}
\rho^\hs = \frac{\op{1}^\hs}{\cal N} +  \Delta \rho^\hs,
\end{equation}
where the second, traceless term, is sometimes called ``deviation density operator'', and ${\cal N} \equiv  \trs{S}{\op{1}^\hs} = \opc{N}_p^{N}$, with $\opc{N}_p \equiv \trs{S_A}{\op{1}^\hA}$.
Given that decoherence only affects the term $\Delta \rho^\hs$ in  \numeq{eq:rho_inicial_NMR}, the time dependence is contained in the deviation density operator.
With these considerations, \eq{eq:sigma_elem} can be used to describe the decoherence dynamics of the deviation density matrix elements $\Delta\sigma_{\mA,\nA}(t)$, with an initial condition $\Drhoini$, by just replacing
\[\sigma_{\mA,\nA}(t) \rightarrow\! \Delta\sigma_{\mA,\nA}(t),\;\;\rhoini \rightarrow \Drhoini.\]

In consistence with the idea of local observables addressed in this work, we write a general initial state in terms of local traceless operators as
\begin{equation}\label{eq:Drho_inicial}
\Delta \rho^\hs(0) = \frac{1}{\opc{N}_p^{N-1}} \sum_{A} \op{1}^\hl \otimes \cdots \otimes  \Delta\sigma^\hA(0) \otimes \cdots \otimes  \op{1}^\hN,
\end{equation}
where $\Delta\sigma^\hl(0) = \cdots = \Delta\sigma^\hA(0) = \cdots = \Delta\sigma^\hN(0)$.\\
A typical simple example, in NMR, of an initial state like this is that ensuing a saturating ($\pi/2$) pulse applied to a spin system in the state $\rho_{eq}$. In such case
\begin{equation}\label{eq:Dsigma_inicial}
\Delta\sigma^\hA(0) =  -\frac{\hbar\,\omega_0}{K_B\,T}\,\op{I}_\op{x}^\hA,
\end{equation}
where the $\op{x}$ component of the spin operator of the pair is
$\op{I}_\op{x}^\hA \equiv \op{I}_\op{x}^{(s_{\!A},1)}\otimes\op{1}^{(s_{\!A},2)} + \op{1}^{(s_{\!A},1)}\otimes\op{I}_\op{x}^{(s_{\!A},2)}$.
In this way, the matrix elements of the initial deviation density operator \numeq{eq:Drho_inicial} entering in \eq{eq:sigma_elem} (imposed by the partial trace over $\cbb{\bar S}_A$) are
\begin{equation}\label{eq:Deltarho0_sist_red_sep_mn}
\Drhoini = \frac{1}{\opc{N}_p^{N-1}}\,\bigg[\Delta\sigma_{\mA,\nA}(0) + \delta_{\mA,\nA}\sum_{A' \neq A}\Delta\sigma_{\mAp,\mAp}(0)\bigg].
\end{equation}
It can easily be seen that only the first term of \eq{eq:Deltarho0_sist_red_sep_mn} contributes to the time dependence of
$\Delta\sigma_{\mA,\nA}(t)$ in \eq{eq:sigma_elem} (the second term vanishes if  $\mA \neq \nA$, besides,  the sum $\sum_{\{\!\mAn\!\}} \Drhoini$ of \eq{eq:sigma_elem}, in the case that $\mA = \nA$, amounts to calculating the following trace of a traceless operator
$\sum_{\mAp}\!\Delta\sigma_{\mAp,\mAp}(0) = \trs{S_{A'}}{\Delta\sigma^\hAp(0)} = 0$).
Then, by replacing \eq{eq:Deltarho0_sist_red_sep_mn} in \eq{eq:sigma_elem} we obtain
\begin{equation}\label{eq:Dsigma_boson-spin_hatGA}
\begin{split}
\Delta\sigma_{\mA,\nA}(t) &= \Delta\sigma_{\mA,\nA}(0)\;e^{-\ic\frac{\Omega_0(d)}{2}\left(\kappa_{\mA}  - \kappa_{\nA}\!\right)\,t}\\
&\quad\times e^{-\frac{9}{4}\left(\frac{\Omega_0(d)}{d}\!\right)^2\left[\left(\kappa_{\mA} - \kappa_{\nA}\!\right)^2\gamma(\beta,t)\,
+\,\ic\,\left(\kappa_{\mA}^2 - \kappa_{\nA}^2\!\right)\,\varepsilon(t)\right]}\,\widehat{G}_A(t),
\end{split}
\end{equation}
where we defined
\begin{equation}\label{eq:fdeco_hatGA}
\widehat{G}_A(t) \equiv \frac{1}{\opc{N}_p^{N-1}}\sum_{\{\!\mAn\!\}}
e^{-\ic\frac{9}{4}\left(\frac{\Omega_0(d)}{d}\!\right)^2\left(\kappa_{\mA} - \kappa_{\nA}\!\right)\,\widehat{\chi}_{A}^{\{\!\mAn\!\}}\!(t)}.
\end{equation}
Notice that $\widehat{G}_A(t)$ involves the sum over the quantum numbers of the states $\{\mAn\}$ (from the partial trace) and also contains a sum over the sites $A'$ within the variable $\widehat{\chi}_{A}^{\{\!\mAn\!\}}$. Thereby, $\widehat{G}_A(t)$ involves the eigenvalues of all the pairs or partition elements. In other words, this function links the decoherence dynamics of pair $A$ with all the other pairs, even though the pairs were assumed as non-interacting in $\HamS$ and $\HamI$.
It now remains to analyze the time dependence of $\Delta\sigma_{\mA,\nA}(t) $ given by the functions $\gamma(\beta,t), \varepsilon(t)$ and $\widehat{G}_A(t)$ in the limit of a large system.\\
It is worth to remark that $\widehat{G}_A(t)$ also depends on the geometric distribution of the pairs surrounding pair $A$ through its dependence on $\widehat{\chi}_{A}^{\{\!\mAn\!\}}$. However, neglecting border effects and under the reasonable assumption that each pair sees the same distribution as the rest of the surrounding pairs, the function $\widehat{G}_A(t)$ and consequently the reduced density matrix \numeq{eq:Dsigma_boson-spin_hatGA} become independent of index $A$.
Therefore, the expectation value $\langle \op{O}_A^\hs\rangle(t)$ of a local observable (like \eq{eq:obs_partS}) will also be the same for each pair, and a measurement on the whole sample becomes $N$ times that of one pair or partition element (see \eq{eq:redu1})
\begin{equation}\label{eq:obs_N_OA}
\langle \op{O}^\hs \rangle(t) = N\,\langle\op{O}_A^\hs\rangle(t).
\end{equation}
Notice also that the observables in NMR experiments are represented by traceless operators, e.g. $\op{I}_\op{x}^\hs \equiv \sum_A \op{1}^\hl\otimes\cdots\otimes \op{I}_\op{x}^\hA \otimes\cdots\otimes\op{1}^\hN$; thus, their expectation values only depend on the deviation matrix.

\subsection{Estimation of the decoherence time scale} \label{sec:app_calc_tdeco}

In order to estimate the decay time implied by decoherence in \eq{eq:Dsigma_boson-spin_hatGA}, we introduce the following model:
\renewcommand{\theenumi}{\arabic{enumi}}
\begin{enumerate}
    \item \label{assum_latt-or} Let us consider the sublattice formed by the observed spin 1/2 pairs within a crystal sample. For simplicity, we assume that the pair axis ($\hat{d}$) coincides with the $\hat{\csl{a}}$-axis of the sublattice unit cell, and call $\csl{a}$ to the distance between pairs (in this work, we consider strictly non-interacting pairs and thus do not need specifying more details on the lattice symmetry).

    \item \label{assum_a-bigg} The lattice parameter $\csl{a}$ is larger than the pair size $d$, so that
        $\sin\left(\pi\,d/(2\,\csl{a})\right) \simeq \pi\,d/(2\,\csl{a})$ and $\cos\left(\pi\,d/(2\,\csl{a})\right) \simeq 1$.

    \item \label{assum_el-cte-intra} The atomic displacements perpendicular to $\hat{d}$ are much smaller than in parallel directions and can, to a good approximation, be neglected.

    \item \label{assum_el-cte-pair_bigg} The elastic constant (stiffness) of a pair is much larger than between different pairs.\\	

        Under these reasonable conditions we can assume that the mechanical waves propagate mainly along the equilibrium intra-pair direction, that is
        $\vec{k} = k\,\hat{d}$, with
        \[k \equiv \frac{2\pi}{N_1 \csl{a}}\,q \quad \left(q=0,\pm 1,\pm 2,\cdots,\pm \frac{N_1}{2}\;\,\text{or}\;\,\pm \frac{N_1-1}{2}\right),\]
        and $N_1 \sim \sqrt[3]N$ the number of pairs along $\hat{d}$.\\

    \item \label{assum_N1-inf} $N$ is large, so we can replace the sum over $\vec{k}$ by
        \begin{equation}\label{eq:sumk_1D_int}
            \sum_{\bf k} \equiv \sum_{\vec{k}}\sum_l\,
        \rightarrow\,\frac{N \csl{a}}{2\pi}\sum_l\int_{-k_{\!M}}^{k_{\!M}}dk, \quad {\rm with }\: k_{\!M} = \pi/\csl{a}.
        \end{equation}
        Notice that due to assumptions \ref{assum_latt-or} -- \ref{assum_el-cte-pair_bigg} the integral over $k$ is unidimensional but the constant reflects integration over the plane perpendicular to $\hat{d}$.

    \item \label{assum_param} For the sake of concreteness and with the aim of comparing with the experiment, we consider the case of dihydrated calcium sulphate (gypsum) single crystal. Then the spin pairs involved in our model will be those of $^1$H nuclei at hydration water molecules having the same intra-pair axis orientation, and we adopt the following realistic values for the constants involved in the decoherence function:
        \begin{itemize}
            \item $d = 0.153\un{nm}$. We use the hydrogen distance from ref.\cite{Cole74}.
            \item $\csl{a} = 0.8\un{nm}$ corresponds to the distance between equally oriented H$_2$O molecules on the same molecular plane within a unit cell, obtained from crystallographic data \cite{Cole74} (distance between oxygen atoms ${\rm O_W}$ or ${\rm O'_W}$ in Fig.1 of ref.\cite{Cole74}). Notice that these values of $d$ and $\csl{a}$ satisfy assumption \ref{assum_a-bigg}.
            \item $\vs = 4570\un{m/s}$  is the speed of sound in the sample, and its experimental value is obtained from ref.\cite{Dalui96}.
            \item $T = 300\un{K}$.
        \end{itemize} 	

    \item \label{assum_omega} The dispersion relation $\omega_{\bf k}(k)$ for acoustic phonons can be well approximated by
        \begin{equation}\label{eq:rel_disp_wk_acustico}
            \omega_{\bf k} = \frac{2\,\vs}{\csl{a}} \abs{\sin\left(k\,\csl{a}/2\right)} \sim \vs \abs{k}.
        \end{equation}
        These assumptions may have some effect when evaluating the sums (or integrals) \numeq{eq:Fga_Fep_Fze_boson-spin} and \numeq{eq:Fze_boson-spin} for large $k$, however, the frequency dependence in such equations goes as $\omega_{\bf k}^{-2}$ which makes the larger $k$ region less significant.

    \item \label{assum_no-optical} We neglect optical modes because their frequencies are greater than the maximum acoustic mode frequency, and thus contribute negligibly to \eqs{eq:Fga_Fep_Fze_boson-spin} and \numeq{eq:Fze_boson-spin} (see details in \apen{app:sys-env_coup_coef}).
\end{enumerate}

\apen{app:calc_Deco_func_1Dk} shows  the use of these assumptions to calculate functions $\gamma(\beta,t)$, $\varepsilon(t)$ and $\zeta_{A,A'}(t)$, defined in  \numeq{eq:Fga_Fep_Fze_boson-spin} and \numeq{eq:Fze_boson-spin}; it also shows that they adopt the following expressions
\begin{subequations}\label{eq:Fga_Fep_Fze_boson-spin_aprox}
\begin{align}
\gamma(T,t) &\simeq d^2\,\frac{K_B\,T\,\csl{a}}{4\,\vs^{\,3}\,m_p}\,t,\label{eq:Fga_boson-spin_aprox}\\
\varepsilon(t) &\simeq -d^2\,\frac{\hbar}{4\,\vs^{\,2}\,m_p}\,t,\label{eq:Fep_boson-spin_aprox}\\
\zeta_{A,A'}(t) &\simeq d^2\,\frac{\hbar\,\csl{a}}{2\,\vs^{\,3}\,m_p}\,\left[\frac{\varphi_{A,A'}(t)}{2\pi} -\frac{\vs}{\csl{a}}\,t\,\opc{S}_{A,A'}\right], \label{eq:Fze_boson-spin_aprox}
\end{align}
\end{subequations}
with
\begin{equation}\label{eq:Sinc_SAAp}
\opc{S}_{A,A'} \equiv \frac{\sin\left[\pi\,r_{A,A'}\cos\left(\theta_{A,A'}\right)/\csl{a}\,\right]}
{\pi\,r_{A,A'}\cos\left(\theta_{A,A'}\right)/\csl{a}}.
\end{equation}
The function $\varphi_{A,A'}(t)$ in \numeq{eq:Fze_boson-spin_aprox} is defined in \eq{eq:def_vphi}; it depends on the distance between the pairs $A, A'$ along the wavevector, and its time dependence is shown in \fig{fig:Fig_5}.

Let us now focus on the dependence of $\widehat{\chi}_{A}^{\{\!\mAn\!\}}(t)$ on the state of all the spin pairs in the complementary space $\shSAn$. Using  \numeq{eq:Fze_boson-spin_aprox} in \eq{eq:Fchi_boson-spin}, we may write
\begin{equation}\label{eq:Chihat_boson-spin_aprox}
\widehat{\chi}_{A}^{\{\!\mAn\!\}}(t) \simeq d^2\,\frac{\hbar\,\csl{a}}{2\,\vs^{\,3}\,m_p}
\left[\frac{1}{2}\,X_{A}^{\prime}\!(t) - \frac{\vs}{\csl{a}}\,t\,X_{A} \right]
\end{equation}
where we defined the variables
\begin{subequations}\label{eq:XXp_A}
\begin{align}
X_{A}^{\prime}(t) &\equiv \sum_{A' \neq A}\kappa_{\mAp}\,\varphi_{A,A'}(t)/\pi, \label{eq:Xp_A}\\
X_{A} &\equiv \sum_{A' \neq A}\kappa_{\mAp}\,\opc{S}_{A,A'}, \label{eq:X_A}
\end{align}
\end{subequations}
which depend on the quantum numbers ${\{\mAn\}}$ and on the geometric distribution of spin pairs encompassed in $\varphi_{A,A'}(t)$ and $\opc{S}_{A,A'}$.
In this way, after \numeq{eq:Chihat_boson-spin_aprox} the function  $\widehat{G}_A(t)$, defined in \eq{eq:fdeco_hatGA} becomes
\begin{equation}\label{eq:fdeco_hatGA_cota}
\begin{split}
\widehat{G}_A(t) &\simeq \frac{1}{\opc{N}_p^{N-1}}
\sum_{\{\!\mAn\!\}} e^{-\ic\,2\pi \nuD\left(\kappa_{\mA} - \kappa_{\nA}\right)\,\frac{\csl{a}}{\vs}\,X_{A}^{\prime}(t)}
\;e^{\,\ic\,4\pi \nuD\left(\kappa_{\mA} - \kappa_{\nA}\right)\,t\,X_{A}},
\end{split}
\end{equation}
where
\begin{equation}\label{eq:def_nu_vep}
\nuD \equiv \frac{9\,\Omega_0^{\,2}(d\,)\,\hbar}{32\;\pi\,\vs^{\,2}\,m_p}.
\end{equation}
Notice that only $X_{A}^{\prime}(t)$ and $X_{A}$ in \eq{eq:fdeco_hatGA_cota} depend on the quantum numbers $\{\mAn\}$, thus we can replace the sum over $\{\mAn\}$ by a sum over these variables, that is
\begin{equation}\label{eq:sust_sum_bar-a_chi_chip}
\sum_{\{\!\mAn\!\}}\,\rightarrow\,\sum_{X_{A}^{\prime}\!(t),\,X_{A}}\,\widehat{\alpha}\big(X_{A}^{\prime}(t),X_{A}\big),
\end{equation}
where the factor $\widehat{\alpha}\big(X_{A}^{\prime}(t),X_{A}\big)$ represents the number of configurations $\{\mAn\}$ that yield the same values for $X_{A}^{\prime}(t)$ and $X_{A}$. We can see from the definitions \numeq{eq:Xp_A} and \numeq{eq:X_A} that their dependence on $\{\mAn\}$ is quite different. In fact, $\varphi_{A,A'}(t)$ and $\opc{S}_{A,A'}$ are different functions of $A'$, besides, the former depends on time while the latter does not. These facts support assuming $X_{A}^{\prime}(t)$ and $X_{A}$ as independent variables, which means that the number of total configurations can be written as the product
\begin{equation}\label{eq:sep_alpha_chi}
\widehat{\alpha}\big(X_{A}^{\prime}(t),X_{A}\big) = \alpha^{\prime}\big(X_{A}^{\prime}(t)\big)\,\alpha(X_{A}),
\end{equation}
that satisfy the following constraints
\begin{equation}\label{eq:N_sum_chi}
\opc{N}_{X^{\prime}} = \sum_{X_{A}^{\prime}\!(t)}\alpha^{\prime}\big(X_{A}^{\prime}(t)\big),\quad
\opc{N}_{X} = \sum_{X_{A}}\alpha(X_{A}),
\end{equation}
and
\begin{equation}\label{eq:rel_N_chi}
\opc{N}_p^{N-1} = \sum_{\{\!\mAn\!\}} 1 = \opc{N}_{X^{\prime}}\,\opc{N}_{X}.
\end{equation}

Let us now recognize that the sum over ${\{\mAn\}}$ in $\widehat{G}_A(t)$ runs over  all the possible combinations of quantum numbers  $\{\ml,\cdots,\mi,\cdots,\mN\}$ with $i \neq A$, so it consists of a macroscopic number of terms. Hence, the variables  $X_{A}^{\prime}(t)$ and $X_{A}$ can be assumed continuous and the following sums can be replaced by integrals
\[\sum_{X_{A}^{\prime}\!(t)} \rightarrow \int dX_{A}^{\prime}\;p^{\prime}\big(X_{A}^{\prime}(t)\big), \qquad \sum_{X_{A}}\rightarrow \int dX_{A}\;p(X_{A}),\]
with the corresponding density functions
\[p^{\prime}\big(X_{A}^{\prime}\!(t)\big) \equiv \alpha^{\prime}\big(X_{A}^{\prime}(t)\big)/\opc{N}_{X^{\prime}},\qquad
p(X_{A}) \equiv \alpha(X_{A})/\opc{N}_{X}.\]
Then, we can write \eq{eq:fdeco_hatGA_cota} as
\begin{equation}\label{eq:fdeco_hatGA_GA_GAp}
\widehat{G}_A(t) \simeq G_A^{\,\prime}(t)\;G_A(t),
\end{equation}
where
\begin{subequations}\label{eq:fdeco_GAp_GA}
\begin{equation}\label{eq:fdeco_GAp}
G_A^{\,\prime}(t) \equiv \int dX_{A}^{\prime}(t)\;p^{\prime}\big(X_{A}^{\prime}(t)\big)\;
e^{-\ic\,2\pi \nuD\left(\kappa_{\mA} - \kappa_{\nA}\right)\,\frac{\csl{a}}{\vs}\,X_{A}^{\prime}\!(t)},
\end{equation}
\begin{equation}\label{eq:fdeco_GA}
G_A(t) \equiv \int dX_{A}\;p(X_{A})\;e^{\,\ic\,4\pi\nuD\,\left(\kappa_{\mA} - \kappa_{\nA}\right)\,t\,X_{A}}.
\end{equation}
\end{subequations}
It is worth to note that functions \numeq{eq:fdeco_GAp_GA} become the Fourier transforms of the probability densities $p'$ and $p$ whenever the integration limits can be extended to $\pm\infty$; and that $G_A^{\,\prime}(t)$ and $G_A(t)$ will be decaying functions for any $L^1$ integrable density.
Accordingly, there is a direct link between the distribution of eigenvalues of $\op{H_I}$ and the decay functions involved in the decoherence process. This feature appears as general characteristic of adiabatic decoherence in various problems. For example, the decoherence rate was also associated with the distribution of interaction Hamiltonian eigenvalues in the case of nematic liquid crystals \cite{SegZam11}.
We may call $G_A^{\,\prime}(t)$ and $G_A(t)$ the \emph{specific decoherence functions} associated with variables $X_{A}^{\prime}(t)$ and $X_{A}$.

Finally, the expression for the pair-reduced density matrix elements in the case of non-interacting spin pairs comes after replacing the approximate functions \numeq{eq:Fga_Fep_Fze_boson-spin_aprox} and \numeq{eq:fdeco_hatGA_GA_GAp} in \eq{eq:Dsigma_boson-spin_hatGA}, that is
\begin{equation}\label{eq:rho_boson-spin}
\begin{split}
\Delta\sigma_{\mA,\nA}(t) &= \Delta\sigma_{\mA,\nA}(0)\;e^{\,\ic\,2\pi\nu_0\,\left(\kappa_{\mA} - \kappa_{\nA}\right)\,t}\\
&\quad\times e^{-\left(\kappa_{\mA} - \kappa_{\nA}\right)^2\,t/\tau_{\gamma}}\,
e^{\,\ic\,2\pi\nuD\,\left(\kappa_{\mA}^2 - \kappa_{\nA}^2\right)\,t}\,G_A^{\,\prime}(t)\,G_A(t),
\end{split}
\end{equation}
with $\nuD$ from \numeq{eq:def_nu_vep} and
\begin{equation}\label{eq:def_nu-tau}
\nu_0 \equiv -\frac{\Omega_0(d\,)}{4\pi}, \qquad \tau_{\gamma}^{-1} \equiv \frac{9\,\Omega_0^{\,2}(d\,)\,K_B\,T\,\csl{a}}{16\;\vs^{\,3}\,m_p}.
\end{equation}

At this point we can already estimate the time constants involved in \eq{eq:rho_boson-spin}.
We first evaluate the dipolar coupling using \eq{eq:def_Omega0} with $d$ as in assumption \ref{assum_param}, and set
$\theta_{\hat{d},\hat{z}} = 0^\circ$, to obtain $\Omega_0(d\,) \simeq  -210.7\times10^{3}\un{s}^{-1}$. Then, the constants adopt the following values
\begin{equation}\label{eq:const_frec_Fga_Fep_Fze_aprox_val}
\nuD \simeq 1.2 \times 10^{-8}\un{kHz},\quad \nu_0 \simeq 16.8\un{kHz},\quad \tau_{\gamma} \simeq 1926\un{s}.
\end{equation}
The fact that  $\nuD\ll \nu_0$ allows us to assume
\begin{equation}\label{eq:aprox_exp_nu_vep}
e^{\,\ic\,2\pi\nuD\,\left(\kappa_{\mA}^2 - \kappa_{\nA}^2\right)\,t} \simeq 1
\end{equation}
in the second row of \eq{eq:rho_boson-spin}.
Besides this, since  $\abs{\kappa_{\mA} - \kappa_{\nA}}_{max} = 3$, the minimum decay constant of the exponential containing
$\tau_{\gamma}$ is
\begin{equation}\label{eq:tau_G_min}
\tau_{\gamma(min)} = \tau_{\gamma}/\abs{\kappa_{\mA} - \kappa_{\nA}}_{max}^2 \simeq 214\un{s}.
\end{equation}

It still remains discussing the time dependence introduced by  $G_A(t)$ and $G^{\prime}_A(t)$ in \eq{eq:rho_boson-spin}. To solve \eq{eq:fdeco_GAp_GA} we need an expression for the density functions, then, let us first examine the geometric weights $\varphi_{A,A'}(t)$ and $\opc{S}_{A,A'}$ involved in the variables $X_{A}^{\prime}(t)$ and $X_{A}$.
We see from \eq{eq:Xp_A} that since $\varphi_{A,A'}(t)/\pi$ can essentially be 0 or 1, the number of terms involved in the sum over $A'$ varies with the time and may be less or equal to $N$. Concerning the sinc function $\opc{S}_{A,A'}$ in \eq{eq:X_A}, it also has the effect of reducing the number of terms involved in the sum over $A'$ since we can safely assume that is equal to 1 for a set of pairs where $\opc{S}_{A,A'}$ is close to its maximum value and 0 for the rest of pairs. 	
In view of this feature, we consider that both $X_{A}^{\prime}(t)$ and $X_{A}$ are sums of eigenvalues. In \apen{app:dist_eig} we show that the density function for this kind of variables for large $N$ has a Gaussian form. Then, we assume that $p^{\prime}\big(X_{A}^{\prime}(t)\big)$ and $p(X_{A})$ are Gaussian distributions, with standard deviations $\sigma_{X^{\prime}}(t)$ and $\sigma_{X}$, centered at $\overline{X}^{\,\prime}(t) = \overline{X} = 0$, and that they are narrow enough so we can extend the integration limits to infinity, thus
\begin{subequations}\label{eq:GGp_Gauss}
\begin{align}
G_A^{\,\prime}(t) &\simeq e^{-\left[\sqrt{2}\,\pi\nuD\,\left(\kappa_{\mA} - \,\kappa_{\nA}\right)\,\sigma_{X^{\prime}}(t)\;\csl{a}/\vs\right]^2},\label{eq:intChip_Gauss}\\
{\rm and} \quad G_A(t) &\simeq  e^{-\left[\left(\kappa_{\mA} - \,\kappa_{\nA}\right)\,t/\tau_{X}\right]^2},\label{eq:intChi_Gauss}
\end{align}
\end{subequations}
where we introduced the characteristic time
\begin{equation}\label{eq:tau_chi}
\tau_{X} \equiv \left[2\sqrt{2}\,\pi\nuD\,\sigma_{X}\right]^{-1}.
\end{equation}
In this way, we obtain an expression of the decoherence rate in terms of the physical constants of the model
\begin{equation}\label{eq:tau_chi_inv}
\frac{1}{\tau_{X}} = \frac{9}{8\sqrt{2}}\,\frac{\Omega_0^{\,2}\,\hbar\,\sigma_{X}}{\vs^{\,2}\,m_p}
= \sqrt{2}\,\pi^2\,\frac{\hat{\nu}_0^2\,\hbar\,\sigma_{X}}{\vs^{\,2}\,m_p},
\end{equation}
with the dipolar frequency $\hat{\nu}_0 \equiv 3\abs{\nu_0} = 3\,\Omega_0/(4\pi)$.\\

Finally, in order to estimate the characteristic time scale of the functions $G_A(t)$ and $G^{\prime}_A(t)$, we need to fix some criteria related to the particular model system:
\renewcommand{\theenumi}{\Roman{enumi}}
\begin{enumerate}
    \item \label{crit_latt-or}
        In a generic lattice with parameters $\csl{a},\csl{b},\csl{c}$ the vector joining pairs (sites) $A$ and $A'$ can be written
        \[r_{A,A'} = \sqrt{(q_x\,\csl{b})^2 + (q_y\,\csl{c})^2 + (q_z\,\csl{a})^2},\] with $q_x, q_y, q_z \in \cbb{Z}$.

    \item \label{crit_NX} Let $\hat{d} \parallel \hat z$, then $r_{A,A'}\cos\left(\theta_{A,A'}\right)/\csl{a} = q_z$ (due to assumption \ref{assum_latt-or}).
        This implies that \eq{eq:Sinc_SAAp} turns into the sinc function $\opc{S}_{A,A'} = \sin\left(\pi q_z\right)/\left(\pi q_z\right)$, which is zero
        for every integer $q_z$ except for $q_z = 0$, that is, for the sites lying on a plane perpendicular to $\hat{d}$. In other words, only a two-dimensional set of sites $A'$ contributes to the three-dimensional sum in the definition of $X_A$ (see \eq{eq:X_A}), then if the complete sample has, say $N \approx 1\times10^{23}$ sites, only $N_{X}\approx N^{2/3} \simeq 2.15\times10^{15}$ of them contribute to $X_{A}$.

    \item \label{crit_sigma} Using the results of \apen{app:dist_eig}, the width of $p(X_{A})$ is $\sigma_{X} = \sqrt{3 N_{X}/2} \simeq 5.68\times10^{7}$.

    \item \label{crit_NXp} Concerning $\sigma_{X^{\prime}}(t)$, let us appraise the maximum effectiveness of $G_A^{\,\prime}(t)$ on the attenuation of $\Delta\sigma_{\mA,\nA}(t)$. We consider that all the pairs of the sample contribute to $X_{A}^{\prime}(t)$ (see \eq{eq:Xp_A}). This assumption is supported by the dependence of $\varphi_{A,A'}(t)$ on time, which is shown in \fig{fig:Fig_5} (\apen{app:calc_Deco_func_1Dk}). We can assume that over a short time scale (compared with the NMR experiment) the width becomes constant $\sigma_{X^{\prime}}(t) = \sqrt{3N/2} \simeq 3.87\times10^{11}$ (see \apen{app:dist_eig}).
\end{enumerate}

Using criteria \ref{crit_latt-or}, \ref{crit_NX}, and \ref{crit_sigma}, we can estimate the characteristic time involved in $G_A(t)$ as
\begin{equation}\label{eq:tau_U_num}
\tau_{X} \simeq 165\un{\mu s},
\end{equation}
which is several orders of magnitude smaller than the characteristic time $\tau_{\gamma}$ in \numeq{eq:tau_G_min}.\\
Based on \ref{crit_NXp} we get that the minimum value of \numeq{eq:intChip_Gauss} is very close to unity, this is
\begin{equation}\label{eq:intChip_Gauss_val}
G_A^{\,\prime}(t)|_{min} \simeq 1.
\end{equation}

In \fig{fig:Fig_2}(a)  we show the variation of $\tau_{X}$ as a function of the sample size $N$ and the speed of sound $\vs$ (with the other parameters as in assumption \numeq{assum_param} in \sect{sec:app_calc_tdeco}). We see that $\tau_{X}$ varies smoothly  in spite of the rather wide parameter range ($10^{21} \leq N \leq 10^{25}$ and $2000\un{m/s} \leq \vs \leq 8000\un{m/s}$).

\insertfig{ht}{fig:Fig_2}{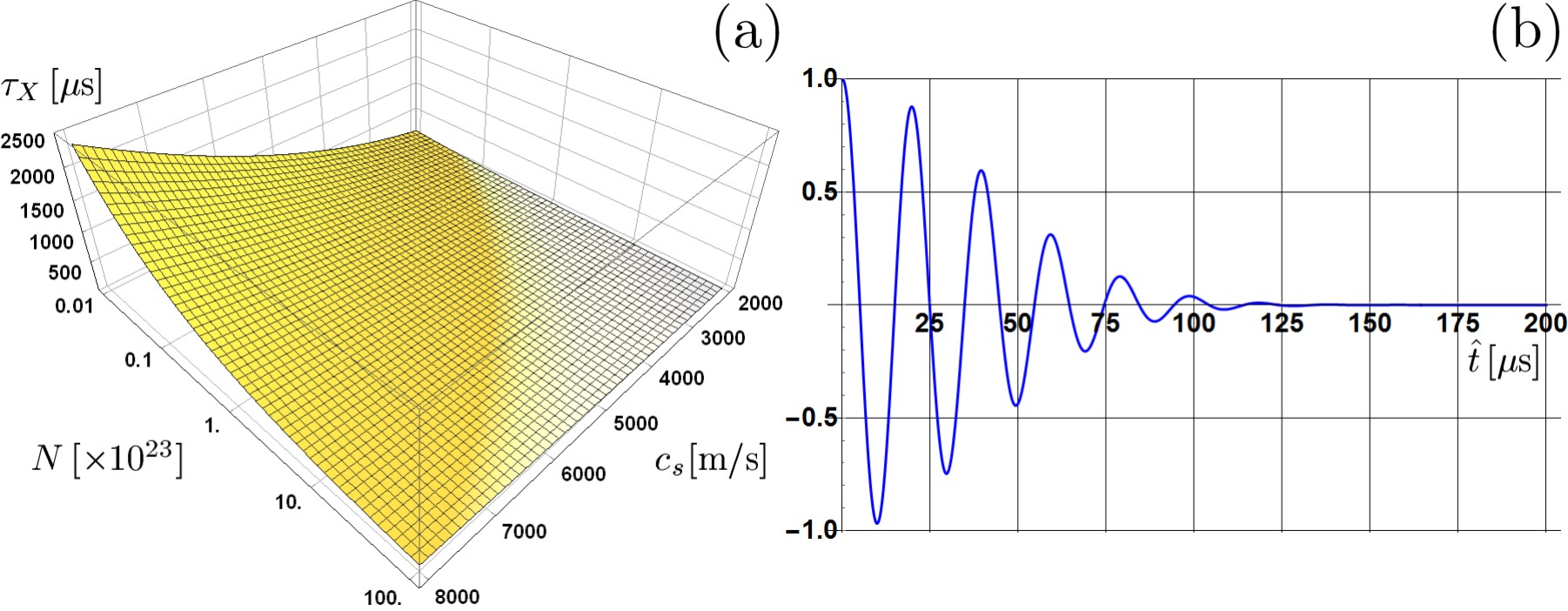}{13.5}{5.3}{(a) Decay time $\tau_{X}$ dependence on the number of pairs $N$ (log scale) and  the speed of sound $\vs$, for the parameters: intra-pair distance $d = 0.153\un{nm}$, lattice distance $\csl{a} = 0.8\un{nm}$, and wavevector along the intra-pair direction. The extreme values are $\tau_{X(max)} \simeq 2343\un{\mu s}$ (for $N = 10^{21}$ and $\vs = 8000\un{m/s}$), and $\tau_{X(min)} \simeq 6.8\un{\mu s}$ (for $N = 10^{25}$ and $\vs = 2000\un{m/s}$).
(b) Real part of the time evolution of the reduced density matrix elements \numeq{eq:Deltarho_boson-spin_aprox_fin} for $\Delta\sigma_{\mA,\nA}(0) = 1$, i.e. $\cos\left(2\pi\,\hat{\nu}_0\,\hat{t}\,\right)\exp\big[\!-\!\left(\,\hat{t}\,/\,\hat{\tau}_{X}\right)^2\big]$, 	with $\hat{t}$ the normalized time, $\hat{\nu}_0 = 50.4\un{kHz}$ and $\hat{\tau}_{X} = 55\un{\mu s}$.}

In this way, we evaluated the magnitudes that determine the time dependence of the  reduced density matrix elements of \eq{eq:rho_boson-spin}. Given the approximations \numeq{eq:aprox_exp_nu_vep} and \numeq{eq:intChip_Gauss_val} as well as the fact that $\tau_{\gamma(min)} \gg \tau_{X}$ (compare \numeq{eq:tau_G_min} and \numeq {eq:tau_U_num}), we may finally write the reduced density matrix elements in terms of the initial state, the dipolar frequency $\nu_0$ and the decay time $\tau_{X}$ as
\begin{equation}\label{eq:Deltarho_boson-spin_aprox_fin}
\Delta\sigma_{\mA,\nA}(t) \simeq \Delta\sigma_{\mA,\nA}(0)\;e^{\,\ic\,2 \pi \nu_0\left(\kappa_{\mA} - \,\kappa_{\nA}\right)\,t}\;
e^{-\left[\left(\kappa_{\mA} - \,\kappa_{\nA}\right)\,t/\tau_{X}\right]^2}.
\end{equation}
Therefore, each element of the pair-reduced density matrix decreases in time with different decay rates given by $\abs{\kappa_{\mA} -\kappa_{\nA}}/\tau_{X}$.
To illustrate the matrix element dynamics, in \fig{fig:Fig_2}(b) we plot the real part of \eq{eq:Deltarho_boson-spin_aprox_fin}
for $\Delta\sigma_{\mA,\nA}(0) = 1$, using the constants and approximations declared above. The abscissa is a normalized time $\hat{t} \equiv \abs{\kappa_{\mA} - \kappa_{\nA}}\,t/3$. The time evolution of the different matrix elements can be visualized by just multiplying $\hat t$ by
$3/\abs{\kappa_{\mA} - \kappa_{\nA}}$.

Equation \numeq{eq:Deltarho_boson-spin_aprox_fin} represents the reduced dynamics of a spin pair in a model where the basic units do not interact directly with each other; however, it keeps complete information about the collective spin pair correlation through the common phonon bath.
Though system Hamiltonian $\HamS$ used in this model may not be sufficient to account for the spectroscopic aspects of real solids, the decoherence time scale depends on the eigenvalues of $\HamI$ and not on the particular model used in  $\HamS$.
The estimated value of $\tau_{X}$ gives the decoherence time scale predicted by the pair-phonon model. The fact that it is consistent with the time scale of signal attenuation in solid-state NMR experiments \cite{SkrebZaripov_99,DomZamSegGzz17,LevsHirsc_PhA2000,Kroj_Suter06} indicates that the proposed system-environment coupling is a plausible decoherence mechanism.	
To compare our result with actual data, in the next section, we deduce the reduced density matrix and the signal amplitude corresponding to a paradigmatic NMR reversion experiment within the open-quantum approach.

\section{Decoherence in a reversion experiment} \label{sec:din_rev}

This section applies the methodology developed so far to evaluate decoherence within a paradigmatic reversion experiment called  ``magic-echo''  (ME) \cite{RhimPinesVaugh_71}. We first outline the experiment within the spirit of  open-quantum systems. Then, we use the strategy of \sects{sec:din_sist-boson} and \nsect{sec:app_pair_phon} to write the corresponding reduced density matrices and finally calculate the outcome ME signal amplitude  (with decoherence) and compare its decoherence rate with observed behavior.

It is nowadays accepted that the coupling between an observed quantum system and its environment poses an unbeatable limit to reversibility
\cite{MorelloStamp_06,Stamp_06,Petitjean_PRL06-AdPh09}.
A variety of clever ``time-reversal'' NMR experiments have the effect of compensating the Liouvillian evolution under $\HamS$ to isolate (or at least spotlight on) the effect of decoherence.
Much experimental work has been devoted to reduce imperfections in reversion experiments, evaluate the effect of non-idealities (high-field approximation, rf field homogeneity, pulse shape), or even mitigate decoherence. In addition, they recognize that an insurmountable natural limitation exists even for any hypothetical unlimited improvement of the reversal \cite{SkrebZaripov_99,SanchezCapellaro_PRL20,Buljubasichetal_JCP15,AlvarezSuter_RevModP16,Kroj_Suter06}.
However, there is still no consensus on its origin. To validate our proposal, we calculate the outcome of a theoretical ME sequence, particularize it to a model system and compare it with experimental data.

\subsection{Evolution operator of an open system in the magic-echo sequence} \label{sec:U_ME}

The experimental scheme of reversion experiments generally consists of a period where the system evolves ``forward'', and a period where the system evolves ``backward'' undoing part of the forward evolution. Particularly in the ME pulse sequence, the backward evolution is designed to revert (or at least minimize) evolution under the secular dipolar Hamiltonian. The NMR signal amplitude measured after the sequence (also called magic echo) witnesses any coherence loss, either due to experimental causes (defects, evolution under the non-secular part of the dipolar Hamiltonian) or to the irreversible loss due to system-environment coupling. Figure \ref{fig:Fig_3} shows a diagram of the ME sequence. The first $(\pi/2)$-pulse leaves the spin system in the state described by \eq{eq:Dsigma_inicial}, after that comes a \textbf{free-evolution} period during the (forward) time $t_F$ under the forward-evolution operator
\begin{equation}\label{eq:UF}
\op{U}_F(t_F) = e^{-\ic\HamS\,t_F}\,e^{-\ic\left(\HamI + \HamE\right)\,t_F},
\end{equation}
just as in \eq{eq:op_evol_full_ad}.

\insertfig{ht}{fig:Fig_3}{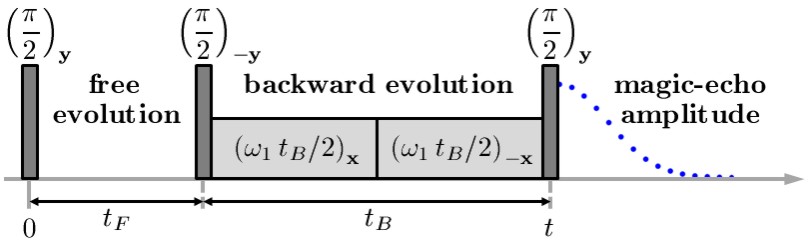}{10}{3}{Pulse sequence of the magic echo experiment. The ``backward'' evolution block of duration $t_B=2\,t_F$ reverts the free, or ``forward'', evolution under the secular dipolar Hamiltonian during $t_F$. The magic-echo amplitude measured after the last $(\pi/2)$-pulse is attenuated due to decoherence during $t_F+t_B$.}

The \textbf{backward evolution} is achieved by irradiating the sample during time $t_B$ with two long soft pulses of amplitude $\omega_1$ (in frequency units) and alternating phases $\op{x}$, $-\op{x}$, which are ``sandwiched'' by two $\pi/2$ hard pulses of phases $\op{y}$ and $-\op{y}$. 
The backward-evolution operator can be expressed as
\begin{equation}\label{eq:UB_sep}
\op{U}_B(t_B) \equiv \op{U}^{(-\op{x})}_B(t_B/2)\,\op{U}^{(+\op{x})}_B(t_B/2),
\end{equation}
with the evolution of each soft-pulse block (in a rotating frame on resonance) defined as
\begin{equation}\label{eq:UB_x}
\op{U}^{(\pm\op{x})}_B(t_B/2) \equiv \op{R}_\op{y}(\pi/2)\,e^{-\ic\,\left(t_B/2\right)\left[\pm\omega_1 \op{I}_\op{x} + \HamS +
	\left(\HamI + \HamE\right)\right]}\,\op{R}_\op{-y}(\pi/2),
\end{equation}
where $\op{R}_\op{\pm y}(\pi/2)$ are the hard pulses that rotate the exponential operator through $\pi/2$ around the $y$-axis (thus $\op{R}_\op{-y}(\pi/2)\,\op{R}_\op{y}(\pi/2) = \op{1}$ between the soft-pulse blocks),
$\pm\omega_1 \op{I}_\op{x}$ is the Hamiltonian of the rf pulse of length $t_B/2$, phase $\pm\op{x}$, and intensity $B_1$, so that $\omega_1 \equiv \gamma_p\,B_1$.

The novelty in \eq{eq:UB_x} is that we consider the spin system as open, since evolution under $\left(\HamI + \HamE\right)$ accounts for the coupling with the environment. Using the Hamiltonians defined in \eqs{eq:def_Ham_sys_onres}, \numeq{eq:def_Ham_Ia0_intra} and \numeq{eq:Ham_bos}, we may write
\begin{equation}\label{eq:UB_x_rot}
\op{U}^{(\pm\op{x})}_B(t_B/2) = e^{-\ic\,\left(t_B/2\right)\left[\pm\omega_1 \op{I}_\op{z} + \fB\eB\HamS + \HamS^\op{ns} +
	\left(\fB\eB\HamI + \HamI^\op{ns} + \HamE \right) \right]},
\end{equation}
where  $\op{I}_\op{z} = \op{R}_\op{y}(\pi/2)\,\op{I}_\op{x}\,\op{R}_\op{-y}(\pi/2)$, and the constant $\fB$ represents the reversion block efficiency and takes the value $\fB = - 1/2$ in a perfect ME experiment. Rotation of $\HamS$ and $\HamI$ involve the identity
\[\op{R}_\op{y}(\pi/2)\,\op{T}_{2,0}\,\op{R}_\op{-y}(\pi/2) = -\frac{1}{2}\op{T}_{2,0} + \sqrt{\frac{3}{8}}\left(\op{T}_{2,+2} + \op{T}_{2,-2}\right),\]
where $\op{T}_{2,0}$ becomes multiplied by the factor $- 1/2$ (notice the minus sign); thus, $\fB =-1/2$.
The term involving  $\op{T}_{2,\pm2}$ gives rise to the non-secular Hamiltonians $\HamS^\op{ns}$ and $\HamI^\op{ns}$.

Setting a high-intensity rf field $B_1$ (large $\omega_1$) allows using the high-field approximation \cite{slichbook,abragol82} in \eq{eq:UB_x_rot} to discard the dynamics produced by $\HamS^\op{ns}$ and $\HamI^\op{ns}$ (they do not commute with $\op{I}_\op{z}$).
It is worth mentioning that different authors test the experimental validity of this condition by analyzing the reversion efficiency at increasing pulse amplitude $B_1$ \cite{SkrebZaripov_99,DomZamSegGzz17,LevsHirsc_PhA2000}. In different samples, the ability of reversion increases with the pulse amplitude at low $B_1$  and reaches a clear limit beyond which it saturates. The maximum characteristic attenuation time defines the experimental decoherence time.

In this way, using the adiabatic condition, we may write
\begin{equation}\label{eq:UB_x_rot_hi}
\op{U}^{(\pm\op{x})}_B(t_B/2) = e^{\mp\ic\,\omega_1 \op{I}_\op{z}\,t_B/2}\,e^{-\ic\,\fB\eB\HamS\,t_B/2}\,e^{-\ic\left(\fB\eB\HamI + \HamE\right)\,t_B/2},
\end{equation}
and using \numeq{eq:UB_x_rot_hi} in \eq{eq:UB_sep} we  obtain
\begin{equation}\label{eq:UB}
\op{U}_B(t_B) = e^{-\ic\,\fB\eB\HamS\,t_B}\,e^{-\ic\left(\fB\eB\HamI + \HamE\right)\,t_B},
\end{equation}
where the evolution introduced by the long soft-pulses cancels.
Finally, the combined forward-backward  evolution operator of an open quantum system is
\begin{equation}\label{eq:U_delta}
\op{U}_\fB(t_F,t_B) \equiv \op{U}_B(t_B)\,\op{U}_F(t_F) = e^{-\ic\HamS\,\left(t_F + \fB\,t_B\,\right)}\,
e^{-\ic\left(\fB\eB\HamI + \HamE\right)\,t_B}\,e^{-\ic\left(\HamI + \HamE\right)\,t_F},
\end{equation}
where we can see that setting $t_B = 2\,t_F$ and $\fB = - 1/2$, cancels the evolution under $\HamS$. Also, notice that evolution under $\HamI$ in the forward dynamics has an opposite sign than in the backward period while  $\HamE$ has the same sign (forward).  Besides, since $\HamI$ and $\HamE$ do not commute, evolution under $\HamI$ cannot be suppressed.

In a closed spin system the backward period, would counteract the free evolution exactly. On the contrary, in an open quantum system the effect of  system-environment correlation cannot be winded back. Thus, decoherence and irreversibility derive from considering both the system and the environment as quantum objects and from the impossibility of controlling the environment variables.
The observable effect of evolution under $\op{U}_\fB(t_F,t_B)$ is that the reverted signal amplitude after the ME sequence becomes attenuated compared to the FID amplitude after a single $\pi/2$ pulse.

The evolution operator \numeq{eq:U_delta} is a good representation of an ideal ME sequence with experimental conditions set to drastically minimize error sources and the influence of possible non-refocused interactions. The decoherent dynamics under $\op{U}_\fB(t_F,t_B)$ should then give a very close description of the inherent decay expected in a refocusing experiment.

\subsection{Decoherence under reversion} \label{app:deco_rev_exp}

In this section, we use the evolution operator \numeq{eq:U_delta} to calculate the decoherence function in a reversion experiment.
According with the properties of $\HamI$ and $\HamE$ defined in \sect{sec:Hamiltonians}, we write the action of $\op{U}_\fB(t_F,t_B)$ on the spin states (see \eq{eq:OpEvol_eigen}) as
\begin{equation}\label{eq:U_delta_eigen}
\op{U}_\fB(t_F,t_B)\ket{m} = e^{-\ic E_m\,\left(t_F + \fB\,t_B\,\right)}\,
\prod_{\bf k} e^{-\ic\left(\op{M} + \fB\eB\op{J}_m\right)_{\bf{k}}^\he t_B}\,e^{-\ic\left(\op{M} + \op{J}_m\right)_{\bf{k}}^\he t_F}\ket{m}.
\end{equation}
With the same procedure used in \eq{eq:rho_sist_red_ad}  (\sect{sec:din_dens_red_obs}), the density matrix elements reduced over the environment variables after the forward-backward dynamics, in the eigenbasis of \eq{eq:U_delta_eigen}, are
\begin{equation}\label{eq:rho_sist_red_Ud}
\begin{split}
\rho_{m,n}(t_F,t_B) &= \trs{E}{\bra{m} \op{U}_\fB(t_F,t_B)\,\rho(0)\,\op{U}_\fB^{\dagger}(t_F,t_B) \ket{n}}\\
&= \rho_{m,n}(0)\;e^{-\ic(E_m - E_n)\,\left(t_F + \fB\,t_B\,\right)}\,\prod_{\bf{k}} S_{m,n}^{\bf{k}}(t_F,t_B),
\end{split}
\end{equation}
where the new decoherence function is
\begin{equation}\label{eq:fundeco_Ud}
\begin{split}
S_{m,n}^{\bf{k}}(t_F,t_B) &= \trbs{E}{e^{-\ic\left(\op{M} + \fB\eB\op{J}_m\right)\,t_B}\,e^{-\ic\left(\op{M} + \op{J}_m\right)\,t_F}
	\,\op{\Theta}\\
	&\qquad\qquad\qquad\times e^{\,\ic\left(\op{M} + \op{J}_m\right)\,t_F}\,e^{\,\ic\left(\op{M} + \fB\eB\op{J}_m\right)\,t_B}}_{\bf{k}}^\he.
\end{split}
\end{equation}
We can compute the trace in \eq{eq:fundeco_Ud} using the procedure developed in \apen{app:calc_Int_EC}. The calculation involves now five integrals with the form \numeq{eq:Int_gral_S}. As in \eq{eq:des_Smnkl_EC}, to deal with exponential operators
$e^{\pm\ic\left(\op{M} + \fB\eB\op{J}_m\right)\,t}$, it is convenient to define the displacement operator
$\op{a}_m \equiv \op{b} + \fB\,\lambda_m/\omega$ with eigenstates $\ket{\cur{z}_m}$ (which may differ from $\ket{z}$ at most by a phase factor, similar to \numeq{eq:prop_eigen_b-am}) and eigenvalues $\cur{z}_m \equiv z + \fB\,\lambda_m/\omega$.
This displacement operator allows writing (similar to \eq{eq:def_Ndesp})
\begin{equation}\label{eq:def_Ndesp_delta}
\op M + \fB\eB\op J_m  = \omega \op{\widehat{N}}_m - \fB^2 \frac{\left| \lambda_m\right|^2 }{\omega},
\end{equation}
which leads to the relation
\begin{equation}\label{eq:des_exp_gm_ECb_delta}
\bra{z_1}e^{\pm\ic(\op M + \fB\eB\op J_m) t}\ket{z_2} = e^{\mp\ic\,\omega t\,\fB^2\frac{\abs{\lambda_m}^2}{\omega^2}}\,\braket{z_1}{z_2}\,
e^{\left(z_1^*+\fB\frac{\lambda_m^*}{\omega}\right)\left(e^{\pm\ic\,\omega t}-1\right)\left(z_2+\fB\frac{\lambda_m}{\omega}\right)},
\end{equation}
as in \eq{eq:des_exp_gm_ECb}.

After some algebra, we may finally write
\begin{equation}\label{eq:Func_Smnkl_des_delta}
S_{m,n}^{\bf{k}}(t_F,t_B) = e^{-\Gamma_{m,n}^{\bf{k}}(t_F,t_B)}\,e^{-\ic\Upsilon_{m,n}^{\bf{k}}(t_F,t_B)},
\end{equation}
with
\begin{subequations}\label{eq:Func_Gmnkl_Umnkl_delta}
\begin{equation}\label{eq:Func_Gmnkl_delta}
	\Gamma_{m,n}^{\bf{k}}(t_F,t_B) \equiv \bigg\{\frac{\abs{\lambda_{m}-\lambda_{n}}^2}{\omega^2}\,
	\coth\left(\beta\,\omega/2\right)\,\FC(t_F,t_B)\bigg\}_{\bf{k}},
\end{equation}
\begin{equation}\label{eq:Func_Umnkl_delta}
\begin{split}
	\Upsilon_{m,n}^{\bf{k}}(t_F,t_B)
	&\equiv \bigg\{\frac{\left(\lambda_m-\lambda_n\right)
		\left(\lambda_m+\lambda_n\right)^*}{\omega^2}\left[\FS(t_F,t_B)-\omega\left(t_F + \fB^2\,t_B\,\right)\right]\\
	&\;\,-\frac{2\,\Im\left\{\lambda_m\lambda_n^*\right\}}{\omega^2}\bigg[\FC(t_F,t_B)\;
	+\ic\left[\FS(t_F,t_B)-\omega\left(t_F + \fB^2\,t_B\,\right)\right]\bigg]\bigg\}_{\bf{k}},
\end{split}
\end{equation}
\end{subequations}
where
\begin{subequations}\label{eq:Func_Cdelta_Sdelta}
\begin{equation}\label{eq:Func_Cdelta}
\begin{split}
	\FC(t_F,t_B) &\equiv  \left(1-\fB\right)\left[1-\cos\left(\omega\,t_F\right)\right] + \fB\left(\fB-1\right)\left[1-\cos\left(\omega\,t_B\right)\right]\\
	&\quad + \fB\left[1-\cos\left(\omega\,(t_F+t_B)\right)\right],
\end{split}
\end{equation}
\begin{equation}\label{eq:Func_Sdelta}
\begin{split}
	\FS(t_F,t_B) \equiv \left(1-\fB\right)\sin\left(\omega\,t_F\right) + \fB\left(\fB-1\right)\sin\left(\omega\,t_B\right)
	+ \fB\sin\left(\omega\,(t_F+t_B)\right).
\end{split}
\end{equation}
\end{subequations}
It is worth emphasizing some aspects of equations \numeq{eq:Func_Gmnkl_Umnkl_delta}
\begin{itemize}
\item The displacement constants in \eqs{eq:def_Ndesp_delta} and \numeq{eq:def_Ndesp} introduce the complex exponentials with exponents depending linearly on $t$ in \eq{eq:Func_Smnkl_des_delta} (see \numeq{eq:des_exp_gm_ECb_delta} and \numeq{eq:des_exp_gm_ECb}); this is, the linear time terms in \eq{eq:Func_Umnkl_delta}. Such exponentials constitute the essence of function $G_A(t)$ in \eq{eq:fdeco_GA} and, thus, of the emergence of the leading irreversible dynamics.
\item Similar to \eq{eq:Func_Gmnkl_Umnkl}, \eqs{eq:Func_Gmnkl_Umnkl_delta} are also eigen-selective. Therefore, the decoherence function \numeq{eq:Func_Smnkl_des_delta} has the same characteristics discussed at the end of \sect{sec:din_dens_red_obs}.
\item Equations \numeq{eq:Func_Gmnkl_Umnkl_delta} depend on $\fB$, which in principle can be assigned any value, depending on the particular reversion scheme.
\end{itemize}
For example, by setting $\fB = 1$ we recover the case of free evolution without reversion. In this case, the compound evolution operator is the product
\begin{equation}\label{eq:Udelta_1}
\op{U}_{\fB = 1}(t_F,t_B) = \op{U}_F(t_B)\,\op{U}_F(t_F) = \op{U}_F(t_F + t_B).
\end{equation}
which is the same as \numeq{eq:op_evol_full_ad}, valued in $t = t_F + t_B$.
Therefore, we obtain the same functions as \eqs{eq:Func_Gmnkl_Umnkl}, showing that decoherence in \numeq{eq:Func_Smnkl_des_delta} is additive.
	
The next section is devoted to the case of interest, the ME experiment, where $\fB = -1/2$.

\subsection{Reversion dynamics in the pair-phonon model} \label{sec:sigma_ME}

Let us consider the case $\fB = -1/2$ and $t_B = 2\,t_F$, and use $\HamI$ as in \eq{eq:def_Ham_Ia0_intra}.
Setting $t = t_F + t_B = 3\,t_F = 3\,t_B/2$
in \numeq{eq:Func_Cdelta_Sdelta} and \numeq{eq:Func_Gmnkl_Umnkl_delta},
we may write the density matrix reduced over the environment \numeq{eq:rho_sist_red_Ud} as a function of the total time in the ME sequence as
\begin{equation}\label{eq:rho_sist_red_Ud_B}
    \rho_{m,n}^\sB(t) = \rho_{m,n}(0)\;\prod_{\bf{k}} S_{m,n}^{\sB\,\bf{k}}(t),
\end{equation}
where we use a triangle with a vertex pointing backwards to denote the ME reverted dynamics.
Note that \eq{eq:rho_sist_red_Ud_B} differs from \eq{eq:rho_sist_red_ad} in the absence of the factor with exponent $\left(E_{\mA}\! - E_{\nA}\right)$, canceled by the reversion procedure.
The corresponding decoherence function is
\begin{equation}\label{eq:Func_Smnkl_des_delta_B}
    S_{m,n}^{\sB\,\bf{k}}(t) = e^{-\Gamma_{m,n}^{\sB\,\bf{k}}(t)}\,e^{-\ic\Upsilon_{m,n}^{\sB\,\bf{k}}(t)},
\end{equation}
with $\Gamma_{m,n}^{\sB\,\bf{k}}(t)$ and $\Upsilon_{m,n}^{\sB\,\bf{k}}(t)$ as in \numeq{eq:Func_Gmnkl_Umnkl_delta} but replacing
\[\left(t_F + \fB^2\,t_B\,\right) \,\rightarrow\, t/2,\quad \FC(t_F,t_B) \,\rightarrow\, \FC_\sB(t),\quad \FS(t_F,t_B) \,\rightarrow\, \FS_\sB(t),\]
which involve
\begin{subequations}\label{eq:Func_Cdelta_Sdelta_B}
	\begin{equation}\label{eq:Func_Cdelta_B}
	\begin{split}
	\FC_\sB(t) =  \frac{3}{2}\left[1-\cos\left(\omega\,t/3\right)\right] + \frac{3}{4}\left[1-\cos\left(2\,\omega\,t/3\right)\right]
	- \frac{1}{2}\left[1-\cos\left(\omega\,t\right)\right],
	\end{split}
	\end{equation}
	\begin{equation}\label{eq:Func_Sdelta_B}
	\begin{split}
	\FS_\sB(t) = \frac{3}{2} \sin\left(\omega\,t/3\right) + \frac{3}{4}\sin\left(2\,\omega\,t/3\right)
	- \frac{1}{2}\sin\left(\omega\,t\right).
	\end{split}
	\end{equation}
\end{subequations}
	
Using the same procedure of \sect{sec:dyn_sigmaA}, we may now write the reduced density operator
$\sigma^{\sB\hA}(t) = \trs{\oS_A}{\rho^{\sB\hs}(t)}$ relevant for calculating the expectation value of a local observable. In \apen{app:rev_p-p_model} we derive the reduced density matrix for the ME experiment, applied to the pair-phonon model of \sect{sec:app_sysm_sigma} and obtain
\begin{equation} \label{eq:Deltarho_boson-spin_aprox_fin_B}
    \Delta\sigma^\sB_{\mA,\nA}(t) \simeq \Delta\sigma_{\mA,\nA}(0)\;e^{-\left[\left(\kappa_{\mA} - \,\kappa_{\nA}\right)\,t/\tau^\sB_{X}\right]^2},
\end{equation}
with $\tau^\sB_{X} \equiv 2\,\tau_{X}$ and $\tau_{X}$ defined by \eq{eq:tau_chi} (or \eq{eq:tau_chi_inv}).
This equation is the ME equivalent to the free-evolution  \eq{eq:Deltarho_boson-spin_aprox_fin}. We see that the dynamics under the ME experiment lack the phase factor due to the closed evolution under $\HamS$, while attenuation due to decoherence is slower than during free-evolution.

Synthesizing, after \eq{eq:U_delta}, we concluded that the ME experiment has the effect of reverting (or undoing) the Liouvillian evolution under $\HamS$. On the contrary, the open quantum character of the system introduces an evolution that cannot be reverted by acting only on the system variables.
This irreversible behaviour stems from the dependence on $\fB^2$ in  \eq{eq:Func_Umnkl_delta}, since setting a negative $\fB$ cannot cancel the factor $(t_F + \fB^2\,t_B)$.
This restriction precludes a complete reversion of the forward evolution in an open quantum system. Decoherence due to the coupling with the environment will always be present, even in an ideal perfect reversion experiment; that is, irreversibility manifests as an inherent property of the adiabatic quantum decoherence in open quantum systems.

\subsection{Calculation of the ME signal amplitude and comparison with experiment} \label{sec:ME_amp_exp}

The short-time scale spin dynamics in NMR experiments, that is, the free induction decay signal (FID) or its spectrum, is generally well described by the Liouvillian evolution under a Hamiltonian $\HamS$ that accounts for the spin-spin interactions in detail. Conversely, the experiments show that the irreversible decoherence manifests over the intermediate time scale, longer than the FID, where we may expect that the system-environment correlation (openness) plays a decisive role.
Equation \numeq{eq:Deltarho_boson-spin_aprox_fin_B} shows that decoherence is independent of the particular form of $\HamS$, provided it satisfies the adiabatic condition. Since our goal is to explain the signal attenuation within the decoherence time scale, we need to describe only the signal amplitude and not its spectroscopic details.
In \sect{sec:sigma_ME} we concluded that irreversibility proceeds from the system correlation with the environment; thus, the very simple model we use for $\HamS$ is not an obstacle for obtaining a good estimation of the decoherence time within the pair-phonon model. The system-environment Hamiltonian $\HamI$ considered in this approach is simple enough to allow an analytical description of decoherence by cause of opening the system, yet encompassing characteristics of a real system. In this section we show that it also provides a decoherence rate consistent with those experimentally observed.

The magic-echo signal is the magnetization after the last pulse of the ME sequence.
Setting the times so that $t_B = 2\,t_F$, the maximum signal amplitude occurs immediately after the pulse.
If the system were isolated, an ideal sequence (free from experimental misadjustments) would completely reverse the dynamics, and the ME signal amplitude would be identical to one after a single $(\pi/2)$ pulse. That is, it would not depend on $t$. Therefore, measuring the amplitude just after the ME sequence as a function of the total time $t_F+t_B$ can be used as a witness of decoherence.

In the scheme of this work, we can use \eq{eq:Deltarho_boson-spin_aprox_fin_B} to calculate the NMR signal amplitude after the ME sequence as the expectation value of $\op{I}_\op{x}$.

The initial-state after a single $(\pi/2)$ pulse is represented by   \eq{eq:Dsigma_inicial}, then the matrix elements  $\Delta\sigma_{\mA,\nA}(0)$ involved in \eq{eq:Deltarho_boson-spin_aprox_fin_B}  are
\begin{equation}\label{eq:Dsigma0_exp}
\Delta\sigma_{\mA,\nA}(0) = -\frac{\hbar\,\omega_0}{K_B\,T}\,\bra{\mA}\op{I}_\op{x}^\hA\ket{\nA}.
\end{equation}
In the eigenbasis \numeq{eq:def_mA_kappaA}, the non-zero elements are
\begin{equation}\label{eq:elem_Ix}
\begin{split}
\bra{1,1}\op{I}_\op{x}^\hA\ket{1,0} = \bra{1,-1}\op{I}_\op{x}^\hA\ket{1,0} = 1/\sqrt{2}\,,\quad
\text{with}\;&\left(\kappa_{\mA} - \,\kappa_{\nA}\right) = 3,\\
\bra{1,0}\op{I}_\op{x}^\hA\ket{1,1} = \bra{1,0}\op{I}_\op{x}^\hA\ket{1,-1} = 1/\sqrt{2}\,,\quad
\text{with}\;&\left(\kappa_{\mA} - \,\kappa_{\nA}\right) = -3.
\end{split}
\end{equation}
We now use \eq{eq:redu1} to calculate the expectation value of $\op{I}_\op{x}$ in a partitionable system of $N$ identical spin pairs
\begin{equation}\label{eq:med_Ix}
\langle \op{I}_\op{x} \rangle(t) = N\,\trs{S_A}{\op{I}_\op{x}^\hA\;\Delta\sigma^\sB(t)}
= N\!\sum_{\mA,\nA}\!\Delta\sigma^\sB_{\mA,\nA}(t)\;\bra{\nA}\op{I}_\op{x}^\hA\ket{\mA}.
\end{equation}

Since we are interested in the signal amplitude immediately after the last ME pulse, we set $t= t_F+t_B$ and write
\begin{equation}\label{eq:med_Ix_ME}
\langle \op{I}_\op{x} \rangle(t_F+t_B) = -\frac{\hbar\,\omega_0\,N}{K_B\,T}\!\sum_{\mA,\nA}\!\abs{\bra{\mA}\op{I}_\op{x}^\hA\ket{\nA}}^2\,
e^{-\left[3 \,(t_F+t_B)/\tau_{X}^\sB \right]^2}.
\end{equation}
That is, the normalized signal amplitude in a ME experiment is
\begin{equation}\label{eq:med_SE_dec}
\mathcal{M}_{M\!E}(t_F+t_B) = e^{-\left[\,(t_F+t_B)/\hat{\tau}^\sB_X\right]^2},
\end{equation}
which can be used to measure the decoherence time $\hat{\tau}^\sB_X \equiv \tau^\sB_X/3 = 2\,\tau_X/3$.
This procedure was used, for example, in ref.\cite{DomZamSegGzz17} to measure the decoherence time, defined as $\tau_{\rm exp}$, in a single-crystal sample of gypsum.
The physical constants used in our model system, that bear semblance with a gypsum single-crystal, allow comparing our estimation of $\hat{\tau}^\sB_X$ with the measured decoherence times $\tau_{\rm exp}$.
This is shown in \fig{fig:Fig_4}, where the dots correspond to $\tau_{\rm exp}$ as a function of the dipolar frequency $\hat{\nu}_0$ (in \un{kHz}). The dipolar couplings were varied by rotating the angle between the [100] crystal plane and the static magnetic field.

The theoretical variation of $\hat{\tau}^\sB_X$ vs. $\hat{\nu}_0$ (solid line) is calculated for $\{\vs,N\}$ = $\{4570\un{m/s}$, $4.7\times10^{22}\}$, where the value of $N$ was chosen to fit with the highest frequency measured value.
The shaded zone represents the family of $\hat{\tau}^\sB_X$ curves in the  $\{\vs,N\}$ range, limited by $\{2800\un{m/s}$, $8\times10^{21}\}$ (lower dashed curve) and $\{5500\un{m/s}$, $8\times10^{22}\}$ (upper dashed curve).

We see that the magnitude of the estimated decoherence rate is in excellent agreement with that observed in NMR within the higher frequency interval.
Though we do not aim to fit the data, we draw the solid line in \fig{fig:Fig_4} to show the consistency of our model. There we see that by choosing $N$ so that  $\hat{\tau}^\sB_X$ coincides with the highest frequency data point, a trend is obtained that agrees with the experiment within the  frequency range where the picture of non-interacting pairs is adequate for describing the spin distribution in the actual sample.
It is reasonable to expect the non-interacting pair assumption for $\HamI$ to be unwarranted as the intra-pair dipole energy decreases versus the inter-pair energy.

\insertfig{ht}{fig:Fig_4}{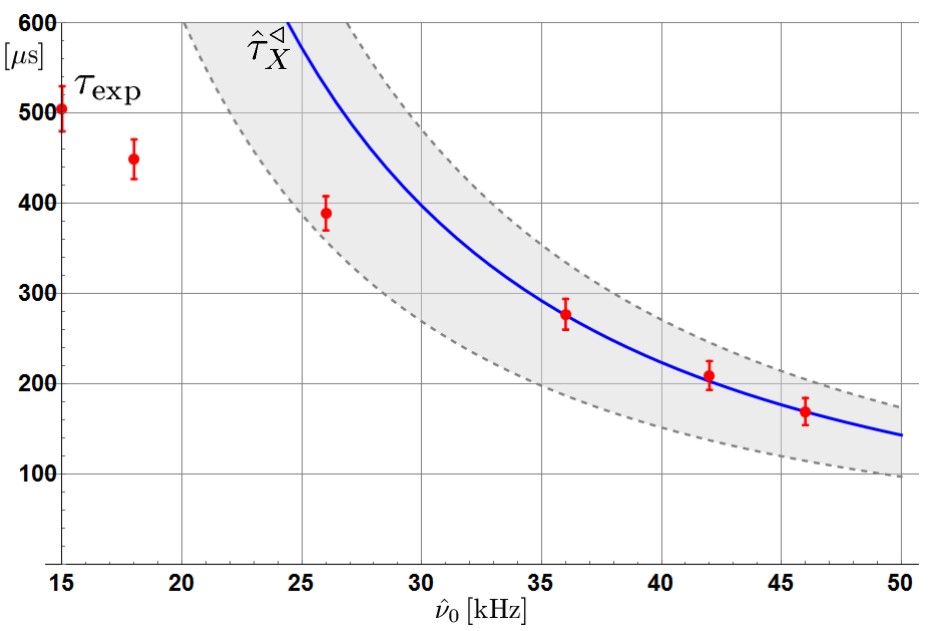}{8}{5.4}{Comparison between the decay constant $\tau_{\rm exp}$ (dots) measured in a single-coherence time-reversal experiment (from ref.\cite{DomZamSegGzz17}) and the decoherence time $\hat{\tau}^\sB_X = \tau^\sB_X/3$ (solid line) predicted by the theory, as a function of the dipolar frequency $\hat{\nu}_0$. The variation of $\hat{\tau}^\sB_X$ is calculated for $\{\vs,N\} = \{4570\un{m/s}, 4.7\times10^{22}\}$ (solid line), where $N$ is chosen to fit with the highest frequency measured value. Dashed curves show the variation of $\hat{\tau}^\sB_X$ for $\{\vs,N\} = \{2800\un{m/s}, 8\times10^{21}\}$ (lower) and $\{\vs,N\} = \{5500\un{m/s}, 8\times10^{22}\}$ (upper). Notice the agreement within the frequency range where the pairs can be described as non-interacting.}

\section{Discussion} \label{sec:Disc}

We studied the dynamics of decoherence in the adiabatic regime, for a system consisting of a partition of many equivalent elements with no direct interaction, in contact with a common bosonic environment.
This proposal allowed an analytic description of the irreversible dynamics that arises when the only correlation between partition elements comes through their coupling with the common boson bath.
The approach represents a further step in the discussion pioneered by the traditional spin-boson model. Inclusion of the quantum many-body character through an interaction Hamiltonian that interweaves system and environment variables is a critical feature of our contribution.
The derivation does not assume coarse-graining nor uses a master equation and is valid within an intermediate time scale where dissipation effects are not yet manifest.

In this work, we restricted the analysis to local operators compatible with the observed system anisotropy, whose expectation values come in terms of a density operator reduced to a single, representative partition element. Such a reduction implies no additional simplifying hypotheses; instead, it is a natural consequence of the partitionable form of the studied model.
It implies the passage from a description in terms of the ``grand'' density operator $\rho^\hs(t)$ to a local or ``condensed'' density operator $\sigma^\hA(t)$ that retains track of the existence and dynamics of the rest of the solid, contained initially in each element of the grand matrix.
Consequently, the local-observable expectation values reflect the many-body correlations, despite being a sum of contributions from individual partition elements.

A novel feature of this approach is the relevance of the phase function $\Upsilon(t)$ in the condensed density matrix dynamics.
This contribution generates attenuation due to the superposition of complex exponential factors and is unconnected from the thermal state of the environment, in contrast with the usual spin-boson model.
Remarkably, the attenuation implied by $\Upsilon(t)$ arises even when $\HamI$ does not involve direct interaction between the partition elements.
In this sense, we can say that the decoherent dynamics of the partitionable system arises from the interference of all the quantum states that the system can attain.
Hence, it emerges as an intrinsic feature of many-body open quantum systems, pictured by a non-separable interaction Hamiltonian like  \eq{eq:Ham_int_ad_reesc}. We remark that the $\Upsilon(t)$ contribution would be absent if $\HamI$ describes an ensemble of elements coupled to their own boson bath (e.g., single qubits or pairs of qubits) represented by a separable interaction Hamiltonian like \eq{eq:separableHI}.

When treating the particular case of non-interacting dipole-coupled spin-pairs in a phonon bath, the system and system-environment Hamiltonians arise unambiguously from considering the position variables of the dipole coupling as quantum operators and by expressing the small atomic shifts in terms of creation and annihilation operators.
Although the model does not consider direct magnetic interaction between the spin pairs, an indirect pair-coupling emerges mediated by phonons.
Accordingly, the Hamiltonian $\HamI$ allows transitions between the dipolar energy levels of different pairs through phonon creation and annihilation in a many-body interaction schema.
The resulting system-environment coupling coefficients are written in terms of physical magnitudes of the treated systems, and their dependence on the partition element coordinates introduces the correlation between elements.

We observe that $\Upsilon(t)$ does not depend on the bath temperature, and neither does the decay constant $\tau_{X}$ derived from it (see \eq{eq:tau_chi}). This prediction agrees with the experiment \cite{DomZamSegGzz17}.
We might physically relate decoherence process found in this work to the indeterminacy in the spin-bearing nuclei position due to the phonon field.
It is a collective effect associated with the correlation introduced by the low-frequency phonon modes, which makes the acoustic modes more effective than the high-frequency optical modes, contrarily to intuition.
On the contrary, the decay constant $\tau_{\gamma}$, derived from  $\Gamma(t)$, does depend on temperature as in the traditional spin-boson model. In this case, its influence on $\sigma_{m_A,n_A}(t)$ is negligible, which is not surprising since $\Gamma(t)$ represents the spin dynamics of a single pair \cite{footnote_puredephasing}.
However, the thermal process connected with $\tau_{\gamma}^{-1}$ might instead be significant in non-partitionable systems, for example,  where the dynamics allow fast growth of multi-spin correlations driven by the flip-flop terms of the dipolar Hamiltonian. A fast scrambling of information across the observed system due to the direct interaction between elements may enhance  the influence of $\Gamma(t)$.
In other words, the relative importance of $\Gamma(t)$ and $\Upsilon(t)$ in a general case will be determined by both the system and system-environment Hamiltonian structure.

It is worth adding a comment on the influence of the spin network dimensionality on the decoherence rate. In a 1-D lattice the sinc function $\opc{S}_{A,A'}$ is identically zero for $A' \neq A$ and consequently $\tau_X \rightarrow  \infty$. This suggests that lowering the dimension would be a way to protect a system from the decoherence mechanism described by the phase function $\Upsilon(t)$. A similar result was obtained in ref.\cite{MorelloStamp_06} where the effectiveness of `correlated errors' in a multi-qubit system also depends on the system dimension, suggesting this aspect as a general characteristic of decoherence in correlated systems.

A significant result of this work stems from applying the formalism to explain the signal amplitude attenuation in a typical NMR reversion experiment. The open system approach allows showing that the net evolution of the system state at the end of a reversion sequence is driven by $\HamI+\HamE$, which implies that one should expect quantum adiabatic decoherence to manifest even under ideal experimental conditions. In other words, irreversibility manifests as an inherent property of adiabatic quantum decoherence. The formalism allows calculating the signal amplitude after a reversion sequence and describing its decoherence rate in terms of characteristic magnitudes of the system and the environment. The qualitative and quantitative agreement between calculated decoherence rates and those measured on a hydrated salt indicates that the pair-phonon model used for $\HamI$ satisfactorily explains the decoherence mechanism in this kind of sample. Also, the formalism admits future refinements of $\HamI$ to include inter-pair interactions that might extend its applicability.

It is worth recalling that functions $G_A$  in \numeq{eq:fdeco_GA_B} and  \eq{eq:fdeco_GA}, that drive decoherence in both the reverted and the free-evolution regimes, emerge from a phase factor whose exponent depends on the variable $X_A$, is linear on time, and is eigen-selective since it depends on the difference of quantum numbers $(\kappa_{\mA} - \,\kappa_{\nA})$.
This exponential derives from the non-oscillatory term within the real part of the phase function $\Upsilon_{m,n}^{\bf{k}}$, which diverges with time (see \eqs{eq:Func_Umnkl_delta} and \numeq{eq:Func_Umnkl}). In turn, $\Upsilon_{m,n}^{\bf{k}}$ derives ultimately from the fact that $[\HamI,\HamE]\neq0$ which motivate using the displacement transformation \numeq{eq:def_Ndesp_delta} and \numeq{eq:def_Ndesp} applied to the creation and annihilation operators of the boson environment, and the linear time terms in $\Upsilon_{m,n}^{\bf{k}}$ are introduced by the displacement constant.
The decoherence efficiency of $G_A$ is a consequence of the macroscopic character of variable $X_A$, which depends on all the partitions eigenstates.
$X_A$ correlates all the partitions and introduces a phase superposition (the integrals in \eqs{eq:fdeco_GA_B} and \numeq{eq:fdeco_GA}) producing the eigen-selective decay of the density matrix elements.

The topic of irreversible evolution of many-body systems is not addressed in the literature under a unique look. One point of view stands on the hypothesis that the system of interest is a subsystem of a larger one with the same nature, that plays the role of ``the bath''. Thus, the system evolution is driven by self-interactions \cite{Nandkishore2016,gogolin2016}, in the scheme of Gibbs' statistical mechanics.
Another angle, the one considered here, regards the observed system as an open quantum system that undergoes non-unitary evolution.
In our opinion, this viewpoint may also apply to the understanding of current open questions in many-body irreversible processes, as thermalization and spin thermodynamics, where the occurrence of quasi-equilibrium was justified heuristically in terms of processes that develop in isolated systems \cite{deutsch91,goldman1970spin}.
Consistency between calculations and experimental results supports the idea that an energy-conserving decoherence process induced by the system-environment coupling can bring the density matrix to a diagonal-in-blocks form (i.e., a quasi-equilibrium state), in the common eigenbasis of the system and system-environment Hamiltonians, namely the ``preferred basis''. Accordingly, the eigen-selective and irreversible character of adiabatic decoherence explains the occurrence of quasi-equilibrium states in some solids, which was previously justified in terms of spin thermodynamics \cite{Keller_88,Eisendrath_78}. Also, these results are consistent with previous theoretical proposals on the build-up of quasi-equilibrium in nematic liquid crystals \cite{GzzSegZam11,SegZam11,SegZam13,Bonin13}, where the molecular environment also has long-range collective excitations (order fluctuations).

In summary, we see that the model of non-interacting partition elements in contact with a common boson bath captures essential features of the decoherence phenomenon, and also enables a quantitative calculation which opens the possibility of accounting for actual experimental results.
Having found an estimate of the decoherence rate of a real system can be considered an advance in decoherence theory \cite{Stamp_06}.
The finding allows confirming the hypotheses assumed in the theoretical formulation, validating a novel proposal for the microscopic origin of the decoherence of a real system in terms of a  mechanism independent of the thermal properties of the environment.

\section{Acknowledgement}

This work was supported by SECYT, Universidad Nacional de Córdoba. H.H.S. thanks CONICET for financial support.


\begin{thebibliography}{10}
\bibitem{footnote-dephasing_A}
    The secular contribution to transverse relaxation $T_2$ is frequently called
      ``pure dephasing''. See ref.\cite{Bylander-Cory}.
\bibitem{eigenselection}
    The word eigen-selection \cite{SegZam11} is inspired in the term einselection
      (environment-induced selection), previously coined by W. Zurek
      \cite{PazZurek99} when introducing the idea of superselection.
\bibitem{footnote_puredephasing}
    The insignificance of this ``pure dephasing'' contribution is consistent with
      the fact that thermal fluctuations of the spin-phonon interaction in solids
      do not contribute to relaxation dynamics.
\end{thebibliography}

%
%
%

\appendix

\section{Reduced density operator} \label{app:reduc}

\subsection{Reduction over the environment} \label{app:reduc_E}

The density operator reduced over the environment variables derives from
\begin{equation}
\begin{split}
\langle \op{O} \rangle(t) &= \tr{\op{O} \rho(t)} 
= \sum_{m,m';e,e'} \bra{m,e} \op{O}^\hs\ket{m',e'} \bra{m',e'}\rho(t)\ket{m,e}\\
&= \sum_{m,m'} \bra{m}\op{O}^\hs\ket{m'}\bra{m'}\left[\sum_{e}\bra{e}\rho(t)\ket{e}\right]\ket{m}
= \trs{S}{\op{O}^\hs\,\rho^\hs (t)},
\end{split}
\end{equation}
where we use that
\[\bra{m,e}\op{O}^\hs\ket{m',e'} = \bra{m}\op{O}^\hs\ket{m'}\,\delta_{e,e'},\]
and the definition \numeq{eq:rho_sist_red}.\\

\subsection{Reduction over the complementary space} \label{app:reduc_complem}

Each term in \eq{eq:redu1} has the form
\begin{equation}\label{eq:redu2}
\begin{array}{rl}
\langle \op{O}_A^\hs\rangle(t) = \sum_{m,n} \bra{n}\op{O}_A^\hs\ket{m}\bra{m}\rho^\hs(t)\ket{n},
\end{array}
\end{equation}
where owing to the definition \numeq{eq:obs_partS} of a local observable
\begin{equation}\label{eq:redu3}
\bra{n}\op{O}_A^\hs\ket{m} = \bra{\nA}\op{O}_A^\hA\ket{\mA}\,\delta_{\{\mAn,\nAn\}},\\
\end{equation}
with $\delta_{\{\mAn,\nAn\}} \equiv \delta_{\ml,\nl}\cdots\delta_{\mi,\nii}\cdots\delta_{\mN,\nN}$, which is a product of $N - 1$ Krönecker deltas that excludes $\delta_{\mA,\nA}$.
This leads to
\begin{equation}\label{eq:redu4}
\begin{split}
\langle \op{O}_A^\hs\rangle(t) &= \sum_{\mA,\nA} \bra{\nA}\op{O}^\hA\ket{\mA}
\sum_{\{\!\mAn,\nAn\!\}}\bra{m}\rho^\hs(t)\ket{n}\,\delta_{\{\mAn,\nAn\}}\\
&= \sum_{\mA,\nA} \bra{\nA}\op{O}^\hA\ket{\mA} \bra{\mA}\sigma^\hA(t)\ket{\nA}
=  \trs{S_A}{\op{O}^\hA\;\sigma^\hA(t)},
\end{split}
\end{equation}
where the double sum in the second row of \numeq{eq:redu4} is $\sum_{\{\!\mAn,\nAn\!\}} \equiv \sum_{\{\!\mAn\!\}}\sum_{\{\!\nAn\!\}}$, with $\sum_{\{\!\mAn\!\}}\!\equiv \sum_{\ml}\!\!\cdots\sum_{\mi}\!\!\cdots\sum_{\mN}$ running over all the quantum numbers $\mi$ with $i \neq A$.
Similarly to \apen{app:reduc_E},  in \numeq{eq:redu4} we used that
\begin{equation}\label{eq:rho_sist_red_SA_sum}
\begin{split}
\sum_{\{\!\mAn,\nAn\!\}}\! \!\bra{m}\rho^\hs(t)\ket{n}\,\delta_{\{\mAn,\nAn\}}
&= \bra{m_A} \otimes \sum_{\{\!\mAn\!\}} \bra{m_{\bar{A}}} \rho^\hs(t)\ket{m_{\bar{A}}} \otimes \ket{n_A}\\
&\equiv \bra{m_A} \sigma^\hA(t) \ket{n_A} \equiv \sigma_{\mA,\nA}(t).
\end{split}
\end{equation}
In the last row of \eq{eq:rho_sist_red_SA_sum} we define the density operator $\sigma^\hA(t)$ reduced to a single partition element as
\begin{equation}
\sigma^\hA(t) \equiv \trs{\oS_A}{\rho^\hs(t)} = \sum_{\{\!\mAn\!\}} \bra{m_{\bar{A}}} \rho^\hs(t)\ket{m_{\bar{A}}}.
\end{equation}

\section{Partial trace over the environment} \label{app:calc_Smnkl}

In this appendix we calculate the trace over the environment variables used in \eq{eq:rho_sist_red_ad}. We follow the strategy described in ref.\cite{priv98}, and apply it to a more general Hamiltonian, like the one of  \eq{eq:Ham_int_ad_reesc}. In \apen{app:coherent_states} we summarize the coherent states formalism \cite{Glauber63,Knight_Cap3}, and calculate the trace in \apen{app:calc_Int_EC}.

\subsection{Coherent states} \label{app:coherent_states}

Coherent states $\{\ket{z_{\bf{k}}}\}$ are the eigenstates of the annihilation operator $\op{b}_{\bf{k}}$, that is
\begin{equation*}\label{eq:eigen_b}
\op{b}_{\bf{k}}^\he\ket{z_{\bf{k}}} = z_{\bf{k}}\,\ket{z_{\bf{k}}},
\end{equation*}
where $z_{\bf{k}} \in \cbb{C}$. Their expression in terms of  the number states $\{\ket{n_{\bf{k}}}\}$  is \cite{Glauber63,Knight_Cap3}
\begin{equation}\label{eq:def_EC}
\ket{z_{\bf{k}}} \equiv e^{-\frac{1}{2}\abs{z_{\bf{k}}}^2}\sum_{n_{\bf{k}}=\,0}^{\infty} \frac{(z_{\bf{k}})^n}{\sqrt{n!}}\,\ket{n_{\bf{k}}},
\end{equation}
where $n_{\bf{k}}$ is the number of bosons with wavenumber $\vec{k}$ and normal mode $l$.

Coherent states are normal $\braket{z_{\bf{k}}}{z_{\bf{k}}} = 1$, but not orthogonal
\begin{equation}\label{eq:prop_ortog_EC}
\braket{z_{\bf{k}}}{z'_{\bf{k}}} = e^{z^*_{\bf{k}}\,z'_{\bf{k}}-\frac{1}{2}\abs{z_{\bf{k}}}^2-\frac{1}{2}\abs{z'_{\bf{k}}}^2},
\end{equation}
and form an over-complete set that satisfies
$\int\!d^{\,2}\!z_{\bf{k}}\,\ket{z_{\bf{k}}}\bra{z_{\bf{k}}} = \op{1}^\he,$
where the differential is
\begin{equation}\label{eq:def_d2z}
d^{\,2}\!z_{\bf{k}} \equiv \frac{1}{\pi}\,dx_{\bf{k}}\,dy_{\bf{k}},
\end{equation}
with $x_{\bf{k}} \equiv \Re\{z_{\bf{k}}\}$ and $y_{\bf{k}} \equiv \Im\{z_{\bf{k}}\}$ the real and imaginary parts of $z_{\bf{k}}$, respectively.

Thus, the trace of an operator over the environment space can be written as
\begin{equation*}\label{eq:TrB_EC}
\trs{E}{\op{O}^\he} = \int d^{\,2} z_{\bf{k}}\,\bra{z_{\bf{k}}}\op{O}^\he\ket{z_{\bf{k}}}.
\end{equation*}
We use this property to calculate the desired partial trace from \eq{eq:fundeco}
\begin{equation}\label{eq:des_Smnkl_EC}
\begin{split}
S_{m,n}^{\bf{k}}(t) &= \int\!d^{\,2}\!z\,\bra{z}e^{-\ic\left(\op{M} + \op{J}_m\right)t}\,\op{\Theta}\,e^{\,\ic\left(\op{M} + \op{J}_n\right) t}\ket{z}
= \frac{1}{Z}\iiint\!d^{\,2}\!z_0\,d^{\,2}\!z_1\,d^{\,2}\!z_2\\
&\qquad\qquad\times\bra{z_0}e^{-\ic (\op{M} + \op{J}_m) t}\ket{z_1}\bra{z_1}e^{-\beta \omega \op{b}^{\dagger}\op{b}}\ket{z_2}
\bra{z_2}e^{\,\ic (\op{M} + \op{J}_n) t}\ket{z_0},
\end{split}
\end{equation}
where all the operators belong to $\shE$ and we omit the superindex $\he$ as well as  index $\bf k$. 
The partition function $Z$ in \numeq{eq:des_Smnkl_EC} is
\begin{equation}\label{eq:def_Z_EC}
Z \equiv \int\!d^{\,2}\!z\,\bra{z}e^{-\beta \omega \op{b}^{\dagger}\op{b}}\ket{z}.
\end{equation}

Let us now write the matrix elements involved in \numeq{eq:des_Smnkl_EC} and \numeq{eq:def_Z_EC}.  Using the properties \numeq{eq:def_EC} and \numeq{eq:prop_ortog_EC} we write
\begin{equation}\label{eq:des_exp_N_EC}
\bra{z_1}e^{-\beta \omega \op{b}^{\dagger}\op{b}}\ket{z_2} = \sum_{n=0}^\infty\,\bra{z_1}e^{-\beta \omega \op{b}^{\dagger}\op{b}}\ket{n}\braket{n}{z_2}
= \braket{z_1}{z_2}\,e^{z_1^* \left(e^{-\beta \omega}-1\right)\,z_2},
\end{equation}
where we used $\op{b}^{\dagger}\op{b}\ket{n} = n\ket{n}$, the closure relation $\sum_{n=\,0}^\infty\,\ket{n}\bra{n} = \op{1}$, and
\begin{equation*}\label{eq:elem_n_EC}
\braket{n}{z} = e^{-\frac{1}{2}\abs{z}^2}\frac{z^n}{\sqrt{n!}},\quad \text{with}\;\;\braket{z}{n} = \braket{n}{z}^*.
\end{equation*}
In order to write the factors involving $e^{\pm\ic\left(\op{M} + \op{J}_m\right) t}$, it is convenient to use a displaced operator
$\op{a}_m \equiv \op{b} + \lambda_m/\omega$,
such that the coherent states are also eigenstates of $\op{a}_m$ with eigenvalues $\cur{z}_m \equiv z + \lambda_m/\omega$. Let us call $\ket{\cur{z}_m}$ to those eigenstates which may differ from $\ket{z}$  at most by a phase factor, that is
\begin{equation}\label{eq:prop_eigen_b-am}
\ket{\cur{z}_m} = e^{\,\ic\,\phi_m}\,\ket{z},
\end{equation}
then
$\op{a}_m\,\ket{\cur z_m} = \cur{z}_m\,\ket{\cur z_m}$
and we define a modified number operator
$\op{\widehat{N}}_m \equiv \op{a}_m^\dagger\op{a}_m$ with a set of orthogonal eigenstates $\{\ket{\eta_m}\}$, where $\braket{\eta_m}{\eta'_m} = \delta_{\eta_m,\eta'_m}$, so that
\begin{equation*}\label{eq:prop_est_num_am_N}
\op{\widehat{N}}_m\ket{\eta_m} = \op{a}_m^\dagger\op{a}_m\ket{\eta_m} = \eta_m\ket{\eta_m},
\end{equation*}
which satisfy
$\op{a}_m\ket{\eta_m} = \sqrt{\eta_m}\,\ket{\eta_m-1},\;\op{a}_m^\dagger\ket{\eta_m} = \sqrt{\eta_m+1}\,\ket{\eta_m+1},$
with $\op{a}_m\ket{0} = 0$, and the closure relation
\begin{equation*}\label{eq:rel_clau_est_num_am}
\sum_{\eta_m=\,0}^\infty\,\ket{\eta_m}\bra{\eta_m} = \op{1}.
\end{equation*}
With these definitions the eigenstates of $\op a_m$ can be written in terms of the basis $\{\ket{\eta_m}\}$ \cite{Knight_Cap3}, as
\begin{equation*}\label{eq:def_ECam}
\ket{\cur{z}_m} \equiv e^{-\frac{1}{2}\abs{\cur{z}_m}^2}\sum_{\eta_m=\,0}^{\infty} \frac{(\cur{z}_m)^{\eta_m}}{\sqrt{\eta_m!}}\,\ket{\eta_m},
\end{equation*}
and also the exponent $\op M + \op J_m$ may be expressed in terms of the displaced creation and annihilation operators as
\begin{equation}\label{eq:def_Ndesp}
\op M + \op J_m  = \omega \op{ b}^{\dagger} \op{ b} + \lambda_m^* \op{ b} +\lambda_m  \op{ b}^{\dagger}  = \omega \op{\widehat{N}}_m -\frac{\left| \lambda_m\right|^2 }{\omega}.
\end{equation}
Then, we can use \eq{eq:des_exp_N_EC} to calculate the matrix elements
\begin{equation}\label{eq:des_exp_gm_ECam}
\begin{split}
\bra{\cur{z}_{1,m}}e^{\pm\ic (\op M + \op J_m) t}\ket{\cur{z}_{2,m}} &= e^{\mp\ic\,\omega t\,\frac{\abs{\lambda_m}^2}{\omega^2}}
\sum_{\eta_m=\,0}^\infty\,\bra{\cur{z}_{1,m}} e^{\pm\ic\,\omega t\,\op{a}_m^\dagger\op{a}_m}\ket{\eta_m}\braket{\eta_m}{\cur{z}_{2,m}} \\
&= e^{\mp\ic\,\omega t\,\frac{\abs{\lambda_m}^2}{\omega^2}} \braket{\cur{z}_{1,m}}{\cur{z}_{2,m}}\,
e^{\,\cur{z}_{1,m}^* \left(e^{\pm\ic\,\omega t}-1\right)\,\cur{z}_{2,m}}.
\end{split}
\end{equation}
From \numeq{eq:prop_eigen_b-am}, one gets
\begin{subequations}\label{eq:prop_ortog_ECam}
\begin{equation}\label{eq:prop_ortog_ECam_z}
\braket{\cur{z}_{1,m}}{\cur{z}_{2,m}} = \braket{z_1}{z_2}\,e^{\,\ic\,\left(\phi_{2,m}-\,\phi_{1,m}\right)},
\end{equation}
\begin{equation}\label{eq:prop_ortog_ECam_z_exp}
\bra{\cur{z}_{1,m}}e^{\pm\ic (\op M + \op J_m) t}\ket{\cur{z}_{2,m}} =
\bra{z_1}e^{\pm\ic(\op M + \op J_m) t}\ket{z_2}\,e^{\,\ic\,\left(\phi_{2,m}-\,\phi_{1,m}\right)}.
\end{equation}
\end{subequations}
Finally, using that $\cur{z}_{j,m} = z_j + \lambda_m/\omega$ and inserting \numeq{eq:prop_ortog_ECam} in \numeq{eq:des_exp_gm_ECam}, the desired matrix elements become
\begin{equation}\label{eq:des_exp_gm_ECb}
\bra{z_1}e^{\pm\ic(\op M + \op J_m) t}\ket{z_2} = e^{\mp\ic\,\omega t\,\frac{\abs{\lambda_m}^2}{\omega^2}}\,\braket{z_1}{z_2}\,
e^{\left(z_1^*+\frac{\lambda_m^*}{\omega}\right)\left(e^{\pm\ic\,\omega t}-1\right)\left(z_2+\frac{\lambda_m}{\omega}\right)}.
\end{equation}

\subsection{Calculation of the trace} \label{app:calc_Int_EC}

This section aims to calculating the integral proposed in \eq{eq:des_Smnkl_EC}.
By using \eqs{eq:des_exp_N_EC} and \numeq{eq:des_exp_gm_ECb} in \numeq{eq:des_Smnkl_EC}, we get
\begin{equation}\label{eq:Smnkl_Int}
S_{m,n}^{\bf{k}}(t) = e^{\,\left[\abs{\lambda_m}^2\left(e^{-\ic\,\omega t}-1\right)+\abs{\lambda_n}^2\left(e^{\,\ic\,\omega t}-1\right)\right]/\omega^2}
\,e^{\,\ic\,t\,\left(\abs{\lambda_m}^2-\abs{\lambda_n}^2\right)/\omega}\,\Iec_{3}(t)/Z,
\end{equation}
where
\begin{equation}\label{eq:Smnkl_Int_3}
\begin{split}
\Iec_{3}(t) &\equiv \iint\!\!d^{\,2}\!z_2\,d^{\,2}\!z_1\;e^{\,z_1^* z_2\,e^{-\beta\,\omega} - \abs{z_1}^2 - \abs{z_2}^2}\\&\qquad\qquad\qquad\times
e^{\,\left[z_1\,\lambda_m^*\,\left(e^{-\ic\,\omega t}-1\right) + z_2^*\,\lambda_n\,\left(e^{\,\ic\,\omega t}-1\right)\right]/\omega}\,\Iec_{0}(t),
\end{split}
\end{equation}
with
\begin{equation}\label{eq:Smnkl_Int_0}
\Iec_{0}(t) \equiv \int\!\!d^{\,2}\!z_0\,e^{-\abs{z_0}^2 + a_0 z_0 + b_0 z_0^*},
\end{equation}
and
\begin{equation}\label{eq:coef_a_b_0}
a_0 \equiv z_2^*\,e^{\,\ic\,\omega t} + \lambda_n^*\,\left(e^{\,\ic\,\omega t}-1\right)/\omega,\quad
b_0 \equiv z_1\,e^{-\ic\,\omega t} + \lambda_m\,\left(e^{-\ic\,\omega t}-1\right)/\omega.
\end{equation}
The general form of the integrals in \numeq{eq:Smnkl_Int_3} and \numeq{eq:Smnkl_Int_0} is
\begin{equation}\label{eq:Int_gral_S}
\Iec \equiv \int\!\!d^{\,2}\!z\,e^{- r \abs{z}^2 + a z + b z^*}
= \frac{1}{\pi}\int_{-\infty}^{\infty}dx\;e^{- r x^2 + (a+b) x}\int_{-\infty}^{\infty}dy\;e^{- r y^2 + \ic(a-b) y},
\end{equation}
where the coefficients $\{a,b\} \in \cbb{C}$ and $r \in \cbb{R}$.
In \eq{eq:Int_gral_S} we used $z \equiv x + \ic y$, with $\{x,y\} \in \cbb{R}$, and the differential defined in \numeq{eq:def_d2z}.\\
The two factors in \numeq{eq:Int_gral_S} have the general form
\begin{equation}\label{eq:Int_gral_S_real}
\hat{\Iec} \equiv  \int_{-\infty}^{\infty}dv\,e^{- r v^2 + c v},
\end{equation}
with $\{v,r\} \in \cbb{R}$ and $c \in \cbb{C}$.
Setting $c = \alpha - \ic \beta$, with $\{\alpha,\beta\} \in \cbb{R}$,  \eq{eq:Int_gral_S_real} can be written as
\begin{equation}\label{eq:Int_gral_S_real_calc}
\begin{split}
\hat{\Iec} =  \int_{-\infty}^{\infty}dv\,e^{- r v^2 + \alpha v}\,e^{-\ic \beta v}
&= e^{\frac{1}{r}\left(\frac{\alpha}{2}\right)^2}\int_{-\infty}^{\infty}dv\,e^{-r\left(v-\frac{\alpha}{2 r}\right)^2}e^{-\ic \beta v}\\
&= e^{\frac{1}{r}\left[\left(\frac{\alpha}{2}\right)^2-\ic\frac{\alpha\beta}{2}\right]}\int_{-\infty}^{\infty}d\hat{v}\,e^{-r\hat{v}^2}
e^{-\ic \beta \hat{v}},
\end{split}
\end{equation}
where $-r\,v^2 + \alpha\,v = -r\left[v-\alpha/(2\,r)\right]^2 + \left(\alpha/2\right)^2/r$,
and $\hat{v} \equiv v - \alpha/(2\,r)$.
The integral in the last expression in \eq{eq:Int_gral_S_real_calc} is the Fourier transform of a   gaussian function
\begin{equation}\label{eq:Fourier_Gauss}
\int_{-\infty}^{\infty}d\hat{v}\,e^{-r\hat{v}^2}e^{-\ic \beta \hat{v}}
= \sqrt{\frac{\pi}{r}}\,e^{-\beta^2/(4 r)}.
\end{equation}
Then  \numeq{eq:Int_gral_S_real} becomes
\begin{equation}\label{eq:Int_gral_S_real_calc_res}
\begin{split}
\hat{\Iec} = \sqrt{\frac{\pi}{r}}\,
e^{\left(\alpha^2-\beta^2-2\,\ic\,\alpha\,\beta\right)/(4 r)}
= \sqrt{\frac{\pi}{r}}\,e^{\,c^2/(4 r)}
\end{split}
\end{equation}
in all the integrals involved in \numeq{eq:Smnkl_Int}.

We may now use \numeq{eq:Int_gral_S_real_calc_res} in  \numeq{eq:Int_gral_S_real} to see that $\Iec$ in \numeq{eq:Int_gral_S} becomes
\begin{equation}\label{eq:Int_gral_S_res}
\Iec = \frac{1}{r}\,e^{\left[\left(a+b\right)^2-\left(a-b\right)^2\right]/(4 r)} = \frac{1}{r}\,e^{\,a b / r}.
\end{equation}
Also, by setting  $r = 1$ in \eq{eq:Int_gral_S_res} we solve \numeq{eq:Smnkl_Int_0}, which results
\begin{equation}\label{eq:Smnkl_Int_0_res}
\Iec_{0}(t) = e^{\,a_0\,b_0},
\end{equation}
with $a_0$ and $b_0$ as in \numeq{eq:coef_a_b_0}.
Consequently \eq{eq:Smnkl_Int_3} becomes
\begin{equation}\label{eq:Smnkl_Int_3_0}
\begin{split}
\Iec_{3}(t) = e^{\lambda_m\,\lambda_n^*\,\left(e^{\,\ic\,\omega t}-1\right)
\left(e^{-\ic\,\omega t}-1\right)/\omega^2}\Iec_{2}(t),
\end{split}
\end{equation}
where we defined
\begin{equation}\label{eq:Smnkl_Int_2}
\begin{split}
\Iec_{2}(t) \equiv \int\!\!d^{\,2}\!z_2\,
e^{-\abs{z_2}^2 - \left(\lambda_m-\lambda_n\right)\left(e^{\,\ic\,\omega t}-1\right)\,z_2^*/\omega}\,\Iec_{1}(t),
\end{split}
\end{equation}
\begin{equation}\label{eq:Smnkl_Int_1}
\Iec_{1}(t) \equiv \int\!\!d^{\,2}\!z_1\,e^{-\abs{z_1}^2 + a_1 z_1 + b_1 z_1^*} = e^{\,a_1\,b_1},
\end{equation}
with $a_1 \equiv z_2^* + \left(\lambda_m-\lambda_n\right)^*\left(e^{-\ic\,\omega t}-1\right)/\omega$, and $b_1 \equiv z_2\,e^{-\beta\,\omega}$.
Then, $\Iec_{2}(t)$ becomes
\begin{equation}\label{eq:Smnkl_Int_2_res}
\begin{split}
\Iec_{2}(t) = \int\!\!d^{\,2}\!z_2\,e^{-r_2 \abs{z_2}^2 + a_2 z_2 + b_2 z_2^*} =\frac{1}{r_2}\,e^{\,a_2\,b_2/r_2},
\end{split}
\end{equation}
where
$a_2 \equiv \left(\lambda_m-\lambda_n\right)^*\left(e^{-\ic\,\omega t}-1\right) e^{-\beta\,\omega}/\omega$,
$b_2 \equiv - \left(\lambda_m-\lambda_n\right)\left(e^{\,\ic\,\omega t}-1\right)/\omega$, and $r_2 \equiv 1-e^{-\beta\,\omega}$.
By replacing \numeq{eq:Smnkl_Int_2_res} in \eq{eq:Smnkl_Int_3_0}, and after some algebra, the integral in \numeq{eq:Smnkl_Int_3} adopts the expression
\begin{equation}\label{eq:Smnkl_Int_3_res}
\Iec_{3}(t) = \frac{1}{1 - e^{-\beta\,\omega}}\;e^{\,\frac{4}{\omega^2}\sin^2\left(\omega t/2\right)\left[-\abs{\lambda_m-\lambda_n}^2\, e^{-\beta\,\omega}/\left(1-e^{-\beta\,\omega}\right) + \lambda_m\,\lambda_n^*\right]},
\end{equation}
where we used
$\left(e^{\,\ic\,\omega t}-1\right)\left(e^{-\ic\,\omega t}-1\right) = 2\left[1-\cos\left(\omega t\right)\right] = 4\sin^2\left(\omega t/2\right)$.

We can also use \numeq{eq:Int_gral_S_res}, with the coefficients $r = 1-e^{-\beta \omega}$ and $a = b = 0$, to get the well known partition function
\begin{equation}\label{eq:Int_Z}
Z = \int\!d^{\,2}\!z\,e^{-\left(1-e^{-\beta \omega}\right)\abs{z}^2} = \frac{1}{1-e^{-\beta \omega}},
\end{equation}
whose integrand comes from using \numeq{eq:des_exp_N_EC} in \eq{eq:def_Z_EC}.\\
Finally, by replacing $\Iec_{3}(t)$ from \numeq{eq:Smnkl_Int_3_res} and the partition function \numeq{eq:Int_Z} in the decoherence function $S_{m,n}^{\bf{k}}(t)$ of \eq{eq:Smnkl_Int}, we obtain the following expression
\begin{equation}\label{eq:Smnkl_Int_res}
\begin{split}
S_{m,n}^{\bf{k}}(t) &= e^{-\frac{2}{\omega^2}\abs{\lambda_m-\lambda_n}^2\sin^2\left(\omega t/2\right)\coth\left(\beta\omega/2\right)}\\
&\quad \times e^{-\frac{\ic}{\omega^2}\left\{\left(\abs{\lambda_m}^2-\abs{\lambda_n}^2\right)\left[\sin\left(\omega t\right)- \omega t\right]
- 4\Im\left\{\lambda_m\lambda_n^*\right\}\sin^2\left(\omega t/2\right)\right\}}.
\end{split}
\end{equation}
By using the identities
\[\abs{\lambda_m}^2 - \abs{\lambda_n}^2 = \left(\lambda_m-\lambda_n\right)\left(\lambda_m+\lambda_n\right)^* - 2\,\ic\,\Im\left\{\lambda_m\lambda_n^*\right\},\; \text{and} \;\sin^2\left(\omega t/2\right) = \left[1-\cos\left(\omega t\right)\right]/2,\]
in the complex exponential of \numeq{eq:Smnkl_Int_res}, we can write $S_{m,n}^{\bf{k}}(t)$ in the form shown in \eq{eq:Func_Smnkl_des}.

In writing the exponent of \eq{eq:Smnkl_Int_res}, we used the identity
\begin{equation}\label{eq:Smnkl_des_exp_1}
\begin{split}
&\frac{1}{\omega^2}\left[\abs{\lambda_m}^2\left(e^{-\ic\,\omega t}-1\right)+\abs{\lambda_n}^2\left(e^{\,\ic\,\omega t}-1\right)
+\,\ic\,\omega t\left(\abs{\lambda_m}^2 - \abs{\lambda_n}^2\right)\right] =\\
&\qquad \frac{1}{\omega^2} \left[-\ic\left(\abs{\lambda_m}^2 - \abs{\lambda_n}^2\right)
\left[\sin\left(\omega t\right)-\omega t\right] - 2 \left(\abs{\lambda_m}^2 + \abs{\lambda_n}^2\right)\sin^2\!\left(\omega t/2\right)\right],
\end{split}
\end{equation}
with $e^{\pm \ic\,\omega t}-1 =\pm \ic\sin\left(\omega t\right) - 2\sin^2\left(\omega t/2\right)$.
Then, using \eqs{eq:Smnkl_Int_3_res} and \numeq{eq:Int_Z}, and the relations
$2\,e^{-\beta\,\omega}/\left(1-e^{-\beta\,\omega}\right) = \coth\left(\beta\omega/2\right)-1$, and
$\lambda_m\lambda_n^* = \Re\left\{\lambda_m\lambda_n^*\right\} + \ic\,\Im\left\{\lambda_m\lambda_n^*\right\}$,
we write the exponent of $\Iec_{3}(t)/Z$ in \eq{eq:Smnkl_Int} as
\begin{equation}\label{eq:Smnkl_des_exp_2}
\begin{split}
&\frac{4}{\omega^2}\,\sin^2\left(\omega t/2\right)\left[-\abs{\lambda_m-\lambda_n}^2
\frac{e^{-\beta\,\omega}}{1-e^{-\beta\,\omega}} + \lambda_m\,\lambda_n^*\right]\\
&\qquad\qquad\qquad\qquad = \frac{1}{\omega^2}\bigg[-2\abs{\lambda_m-\lambda_n}^2\,\sin^2\left(\omega t/2\right)
\,\left[\coth\left(\beta\omega/2\right) - 1\right]\\
&\qquad\qquad\qquad\qquad\qquad\qquad\qquad + 4\left(\Re\left\{\lambda_m\lambda_n^*\right\} + \ic\,\Im\left\{\lambda_m\lambda_n^*\right\}\right)\,
\sin^2\left(\omega t/2\right)\bigg].
\end{split}
\end{equation}
Finally, in \eq{eq:Smnkl_des_exp_1}, we replace
\begin{equation*}
\abs{\lambda_m}^2 + \abs{\lambda_n}^2 = \abs{\lambda_m-\lambda_n}^2 + 2\Re\left\{\lambda_m\lambda_n^*\right\},
\end{equation*}
then, by adding the result to \eq{eq:Smnkl_des_exp_2} we obtain the exponent in \eq{eq:Smnkl_Int_res}.

\section{Quantization of the intra-pair distances} \label{app:quant_intra-pair_dist}

The aim of this section is to use the quantization of spin positions to write the dipole-phonon interaction within the small displacement approximation.

The dipolar interaction depends on the distance between spins as $r_{ij}^{-3}$, where $r_{ij}$ is the modulus of the distance vector $\vec{r}_{ij} \equiv \vec{r}_j-\vec{r}_i$, and $\vec{r}_i$ is the position vector of the $i$-th spin.
The position vector can in turn be written as $\vec{r}_i = \vec{r}_{0,i} + \vec{u}_i$, the sum of a mean or equilibrium vector $\vec{r}_{0,i}$ and a small displacement $\vec{u}_i$.
With this notation  $\vec{r}_{ij} = \vec{r}_{0,ij}+\delta\vec{u}_{ij}$, where  $\vec{r}_{0,ij} \equiv \vec{r}_{0,j}-\vec{r}_{0,i}$ and  $\delta\vec{u}_{ij} \equiv \vec{u}_j-\vec{u}_i$. Expanding $r_{ij}^{-3}$ in powers of the displacement $\delta\vec{u}_{ij}$ one has \cite{Dolinsek00}
\begin{equation}\label{eq:des_rn3}
\frac{1}{r_{ij}^3} = \frac{1}{r_{0,ij}^3}\left[1-3\,\frac{\delta\vec{u}_{ij}\cdot\vec{r}_{0,ij}}{r_{0,ij}^2}
+\frac{15}{2}\,\frac{\left(\delta\vec{u}_{ij}\cdot\vec{r}_{0,ij}\right)^2}{r_{0,ij}^4}
-\frac{3}{2}\,\frac{\left(\delta u_{ij}\right)^2}{r_{0,ij}^2}+\cdots\right],
\end{equation}
with $\delta u_{ij} \equiv \abs{\delta\vec{u}_{ij}}$ and $r_{0,ij} \equiv \abs{\vec{r}_{0,ij}}$.

Assuming small displacements $\delta u_{ij}/r_{0,ij} \ll 1$, we keep only the first order term in expansion \numeq{eq:des_rn3}, then
\begin{equation}\label{eq:des_rij3_aprox}
\frac{1}{r_{ij}^3} \simeq \frac{1}{r_{0,ij}^3}\left[1-3\,\frac{\delta\vec{u}_{ij}\cdot\hat{r}_{0,ij}}{r_{0,ij}}\right],
\end{equation}
where we defined the unit vector $\hat{r}_{0,ij} \equiv \vec{r}_{0,ij}/r_{0,ij}$.

We now write the displacement $\vec{u}_i$  as a quantum variable
\begin{equation}\label{eq:cuant_u}
\vec{u}_i \equiv \sum_{\bf k} \hat{\epsilon}_{{\bf k},i}\,\csl{u}_{\bf k}\,
e^{\,\ic\vec{k}\cdot\vec{r}_{0,i}}\left(\op{b}_{\bf k} + \op{b}_{-\bf k}^{\dagger}\right)^{\!\he},
\end{equation}
where ${\pm\bf k} \equiv \{\pm\vec{k},l\}$ stands for the phonon wavenumber vector $\pm\vec{k}$ and  normal mode $l$ with frequency $\omega_{\bf k}$;
$\hat{\epsilon}_{{\bf k},i}$ is a unitary polarization vector, and we defined
\begin{equation}\label{eq:def_uk}
\csl{u}_{\bf k} \equiv \sqrt{\frac{\hbar}{2\,\omega_{\bf{k}}\,m_p\,N\,l_{\!M}}},
\end{equation}
with $N$ the number of pairs, $m_p$ the proton mass, and $l_{\!M}$ the maximum number of normal modes, thus $\sum_l \equiv \sum_{l=1}^{l_{\!M}}$.
Using the quantum displacement of \eq{eq:cuant_u}, the scalar product in \numeq{eq:des_rij3_aprox} becomes
\begin{equation*}\label{eq:cuant_delta u}
\delta\vec{u}_{ij}\cdot\hat{r}_{0,ij} = \sum_{\bf k} \left(g^{ij*}\,\op{b} + g^{ij}\,\op{b}^{\dagger}\right)_{\bf k}^\he,
\end{equation*}
where
\begin{equation}\label{eq:def_gk_ij}
g_{\bf k}^{ij} \equiv \csl{u}_{\bf k}\left[\left(e^{-\ic\vec{k}\cdot\vec{r}_{0,j}}\,\hat{\epsilon}_{{\bf k},j}
- e^{-\ic\vec{k}\cdot\vec{r}_{0,i}}\,\hat{\epsilon}_{{\bf k},i}\right)\cdot\hat{r}_{0,ij}\right].
\end{equation}

In the particular case of spin pairs arranged in a lattice as in \fig{fig:Fig_1}, we identify pairs with index $A$ and each spin with the numbers $1$ and $2$.
Then we use the following conversion
\begin{equation*}
\vec{r}_{0,i} \rightarrow \vec{r}_{0,1}^A  = \vec{r}_A - (d/2)\,\hat{d},\quad
\vec{r}_{0,j} \rightarrow \vec{r}_{0,2}^A = \vec{r}_A + (d/2)\,\hat{d},
\end{equation*}
with $\vec{r}_A$ the lattice position of pair $A$, and
$\delta\vec{u}_{ij} \rightarrow \delta\vec{u}_{A}$, $g_{\bf k}^{ij} \rightarrow g_{\bf k}^{A}$,
$\vec{r}_{0,ij} \rightarrow \vec{r}_{0,12} = \vec{r}_{0,2}^A - \vec{r}_{0,1}^A = d\,\hat{d}$,
where $\vec{r}_{0,12}$ is independent of index $A$.
Then, the system-environment coupling coefficient  in \eq{eq:def_gk_ij} becomes
\begin{equation*}\label{eq:def_gk_A_desc}
g_{\bf k}^{A} = e^{-\ic\vec{k}\cdot\vec{r}_A}\,g_{\bf k},
\end{equation*}
with
\begin{equation}\label{eq:def_gk_desc}
g_{\bf k} \equiv \csl{u}_{\bf k} \left[\left(e^{-\ic\vec{k}\cdot\hat{d}\,d/2}\,\hat{\epsilon}_{{\bf k},2}
- e^{\,\ic\vec{k}\cdot\hat{d}\,d/2}\,\hat{\epsilon}_{{\bf k},1}\right)\cdot\hat{d}\,\right].
\end{equation}
Notice that the scalar products in \numeq{eq:def_gk_desc} impose that only normal modes with a projection over the intra-pair axis will have a non-zero coupling coefficient $g_{\bf k}$.\\
In this way, the dipolar coupling  $\Omega_0(r^A_{12})$ in the dipolar Hamiltonian of \eq{eq:def_Ham_Da0_A}
\[\Omega_0(r^A_{12})= \frac{D}{(r^A_{12})^3} \simeq \frac{D}{d^3}\left[1-3\,\frac{\delta\vec{u}_A\cdot\hat{d}}{d}\,\right],\]
where $D \equiv \mu_0\gamma_p^{2}\hbar\,\left[1-3\cos^2\left(\theta_{\hat{r},\hat{z}}\right)\right]/8\pi$,
becomes an operator on the environment Hilbert space, with the expression
\begin{equation}\label{eq:Omega_qq}
\op{\Omega}^{(e)} = \frac{D}{d^3} \left[  \op{1}^\he - \frac{3}{d} \sum_{\bf k} \left(g^{A*}\,\op{b} + g^{A}\,\op{b}^{\dagger}\right)_{\bf k}^\he\right]  .
\end{equation}
Accordingly, the full quantum dipolar Hamiltonian becomes
$\op{H}_\op{D} \equiv \sum_A \op{H}_{\op{D},A}$ with
\begin{equation}\label{eq:HDip_qq}
\op{H}_{\op{D},A}
\equiv \sqrt{6}\left[ \op{1}^\hl \otimes \cdots \otimes \op{T}_{2,0}^\hA \otimes \cdots \otimes  \op{1}^\hN\right]\otimes\op{\Omega}^{(e)}.
\end{equation}

\section{Calculation of functions related with the decoherence} \label{app:calc_Deco_func_1Dk}

This section is dedicated to the calculation of the functions $\gamma(\beta,t)$, $\varepsilon(t)$ and $\zeta_{A,A'}(t)$,  defined in \eqs{eq:Fga_Fep_Fze_boson-spin} and \numeq{eq:Fze_boson-spin}
under the assumptions \ref{assum_latt-or} to \ref{assum_no-optical} listed in \sect{sec:app_calc_tdeco}.

\subsection{System-environment coupling coefficients $g_{\bf k}$} \label{app:sys-env_coup_coef}

We first discuss on the form of $g_{\bf k}$ in the system model represented by assumptions \ref{assum_latt-or} to \ref{assum_no-optical}. Accordingly, we can consider the spin system as a chain of pairs with two phonon normal modes, one acoustic and one optical ($l_{\!M} = 2$); then,
the coupling coefficients \numeq{eq:def_gk_desc} are
\begin{equation}\label{eq:gk_ac_op}
g_{\bf k} \equiv -2\,\csl{u}_{\bf k}\!\times\!
\left\{\begin{array}{l}
\!\ic\,\sin\left(k\,d/2\right)  \equiv g_{{\bf k},a},\;
\text{ acoustic mode with $\hat{\epsilon}_{{\bf k},2} = \hat{\epsilon}_{{\bf k},1} = \hat{d}$}\\
\!\cos\left(k\,d/2\right)  \equiv g_{{\bf k},o},\;
\text{ optical mode with $\hat{\epsilon}_{{\bf k},2} = -\hat{\epsilon}_{{\bf k},1} = \hat{d}$}
\end{array}\right..
\end{equation}
Let us call $\delta_o$ and $\delta_a$ to the contribution of each mode to \eqs{eq:Fga_Fep_Fze_boson-spin} and \numeq{eq:Fze_boson-spin}
\begin{equation}\label{eq:rel_gk_omegak}
	\delta_a \equiv \frac{\abs{g_{{\bf k},a}}^2}{\omega_{{\bf k},a}^2} \simeq \frac{\hbar\,d^2}{4\,\vs^{\,3}\,m_p\,N} \frac{1}{\abs{k}},\qquad
	\delta_o \equiv \frac{\abs{g_{{\bf k},o}}^2}{\omega_{{\bf k},o}^2} \simeq \frac{\hbar}{\omega_o^{\,3}\,m_p\,N},
\end{equation}
where we assume the dispersion relations $\omega_{{\bf k},a} = \vs \abs{k}$ for the acoustic mode and $\omega_{{\bf k},o} = \omega_o$
(i.e., independent of ${\bf k}$) for the optical mode. Then, the contribution of the optical modes may be neglected provided
\begin{equation}\label{eq:rel_ka_op_ac}
\frac{\delta_o}{\delta_a} \ll 1\quad\Rightarrow\quad \abs{k}\csl{a} \ll 2\left(\frac{d}{\csl{a}}\right)^{\!\!2}\left(\frac{\omega_o}{\omega_a}\right)^{\!\!3},
\end{equation}
where $\omega_a \equiv 2\,\vs /\csl{a}$, and $k_{\!M} = \pi/\csl{a}$.
We can obtain an estimate for the ratio $\omega_o/\omega_a$ from Raman and infrared data on the lattice mode frequency spectrum of H$_2$O molecules  \cite{Hass56-Krishnamurthy71-Berenblut71-Iishi79,Seidl69-Sarma98-Anbalagan09}.
Ref.\cite{Zhang16} shows a comprehensive description of the water molecule modes and a calculation of their dispersion relations.
Translation modes, with frequencies in the range $\nu \lesssim 330\un{cm^{-1}}$, present acoustic-like dispersion.
The different molecular motions associated with optical modes with projection on the $\hat{d}$ axis are mainly H--O--H bending and H--O stretching. Their frequency ranges are respectively
$1600\un{cm^{-1}} \lesssim \nu \lesssim 1700\un{cm^{-1}}$ and $3200\un{cm^{-1}} \lesssim \nu \lesssim 3800\un{cm^{-1}}$.
Then, the ratio for these modes is $\omega_o/\omega_a \gtrsim 5$.
Using this data and $d/\csl{a}$ as in assumption \ref{assum_param}, the condition \numeq{eq:rel_ka_op_ac} is valid within the whole $\abs{k}$ range ($\abs{k}\csl{a} \leq \pi$).
There is some evidence
\cite{Hass56-Krishnamurthy71-Berenblut71-Iishi79}
of rotation or libration modes of H$_2$O molecules in gypsum at low frequencies ($380\un{cm^{-1}} \lesssim \nu \lesssim 690\un{cm^{-1}}$); however, they involve displacements that are nearly perpendicular to the $\hat{d}$ axis \cite{Zhang16}, and so we can disregard them in consistence with our model.
Therefore, we can assume $\delta_o/\delta_a \ll 1$ and neglect the optical branch contribution in the calculation of $\gamma(\beta,t)$, $\varepsilon(t)$ and $\zeta_{A,A'}(t)$ in \eqs{eq:Fga_Fep_Fze_boson-spin} and \numeq{eq:Fze_boson-spin}.

\subsection{ Calculation of $\gamma(\beta,t)$, $\varepsilon(t)$ and $\zeta_{A,A'}(t)$} \label{app:aproxcalc_gez}

We can now start with \eqs{eq:Fga_Fep_Fze_boson-spin} and \numeq{eq:Fze_boson-spin} and make the replacements
\begin{itemize}
	\item use $\abs{g_{{\bf k},a}}^2/\omega_{{\bf k},a}^2$ as in \numeq{eq:rel_gk_omegak},
	\item use assumption \ref{assum_N1-inf} (the bath has a dense spectrum) and replace the sums by integrals over $k$ as in \eq{eq:sumk_1D_int},
	\item use the high temperature approximation $\frac{\beta\,\omega_{\bf k}}{2} \ll 1$ in \eq{eq:Fga_boson-spin} and approximate \begin{equation*}\label{eq:aprox_coth}
	\coth\left(\frac{\beta\,\omega_{\bf k}}{2}\right) \simeq \frac{2\,K_B\,T}{\hbar\,\vs}\frac{1}{\abs{k}},
	\end{equation*}
\end{itemize}
to obtain
\begin{subequations}\label{eq:Fga_Fep_Fze_boson-spin_aprox_calc}
	\begin{equation}\label{eq:Fga_boson-spin_aprox_calc}
	\begin{split}
	\gamma(T,t) &\simeq \frac{\hbar\,d^2}{4\,\vs^{\,3}\,m_p\,N}\,\frac{2\,K_B\,T}{\hbar\,\vs}\,
	\frac{N \csl{a}}{2\pi} \int_{-\pi/\csl{a}}^{\pi/\csl{a}}dk\;\frac{\left[1-\cos(\vs\,t\abs{k})\right]}{\abs{k}^2},
	\end{split}
	\end{equation}
	\begin{equation}\label{eq:Fep_boson-spin_aprox_calc}
	\begin{split}
	\varepsilon(t) &\simeq \frac{\hbar\,d^2}{4\,\vs^{\,3}\,m_p\,N}\,\frac{N \csl{a}}{2\pi}
	\int_{-\pi/\csl{a}}^{\pi/\csl{a}}dk\;\frac{\left[\sin(\vs\,t\abs{k})-\vs\,t\abs{k}\right]}{\abs{k}},
	\end{split}
	\end{equation}
	\begin{equation}\label{eq:Fze_boson-spin_aprox_calc}
	\begin{split}
	\zeta_{A,A'}(t) \simeq \frac{\hbar\,d^2}{2\,\vs^{\,3}\,m_p\,N}\,\frac{N \csl{a}}{2\pi}\,\int_{-\pi/\csl{a}}^{\pi/\csl{a}}dk\,\frac{1}{\abs{k}}
	&\big\{\!\cos\left(k\,x_{A,A'}\right)\left[\sin(\vs\,t\abs{k})-\vs\,t\abs{k}\right]\\
	&\;+\sin\left(k\,x_{A,A'}\right)\left[1-\cos(\vs\,t\abs{k})\right]\big\}.
	\end{split}
	\end{equation}
\end{subequations}
In \numeq{eq:Fze_boson-spin_aprox_calc}, we define $x_{A,A'} \equiv r_{A,A'}\cos\left(\theta_{A,A'}\right)$
to  write the scalar product $\vec{k}\cdot\vec{r}_{A,A'}\equiv k\,x_{A,A'}$.
The angle $\theta_{A,A'}$ is measured from $\hat{d}$ to $\vec{r}_{A,A'}$ in the counterclockwise sense.

Notice that the second row of \numeq{eq:Fze_boson-spin_aprox_calc} does not contribute to the integral because it is an odd function of $k$.
On the other hand, the integrands in \eqs{eq:Fga_boson-spin_aprox_calc}, \numeq{eq:Fep_boson-spin_aprox_calc} and the first row of \numeq{eq:Fze_boson-spin_aprox_calc} are even functions of $k$, thus we can change the limits as
$\int_{-\pi/\csl{a}}^{\pi/\csl{a}}dk \rightarrow 2\int_{0}^{\pi/\csl{a}}dk$. Besides, according to assumption \ref{assum_param}, $k_{\!M} \simeq 4\times10^{9}\un{m}^{-1}$, then we can substitute $\int_{0}^{\pi/\csl{a}}dk \rightarrow \int_{0}^{\infty}dk$ whenever the integrands depend as $\abs{k}^{-1}$ or $\abs{k}^{-2}$.
Therefore, we obtain
\begin{subequations}\label{eq:Fga_Fep_Fze_boson-spin_aprox_calc_p}
\begin{equation}\label{eq:Fga_boson-spin_aprox_calc_p}
\begin{split}
\gamma(T,t) \simeq d^2\,\frac{K_B\,T\,\csl{a}}{2\pi\,\vs^{\,4}\,m_p}\,\int_{0}^{\infty}dk\;\frac{\left[1-\cos(\vs\,t\,k)\right]}{k^2},
\end{split}
\end{equation}
\begin{equation}\label{eq:Fep_boson-spin_aprox_calc_p}
\varepsilon(t) \simeq
d^2\,\frac{\hbar\,\csl{a}}{4\pi\,\vs^{\,3}\,m_p}\bigg[\int_{0}^{\infty}dk\;\frac{\sin(\vs\,t\,k)}{k}-\frac{\vs\,\pi}{\csl{a}}\,t\bigg],
\end{equation}
\begin{equation}\label{eq:Fze_boson-spin_aprox_calc_p}
\begin{split}
\zeta_{A,A'}(t) \simeq
d^2\,\frac{\hbar\,\csl{a}}{4\pi\,\vs^{\,3}\,m_p}\,&\bigg\{-2\,\vs\,t\,\int_{0}^{\pi/\csl{a}}dk\,\cos\left(k\,x_{A,A'}\right)\\
&\;+\int_{0}^{\infty}\frac{dk}{k}\big\{\sin\left[k \left(\vs\,t+x_{A,A'}\right)\right]
+\sin\left[k \left(\vs\,t-x_{A,A'}\right)\right]\big\}\bigg\},
\end{split}
\end{equation}
\end{subequations}
(in \eq{eq:Fze_boson-spin_aprox_calc_p} we used the identity for $\sin\left(\psi\right)\cos\left(\phi\right)$).\\

Let us now consider each of Eqs.\numeq{eq:Fga_Fep_Fze_boson-spin_aprox_calc_p},
\begin{description}
\item [\eq{eq:Fga_boson-spin_aprox_calc_p}]
    Using
    \begin{equation}\label{eq:resint_Sinc2}
    \int_{0}^{\infty}dk\,\left[1-\cos(\vs\,t\,k)\right]/k^2 = \pi\,\vs\,t/2
    \end{equation}
    in \numeq{eq:Fga_boson-spin_aprox_calc_p} we obtain the expression
    \[\gamma(T,t) \simeq d^2\,\frac{K_B\,T\,\csl{a}}{4\,\vs^{\,3}\,m_p}\,t.\]

\item [\eq{eq:Fep_boson-spin_aprox_calc_p}]
    Since
    \begin{equation}\label{eq:resint_Sinc}
    \int_{0}^{\infty}dk\,\frac{\sin(\alpha\,k)}{k} = \frac{\pi}{2}\,\sgn(\alpha),
    \end{equation}
    with $\alpha \in \cbb{R}$ and the sign function
    \begin{equation*}\label{eq:def_func_sgn}
    \sgn(\alpha) \equiv \left\{
                         \begin{array}{c}
                         \;\;1,\;\alpha > 0\\
                         \;\;0,\;\alpha = 0\\
                            -1,\;\alpha < 0\\
                         \end{array}
                       \right.,
    \end{equation*}
    the integral in the first term of \eq{eq:Fep_boson-spin_aprox_calc_p}
    takes the  value $\pi/2$ immediately for $t \neq 0$, then, it contributes only a constant phase to $\Delta\sigma_{\mA,\nA}(t) $. It is the second term which defines the linear dependence on $t$ provided
    \begin{equation}\label{eq:t_despint_Fep}
    \frac{\vs\,\pi}{\csl{a}}\,t \gg \frac{\pi}{2}\,\Rightarrow\,t \gg \frac{\csl{a}}{2\,\vs} \simeq 9\times10^{-14}\un{s},
    \end{equation}
    where we used $\csl{a}$ and $\vs$ as in assumption \ref{assum_param}.
    Therefore, in a time scale experimentally accesible to NMR ($t > 1\un{\mu s}$), we can write
    \[\varepsilon(t) \simeq -d^2\,\frac{\hbar}{4\,\vs^{\,2}\,m_p}\,t.\]

\item [\eq{eq:Fze_boson-spin_aprox_calc_p}]
    In the first term of \numeq{eq:Fze_boson-spin_aprox_calc_p} we have
    \begin{equation}\label{eq:resint_Fze_Cos}
    \int_{0}^{\pi/\csl{a}}dk\,\cos\left(k\,x_{A,A'}\right)
    = \frac{\pi}{\csl{a}}\,\frac{\sin\left[\pi\,x_{A,A'}/\csl{a}\,\right]}{\pi\,x_{A,A'}/\csl{a}}
    \equiv \frac{\pi}{\csl{a}}\,\opc{S}_{A,A'},
    \end{equation}
    where $\opc{S}_{A,A'}$ is also defined in \numeq{eq:Sinc_SAAp}.

    The second term of \numeq{eq:Fze_boson-spin_aprox_calc_p} can be analyzed using \numeq{eq:resint_Sinc} for the integral
    \begin{equation}\label{eq:resint_Fze}
    \int_{0}^{\infty}\frac{dk}{k}\,\sin\left[k \left(\vs\,t \pm x_{A,A'}\right)\right] = \frac{\pi}{2}\,\sgn\left(\vs\,t \pm x_{A,A'}\right),
    \end{equation}
    which can take the values $0$ or $\pm\pi/2$, depending on the relative position of  $A$ and $A'$, and on time $t$.
    We find convenient to define the function
    \begin{equation}\label{eq:def_vphi}
    \varphi_{A,A'}(t) \equiv \frac{\pi}{2}\left[\sgn\left(\vs\,t+x_{A,A'}\right) + \sgn\left(\vs\,t-x_{A,A'}\right)\right],
    \end{equation}
    which is $0$ for $t = 0$, while its dependence on $x_{A,A'}$ is shown in \fig{fig:Fig_5} for $t \neq 0$. There we can see that only the pairs $A'$ satisfying $\abs{x_{A,A'}} < \vs\,t$ contribute to $\zeta_{A,A'}(t)$ with a value of $\varphi_{A,A'}(t) = \pi$, and its value is $\pi/2$ at $x_{A,A'} = \pm \vs\,t$.

    Finally, the function $\zeta_{A,A'}(t)$ can be expressed as
    \[\zeta_{A,A'}(t) \simeq d^2\,\frac{\hbar\,\csl{a}}{2\,\vs^{\,3}\,m_p}\,\left[\frac{\varphi_{A,A'}(t)}{2\pi}-\frac{\vs}{\csl{a}}\,t\,\opc{S}_{A,A'}\right].\]

\end{description}

\insertfig{ht}{fig:Fig_5}{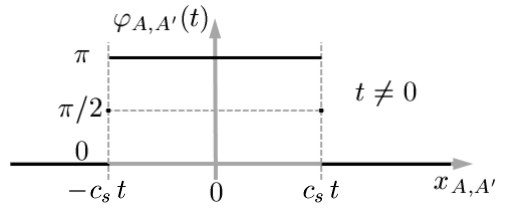}{6.75}{2.8}{Function $\varphi_{A,A'}(t)$ vs. the distance $x_{A,A'}$ for $t \neq 0$.}

\section{Distribution of eigenvalues} \label{app:dist_eig}

This appendix describes the details involved in writing the probability density function $p(X)$ in the continuum limit as a Gaussian function. Though we are not strictly dealing with random variables but with all the possible arrays of $N$ eigenvalues, the density function we seek has the same role than the probability density, and we can refer indistinguishably to one or the other.
	
The discrete variable \mbox{$X= \sum_i^N \kappa_{i}$}, depends on the distribution of eigenvalues of the $N$ equivalent partition elements.
According to \numeq{eq:def_mA_kappaA}, $\kappa_{i}$ can take the values $\{1,0,-2\}$ with multiplicities  $\tilde{\alpha}_{\kappa}= \{2,1,1\}$, respectively.
Then, the density function for each partition element is
\begin{equation*}\label{eq:def_dens_kappa}
p_{\kappa}(\tilde{\kappa}) \equiv \frac{1}{\opc{N}_{\kappa}} \sum_{\kappa} \tilde{\alpha}_{\kappa}\,\delta\left(\tilde{\kappa}-\kappa\right),
\end{equation*}
where the sum runs over all the discrete values of $\kappa$, and $\opc{N}_{\kappa} = 4$ in our case, $\delta\left(\cdot\right)$ is the Dirac delta function.
In order to write the density function of the variable $X$, let us first call $n_1, n_0, n_{-2}$ to the number of partition elements with eigenvalues 1, 0, or -2, and rewrite
\begin{equation}\label{eq:def_X}
X \equiv \sum_{i=1}^N \kappa_i = n_1 - 2\,n_{-2} = 3\,n_1 + 2\,n_0 - 2\,N,
\end{equation}
with $N= n_1+ n_0+ n_{-2}$.  From \numeq{eq:def_X} we note that $-2 N \leq X \leq N$.
In this way, the number of configurations with fixed $n_1, n_0$ is
\begin{equation*}\label{eq:def_alpha_n0n1}
\alpha_{n_0,n_1} \equiv 2^{n_1}\frac{N!}{n_0!\,n_1!\,\left(N-n_0-n_1\right)!}.
\end{equation*}
Therefore, the density function for the variable $X$ is
\begin{equation}\label{eq:def_dens_X_N}
p^{(N)}_x(X) \equiv \frac{1}{\opc{N}_x}\sum_{\{n_0,n_1\} / X} \alpha_{n_0,n_1},
\end{equation}
where the sum runs over all the possible $n_0$ and $n_1$ that satisfy \eq {eq:def_X} for any value of $X$,
and $\opc{N}_x = 4^N$ is the total number of configurations.
In the limit of large $N$ the density function of a variable like our $X$, which is the sum of independent variables $\kappa_{i}$, tends to a  Gaussian function (central limit theorem)
\cite{Reichl_2Ed_ch4-LeCam86}
\begin{equation*}\label{eq:dens_X_Gauss}
p_x(X) = \frac{1}{\sqrt{2\pi}\sigma_x}\;e^{-\left(X - \overline{X}\,\right)^2/(2\,\sigma_x^2)},
\end{equation*}
where the mean value is $\overline{X} \equiv N \langle\kappa\rangle$ and the standard deviation is $\sigma_x \equiv \sqrt{N} \sqrt{\langle\kappa^2\rangle - \langle\kappa\rangle^{\,2}}$, with the $q$-th moments
\begin{equation*}
\langle\kappa^q\rangle \equiv \int_{-\infty}^{\infty}\!\! d\tilde{\kappa}\;\tilde{\kappa}^q\,p_{\kappa}(\tilde{\kappa})
= \frac{1}{\opc{N}_{\kappa}} \sum_{\kappa}\tilde{\alpha}_{\kappa}\,\kappa^q.
\end{equation*}
Using the eigenvalues of our example, we have
$\langle\kappa\rangle = 0$ and $\langle\kappa^2\rangle = 3/2$, thus
\[\overline{X} = 0 \qquad {\rm and} \qquad \sigma_x \equiv \sqrt{3N/2}.\]

\section{Reduced density matrix in the ME sequence} \label{app:rev_p-p_model}

In this section we calculate elements of the reduced density matrix
\begin{equation}\label{eq:sigma_red_SA_B}
\sigma^{\sB\hA}(t) = \trs{\oS_A}{\rho^{\sB\hs}(t)}
\end{equation}
after the ME sequence, applied to the pair-phonon model.

Using $\rho_{m,n}^\sB(t)$ from \eq{eq:rho_sist_red_Ud_B} in the reduction of \numeq{eq:sigma_red_SA_B} over the complementary space $\shSAn$, we have
\begin{equation}\label{eq:rho_sist_red_SA_ad_B}
\sigma^\sB_{\mA,\nA}(t)  = \sum_{\{\!\mAn\!\}}\left[\rho_{m,n}(0)\;S^\sB_{m,n}(t)\right]\!\!\Cmn,
\end{equation}
where
\begin{equation}\label{eq:Func_Smn_B}
S^\sB_{m,n}(t) \equiv \prod_{\bf{k}} S_{m,n}^{\sB\,\bf{k}}(t)
= e^{-\Gamma^\sB_{m,n}(t)}\,e^{-\ic\Upsilon^\sB_{m,n}(t)},
\end{equation}
with
\begin{equation}\label{eq:def_Gmn_Umn_B}
\Gamma^\sB_{m,n}(t) \equiv \sum_{\bf{k}}\Gamma_{m,n}^{\sB\,\bf{k}}(t),\quad\text{and}\quad
\Upsilon^\sB_{m,n}(t) \equiv \sum_{\bf{k}}\Upsilon_{m,n}^{\sB\,\bf{k}}(t).
\end{equation}
As in \sect{sec:app_sigma}, we need the deviation density matrix elements $\Delta\sigma^\sB_{\mA,\nA}(t)$ to calculate the expectation value of local observables in NMR experiments.\\

We can now use the relations \numeq{eq:Func_eigenval_GU} in \numeq{eq:Func_Gmnkl_Umnkl_delta}, with the eigenvalues of the pair-phonon model from \eq{eq:def_lambda-kappa} $\lambda_{\mA} = -\frac{3}{2}\,\frac{\Omega_0(d\,)}{d}\,\kappa_{\mA}$, and the coupling constants $g_{\bf k}^{A} = e^{-\ic\,\vec{k}\cdot\vec{r}_{A}}\,g_{\bf k}$ (from \eq{eq:def_gk_A}) to write
\begin{subequations}\label{eq:Gmn_Umn_boson-spin_B}
	\begin{equation}\label{eq:Gmn_boson-spin_B}
	\Gamma^\sB_{m,n}(t)\Cmn =
	\frac{9}{4}\left(\frac{\Omega_0(d\,)}{d}\right)^2\left(\kappa_{\mA} - \kappa_{\nA}\right)^2\gamma^\sB(\beta,t),
	\end{equation}
	\begin{equation}\label{eq:Umn_boson-spin_B}
	\Upsilon^\sB_{m,n}(t)\Cmn = \frac{9}{4}\left(\frac{\Omega_0(d\,)}{d}\right)^2
	\bigg[\!\left(\kappa_{\mA}^2 - \kappa_{\nA}^2\right) \varepsilon^\sB(t)
	+\left(\kappa_{\mA} - \kappa_{\nA}\right) \widehat{\chi}_{A}^{\,\sB\,\{\!\mAn\!\}}(t)\bigg],
	\end{equation}
\end{subequations}
with
\begin{subequations}\label{eq:Fga_Fep_Fze_boson-spin_B}
	\begin{align}
	\gamma^\sB(\beta,t) &\equiv \sum_{\bf k}\frac{\abs{g_{\bf k}}^2}{\omega_{\bf k}^2}\,
	\coth\left(\beta\,\omega_{\bf k}/2\right)\,\FC^{\bf{k}}_\sB(t),\label{eq:Fga_boson-spin_B}\\
	\varepsilon^\sB(t) &\equiv \sum_{\bf k}  \frac{\abs{g_{\bf k}}^2}{\omega_{\bf k}^2}
	\left[\FS^{\bf{k}}_\sB(t) - \omega_{\bf k}\,t/2 \right],\label{eq:Fep_boson-spin_B}\\
	\widehat{\chi}_{A}^{\,\sB\,\{\!\mAn\!\}}(t) &\equiv \sum_{A' \neq A}\kappa_{\mAp}\,\zeta^\sB_{A,A'}(t),\label{eq:Fchi_boson-spin_B}
	\end{align}
\end{subequations}	
where $\FC^{\bf{k}}_\sB(t)$ and $\FS^{\bf{k}}_\sB(t)$ have the same expressions of \eqs{eq:Func_Cdelta_Sdelta_B}, replacing
$\omega \rightarrow \omega_{\bf k}$, and similarly to \eq{eq:Fze_boson-spin},   $\zeta^\sB_{A,A'}(t)$ adopts the form
\begin{equation}\label{eq:Fze_boson-spin_B}
\zeta^\sB_{A,A'}(t) \equiv 2 \sum_{\bf k} \frac{\abs{g_{\bf k}}^2}{\omega_{\bf k}^2} \left\{\cos\left(\vec{k}\cdot\vec{r}_{A,A'}\right)
\left[\FS^{\bf{k}}_\sB(t) - \omega_{\bf k}\,t/2\right]\right.\!
+ \!\left.\sin\left(\vec{k}\cdot\vec{r}_{A,A'}\right)\FC^{\bf{k}}_\sB(t)\right\}.
\end{equation}

To calculate the sums over the wavenumber vector in \numeq{eq:Fga_Fep_Fze_boson-spin_B} and \numeq{eq:Fze_boson-spin_B} we use assumptions \ref{assum_latt-or} to \ref{assum_no-optical} listed in \sect{sec:app_calc_tdeco}. Hence, following the same procedure that brings \eqs{eq:Fga_Fep_Fze_boson-spin_aprox_calc} to \eqs{eq:Fga_Fep_Fze_boson-spin_aprox_calc_p}
($\int_{-\pi/\csl{a}}^{\pi/\csl{a}}dk \rightarrow 2\int_{0}^{\pi/\csl{a}}dk$, for even integrands; and
$\int_{0}^{\pi/\csl{a}}dk \rightarrow \int_{0}^{\infty}dk$, whenever the integrands depend as $\abs{k}^{-1}$ or $\abs{k}^{-2}$), we have
\begin{subequations}\label{eq:Fga_Fep_Fze_boson-spin_aprox_calc_p_B}
	\begin{equation}\label{eq:Fga_boson-spin_aprox_calc_p_B}
	\begin{split}
	\gamma^\sB(T,t) \simeq d^2\,\frac{K_B\,T\,\csl{a}}{2\pi\,\vs^{\,4}\,m_p}\,\int_{0}^{\infty}dk\;\frac{\FC^{k}_\sB(t)}{k^2},
	\end{split}
	\end{equation}
	\begin{equation}\label{eq:Fep_boson-spin_aprox_calc_p_B}
	\varepsilon^\sB(t) \simeq
	d^2\,\frac{\hbar\,\csl{a}}{4\pi\,\vs^{\,3}\,m_p}\bigg[\int_{0}^{\infty}dk\;\frac{\FS^{k}_\sB(t)}{k}-\frac{\vs\,\pi}{\csl{a}}\,\frac{t}{2}\bigg],
	\end{equation}
	\begin{equation}\label{eq:Fze_boson-spin_aprox_calc_p_B}
	\begin{split}
	\zeta^\sB_{A,A'}(t) \simeq
	d^2\,\frac{\hbar\,\csl{a}}{4\pi\,\vs^{\,3}\,m_p}\,&\bigg[-\vs\,t\,\int_{0}^{\pi/\csl{a}}dk\,\cos\left(k\,x_{A,A'}\right)\\
	&\qquad+ 2\int_{0}^{\infty}\frac{dk}{k}\,\cos\left(k\,x_{A,A'}\right)\,\FS^{k}_\sB(t)\bigg],
	\end{split}
	\end{equation}
\end{subequations}
where
\begin{subequations}\label{eq:Func_Cdelta_Sdelta_Bk}
\begin{equation}\label{eq:Func_Cdelta_Bk}
\begin{split}
    \FC^{k}_\sB(t) =  \frac{3}{2}\left[1-\cos\left(\vs\,t\abs{k}/3\right)\right] + \frac{3}{4}\left[1-\cos\left(2\,\vs\,t\abs{k}/3\right)\right]
    - \frac{1}{2}\left[1-\cos\left(\vs\,t\abs{k}\right)\right],
\end{split}
\end{equation}
\begin{equation}\label{eq:Func_Sdelta_Bk}
\begin{split}
    \FS^{k}_\sB(t) = \frac{3}{2} \sin\left(\vs\,t\abs{k}/3\right) + \frac{3}{4}\sin\left(2\,\vs\,t\abs{k}/3\right)
    - \frac{1}{2}\sin\left(\vs\,t\abs{k}\right).
\end{split}
\end{equation}
\end{subequations}
To solve the integrals in \eqs{eq:Fga_Fep_Fze_boson-spin_aprox_calc_p_B}, we again follow the procedure of \apen{app:aproxcalc_gez}:
\begin{itemize}
\item 
    Using \eq{eq:resint_Sinc2} we write $\int_{0}^{\infty}dk\;\FC^{k}_\sB(t)/k^2 = \pi\,\vs\,t/4$, then
    \begin{equation}\label{eq:Fga_boson-spin_fin_B}
    \gamma^\sB(T,t) \simeq d^2\,\frac{K_B\,T\,\csl{a}}{4\,\vs^{\,3}\,m_p}\,\frac{t}{2}.
    \end{equation}

\item 
    Equation \numeq{eq:resint_Sinc} allows writing
    \[\int_{0}^{\infty}dk\;\frac{\FS^{k}_\sB(t)}{k} = \frac{\pi}{2}\!\left[\frac{3}{2}\,\sgn\left(\frac{\vs\,t}{3}\right)
    \!+\!\frac{3}{4}\,\sgn\left(\frac{2\,\vs\,t}{3}\right)\!-\!\frac{1}{2}\,\sgn\left(\vs\,t\right)\right] \!=\!
    \left\{\begin{array}{l}
        \!\!0,\; t=0. \\
        \!\!7\pi/8,\;  t\neq 0.
    \end{array}\right.\]
    Therefore, if $\;\vs\,\pi\,t/(2\,\csl{a}) \gg 7\pi/8 \;\Rightarrow\; t \gg 7\csl{a}/(4\,\vs) \simeq 3.2\times10^{-13}\un{s}$, we can write
    \begin{equation}\label{eq:Fep_boson-spin_fin_B}
    \varepsilon^\sB(t) \simeq -d^2\,\frac{\hbar}{4\,\vs^{\,2}\,m_p}\,\frac{t}{2}.
    \end{equation}

\item 
    Using \numeq{eq:resint_Fze}, we may write
    \[2\int_{0}^{\infty}\frac{dk}{k}\,\cos\left(k\,x_{A,A'}\right)\,\FS^{k}_\sB(t) = \sum_{n=1}^3 j_n\,\varphi^{(n)}_{A,A'}(t)
    \equiv \varphi^\sB_{A,A'}(t),\]
    where we define
    \[\varphi^{(n)}_{A,A'}(t) \equiv \frac{\pi}{2}\left[\sgn\left(n\,\vs\,t/3+x_{A,A'}\right) + \sgn\left(n\,\vs\,t/3-x_{A,A'}\right)\right],\]
    with $j_1 = 3/2$, $j_2 = 3/4$, $j_3 = -1/2$.
    The function $\varphi^{(n)}_{A,A'}(t)$ is similar to \numeq{eq:def_vphi} with the velocity $\vs$ replaced by $n\,\vs/3$, and the same behaviour shown in \fig{fig:Fig_5}. Using this result and \numeq{eq:resint_Fze_Cos} we get
    \begin{equation}\label{eq:Fze_boson-spin_fin_B}
    \zeta^\sB_{A,A'}(t) \simeq d^2\,\frac{\hbar\,\csl{a}}{2\,\vs^{\,3}\,m_p}\,\left[\frac{\varphi^\sB_{A,A'}(t)}{2\pi}-\frac{\vs}{\csl{a}}\,\frac{t}{2}\,\opc{S}_{A,A'}\right],
    \end{equation}
    and introducing \numeq{eq:Fze_boson-spin_fin_B} in \eq{eq:Fchi_boson-spin_B} we may write
    \begin{equation}\label{eq:Chihat_boson-spin_aprox_B}
    \widehat{\chi}_{A}^{\,\sB\,\{\!\mAn\!\}}(t) \simeq d^2\,\frac{\hbar\,\csl{a}}{2\,\vs^{\,3}\,m_p}
    \left[\frac{1}{2}\,X_{A}^{\sB\,\prime}(t) - \frac{\vs}{\csl{a}}\,\frac{t}{2}\,X_{A} \right],
    \end{equation}
    where $X_{A}$ has the same expression as \eq{eq:X_A} and we defined
    \begin{equation}\label{eq:XXp_A_B}
    X_{A}^{\sB\,\prime}(t) \equiv \sum_{A' \neq A}\kappa_{\mAp}\,\varphi^\sB_{A,A'}(t)/\pi.
    \end{equation}
    It is worth to note that $\varphi^\sB_{A,A'}$ satisfy $-1/2 \leq \varphi^\sB_{A,A'}(t)/\pi \leq 7/4$ (thus its maximum absolute value is $7/4 = 1.75$).

\end{itemize}

Finally, we use \eqs{eq:Fga_boson-spin_fin_B}, \numeq{eq:Fep_boson-spin_fin_B} and \numeq{eq:Chihat_boson-spin_aprox_B}, in \eqs{eq:Gmn_Umn_boson-spin_B} and  follow the same procedure leading to \numeq{eq:rho_boson-spin} in \sect{sec:app_calc_tdeco} to write
\begin{equation}\label{eq:Dsig_boson-spin_B}
\begin{split}
\Delta\sigma^\sB_{\mA,\nA}(t) &= \Delta\sigma_{\mA,\nA}(0)\,
e^{-\left(\kappa_{\mA} - \kappa_{\nA}\right)^2 \frac{t}{2\,\tau_{\gamma}}}\,
e^{\,\ic\,2\pi\nuD\,\left(\kappa_{\mA}^2 - \kappa_{\nA}^2\right)\frac{t}{2}}\,G_A^{\,\sB\,\prime}(t)\,G_A(t/2),
\end{split}
\end{equation}
with the specific decoherence functions
\begin{subequations}\label{eq:fdeco_GAp_GA_B}
\begin{equation}\label{eq:fdeco_GAp_B}
G_A^{\,\sB\,\prime}(t) \equiv \int dX_{A}^{\sB\,\prime}(t)\;p^{\prime}\big(X_{A}^{\sB\,\prime}(t)\big)\;
e^{-\ic\,2\pi \nuD\,\left(\kappa_{\mA} - \kappa_{\nA}\right)\,\frac{\csl{a}}{\vs}\,X_{A}^{\sB\,\prime}(t)},
\end{equation}
\begin{equation}\label{eq:fdeco_GA_B}
G_A(t/2) \equiv \int dX_{A}\;p(X_{A})\;e^{\,\ic\,4\pi\nuD\,\left(\kappa_{\mA} - \kappa_{\nA}\right)\,\frac{t}{2}\,X_{A}},
\end{equation}
\end{subequations}
with $\nuD$ and $\tau_{\gamma}$ as in \eqs{eq:def_nu_vep} and \numeq{eq:def_nu-tau}, respectively.
We can see that $G_A$ in \numeq{eq:fdeco_GA_B} is the same as \eq{eq:fdeco_GA}, valued at half of the total time.
Evaluating parameters in the decoherence functions as in \sect{sec:app_calc_tdeco} and solving the integrals \numeq{eq:fdeco_GAp_GA_B} similarly to \eq{eq:rho_boson-spin}, we can also conclude that the relevant decoherence factor in \numeq{eq:Dsig_boson-spin_B} is given by the function $G_A(t/2)$.
This allows writing the reduced density matrix elements in the ME sequence as
\begin{equation*} 
\Delta\sigma^\sB_{\mA,\nA}(t) \simeq \Delta\sigma_{\mA,\nA}(0)\;e^{-\left[\left(\kappa_{\mA} - \,\kappa_{\nA}\right)\,t/\tau^\sB_{X}\right]^2},
\end{equation*}
with $\tau^\sB_{X} \equiv 2\,\tau_{X}$ and $\tau_{X}$ defined by \eq{eq:tau_chi} (or \eq{eq:tau_chi_inv}).


\end{document}